\documentclass[a4paper,12pt,onehalfspacing]{article}
\usepackage{fancyhdr}

\usepackage[a4paper, left=2.7cm, right=2.7cm, top=3cm]{geometry}

\fancypagestyle{plain}{ 
  \fancyhf{}

  \cfoot{\thepage}
}

\usepackage[utf8]{inputenc}
\usepackage[british]{babel}
\usepackage{csquotes}

\usepackage[pdftex]{graphicx}
\usepackage[
    backend=biber,
    style=chem-angew, articletitle=true,
    maxbibnames=99,
   ]{biblatex}
\let\cite\parencite
\usepackage{hyperref}

\usepackage[format=plain,labelsep=period,bf]{caption}

\usepackage{multirow}
\usepackage{tablefootnote}
\usepackage{threeparttable}
\usepackage{setspace}

\usepackage{amsmath}
\usepackage{soul}

\DeclareFieldFormat[article]{number}{(#1)}

\renewbibmacro*{journal+issuetitle}{%
  \usebibmacro{journal}%
  \setunit*{\addspace}%
  \iffieldundef{series}
    {}
    {\newunit
     \printfield{series}%
     \setunit{\addspace}}%
  \usebibmacro{date}%
  \newunit 
  \printfield{volume}%
  \setunit*{\addspace}%
  \printfield{number}%
  \newunit
}

\DeclareBibliographyDriver{article}{%
  \usebibmacro{bibindex}%
  \usebibmacro{begentry}%
  \usebibmacro{author/translator+others}%
  \setunit{\labelnamepunct}\newblock
  \iftoggle{bbx:articletitle}
    {%
      \usebibmacro{title}%
      \newunit
    }
    {}%
  \usebibmacro{byauthor}%
  \newunit\newblock
  \usebibmacro{bytranslator+others}%
  \newunit\newblock
  \printfield{version}%
  \newunit\newblock
  \usebibmacro{journal+issuetitle}%
  \newunit
  \usebibmacro{byeditor+others}%
  \newunit
  \usebibmacro{note+pages}%
  \newunit\newblock
  \iftoggle{bbx:isbn}
    {\printfield{issn}}
    {}%
  \newunit\newblock
  \usebibmacro{doi+eprint+url}
  \newunit\newblock
  \usebibmacro{addendum+pubstate}%
  \setunit{\bibpagerefpunct}\newblock
  \usebibmacro{pageref}%
  \newunit\newblock
  \usebibmacro{related}%
  \usebibmacro{finentry}%
} 

\begin{document}
\begingroup

\begin{titlepage}

    \begin{figure}[t!]
        \centering
        \includegraphics[scale=0.25]{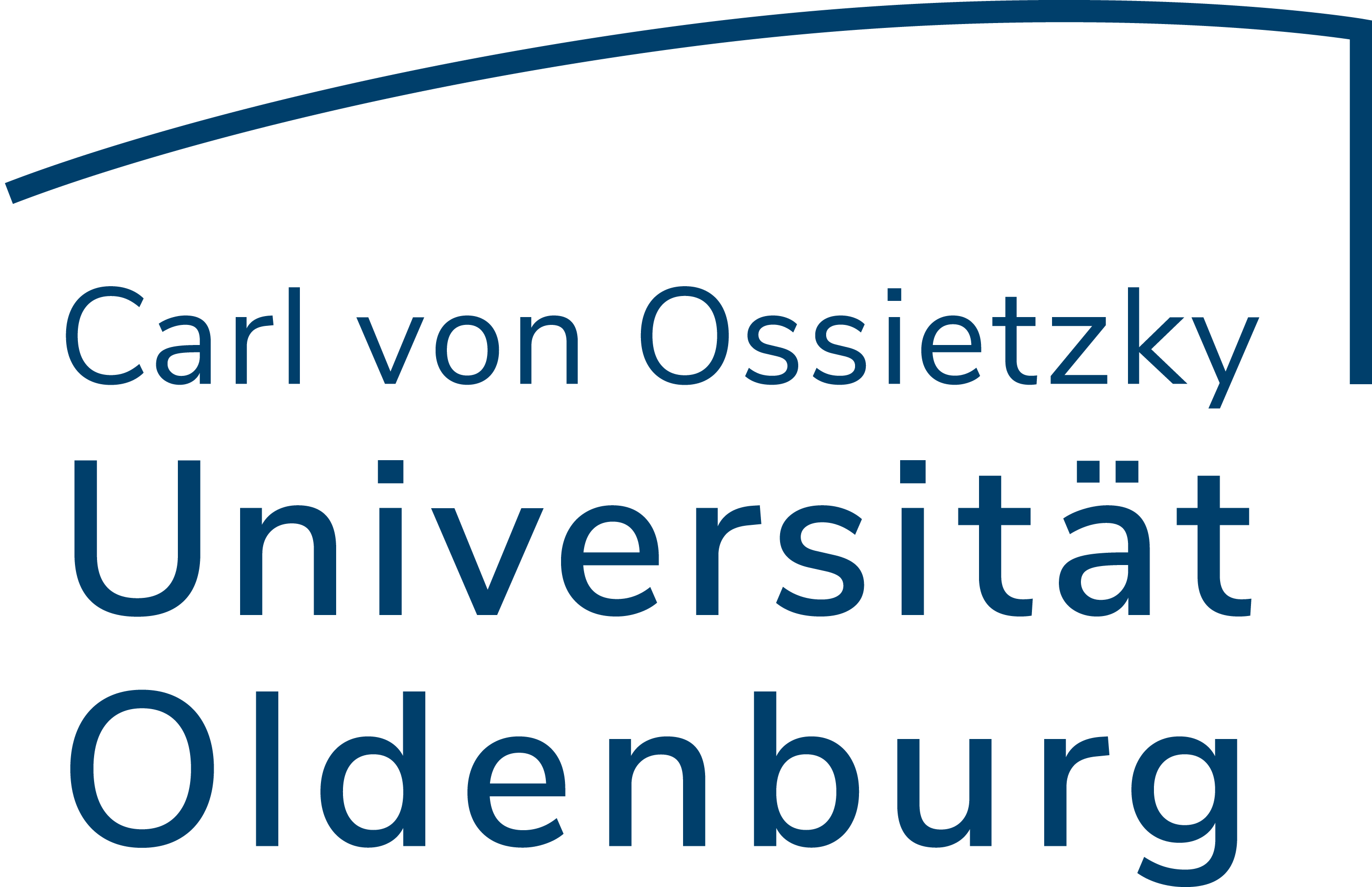}
        \label{fig:my_label}
    \end{figure}
    
    \begin{center}
        Institut für Chemie \\
        \vspace{1cm}
        Bachelorstudiengang Chemie \\
        \vspace{1cm}
        \large\textbf{Bachelorarbeit}\\
        \vspace{1cm}
        \textbf{Calculation of the magnetic properties of quarternary ThMn$_{12}$ - type compounds with Zr as a substitution for Nd}\\
        \vspace{1cm}
        \normalsize
        vorgelegt von: Nico Yannik Merkt\\
        \vspace{1cm}
        Matrikelnummer: 3994107\\
        \vspace{5cm}
        Betreuender Gutachter: Herr Prof. Dr. Thorsten Klüner\\
        \vspace{1cm}
        Zweitgutachter: Herr Dr.-Ing. Halil İbrahim Sözen\\
        \vspace{1cm}
        Oldenburg, 18.10.2023
    \end{center}

\end{titlepage}
\endgroup
\thispagestyle{plain}
\setcounter{page}{2}
\tableofcontents

\newpage
\thispagestyle{plain}
\listoffigures

\newpage
\thispagestyle{plain}
\listoftables

\newpage
\thispagestyle{plain}
\addcontentsline{toc}{section}{Acknowledgements/Danksagung}
\paragraph{\Large{Acknowledgements/Danksagung}}\mbox{}\vspace{0.5cm}\\

    After finishing this work my special thanks go to Prof. Dr. Thorsten Klüner for the fact that I was able to write this thesis in his working group. Furthermore, I appreciate Stephan Erdmann for his help in dealing with the program package VASP, his support with the calculations and the excellent supervision. I would also like to express my gratitude to Dr. Ing. Halil İbrahim Sözen for his supervision and control as second examiner in this thesis. I would also like to thank my family who supported and helped me during my studies. Additionally, I would like to thank in general the working group of Theoretical Chemistry in Oldenburg for their willingness to help with questions and their good reception in the working group. Finally, I would like to appreciate the team of the HPC cluster CARL of the University of Oldenburg and Robert Röhse for their technical help.
\vspace{1cm}

    Mein besonderer Dank gilt Herrn Prof. Dr. Thorsten Klüner dafür, dass ich diese Abschlussarbeit in seinem Arbeitskreis anfertigen konnte. Darüber hinaus möchte ich besonders Stephan Erdmann für seine Hilfe im Umgang mit dem Programmpaket VASP, seiner Unterstützung bei den Berechnungen und für seine hervorragende Betreuung bedanken. Mein Dank gilt dazu auch Dr. Ing. Halil İbrahim Sözen für die Betreuung und der Kontrolle als Zweitprüfer dieser Arbeit. Ebenfalls danke ich meiner Familie, welche mich während des Studiums unterstützt und mir geholfen hat. Auch möchte ich der Arbeitsgruppe der Theoretischen Chemie in Oldenburg für die Bereitschaft zur Hilfe bei Fragen und für die gute Aufnahme im Arbeitskreis danken. Zuletzt möchte ich mich beim Team des HPC-Clusters CARL der Universität Oldenburg und Robert Röhse für die technische Hilfe bedanken.

\newpage
\thispagestyle{plain}
\section{Introduction to Permanent Magnets and Magnetism} \label{1}
    Magnetic materials are substances that have the ability to generate or respond to a magnetic fields. Permanent magnets are components in electric motors and renewable energy technologies, and therefore contributed much to develop these. As a result of their efficient transmission of electrical power, these magnetic materials have become integral parts of developed technologies, such as electric motors, generators, transformers, and wind turbines. As a result of climate change, the demand for these materials is steadily increasing. Therefore, more and more financial and technical efforts are put into their development \cite{1, 2}.

    A lot of this development has focused on the improvement of hard magnetic materials, which are often composed of rare-earth (RE) and transition metal (TM) elements. Due to the resource criticality of RE elements, the focus of research has recently shifted to the reduction of rare-earth elements. This makes the development of new RE-lean or RE-free hard magnets necessary \cite{2, 3}.
    \vspace{0.25cm}

    The first part of this chapter discusses the historical development of magnetic materials and their impact on the world development. After that the current situation on the global market for hard magnetic materials and their most important applications will be given. Several fundamental and basic magnetic properties, such as the magnetic moment \textit{m$_{tot}$}, the magnetisation saturation \textit{M$_{S}$}, the maximum energy product $|BH|_{max}$, the Curie temperature \textit{T$_{C}$}, and the magnetocrystalline anisotropy energy (MAE), will be explained in the last section of chapter 1 to provide readers with a general understanding of the subject matter. 

\subsection{Historical Development of Magnets} \label{1.1}
    The magnet has been known to human beings since the Iron Age \cite{17}. The magnets known at that time were called lodestones. These magnets are pieces of iron magnetised by the influence of large electrical currents, caused by lightning strikes. The first documented usage was described in ancient China and Greece between the 4$^{th}$ and 6$^{th}$ century \cite{17}. The initial devices to integrate magnets were the compass and the "South Pointer". The "South Pointer" was a Iodestone carved into a spoon, that rotates on a small base to align with the Earth's magnetic field. In China, the "South Pointer" was a tool for the planing of grid-like maps of cities. The compass on the other hand was first discovered as a navigational tool by Shen Kua in 1088 \cite{4} and proved itself useful. In Europe this happened under Alexander Neckham between 1157 and 1217 \cite{5}. These successful usages of the compass include the discovery of America by Christopher Columbus in 1492 and the first circumnavigation of the world by Ferdinand Magellan between 1519 and 1522 \cite{4}. 
    
    In the Middle Ages, magnets had a bad reputation due to superstitions. Nevertheless, many predictions have been made during this time that have been proven true or false over time, such as perpetual motion (proven false) or magnetic levitation (proven true). The understanding of magnets by humankind expanded greatly during this period. 
    
    In 1269, when the Frenchman Petrus Peregrinus de Maricourt examined the floating compass, he discovered the magnetic lines and the poles of a magnet. According to the monograph \textit{De Magnete} by William Gilbert in 1600, the earth is also a great magnet \cite{5}. This is the reason for the alignment of freely rotating magnets.
    
    Later in 1820, Hans-Christian Oerstedt discovered the connection between electricity and magnetism. These two phenomena are related because they are the result of the movement of electrons. This led Michael Faraday to the discovery of the phenomenon of magnetic induction in 1821 and the magneto-optical Faraday effect in 1845, revealing the connection between light and magnetism. Faraday's and Oerstedt's discoveries later inspired James Clerk Maxwell to write the Maxwell's equations, named after him in the mid-19$^{th}$ century. Maxwell's equations still serve as the theoretical basis for the future development of hard magnets up to today \cite{4,5}.
    
\begin{figure}[h]
    \centering
    \includegraphics[scale=0.59]{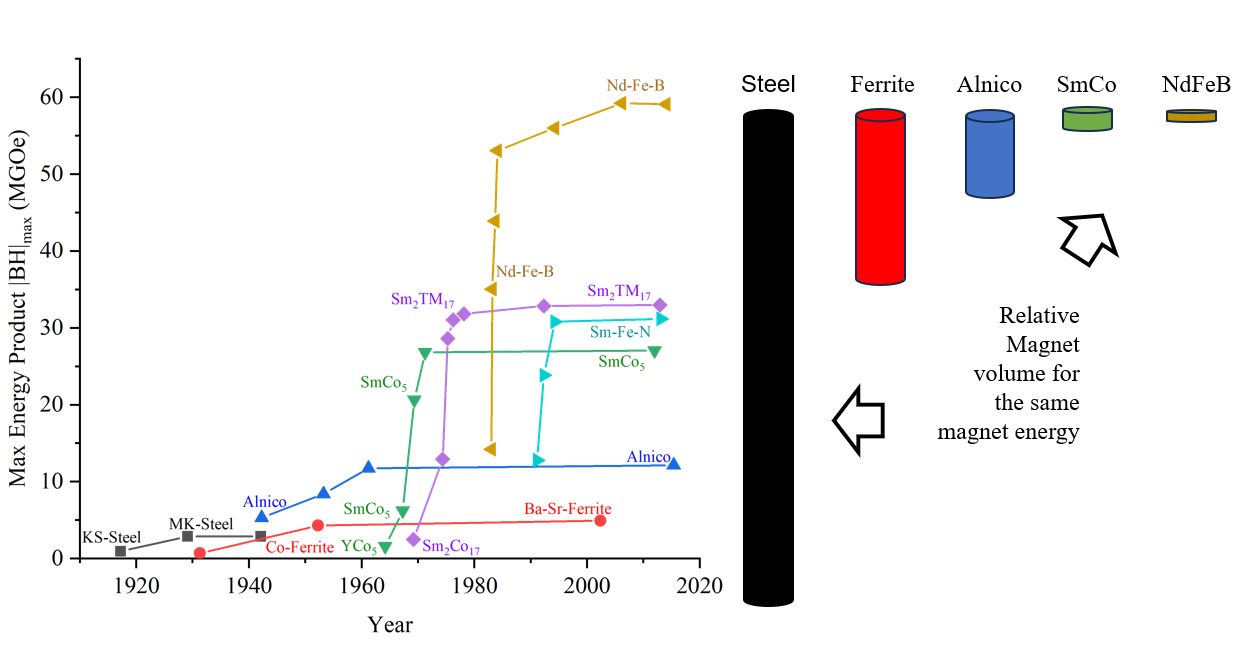}
    \caption{Development of the different types of magnets in the 20$^{th}$ century and the beginning of the 21$^{st}$ century. The origin of the information presented is the source \cite{fig6}.}
    \label{fig:6}
\end{figure}

    Due to limitations in shape, which were solved in 1951 with the discovery of ferromagnetic hexagonal ferrites \cite{5}, no significant successes were achieved until the beginning of the 20$^{th}$ century. Instead, the focus was put mainly on electromagnetism, leading to the development of the first electric motor \cite{5}. From the beginning of the 20$^{th}$ century, there were many magnet types that followed each another in rapid succession. The performance of each new magnet type has steadily improved. The magnets developed during this period are shown in Fig.~\ref{fig:6}. 
    
\newpage

    The 20$^{th}$ century started with the development of steel magnets \cite{13} in 1917. These magnets have a very low maximum energy product $|BH|_{max}$ of 8~kJ/m$^3$ (1~MGOe). Subsequently, steel magnets were improved with tungsten, chromium and an iron-carbon alloy, to suppress the movement of the domain walls at a suitable temperature. Cobalt was then added to increase the energy product and the focus shifted to cobalt alloys. This resulted in the Alnico magnets in 1940 \cite{14}.
    
    Alnico magnets are, as the name suggests, a composition of aluminum (Al), nickel (Ni) and cobalt (Co). After their discovery in 1940, they became the focus of attention because of their good magnetic properties. These include a high Curie temperature $T_C$ (will be explained in Sec.~\ref{1.3}), high remanence Magnetisation $M_r$ and high corrosion resistance. In addition, a smaller volume of material is required to produce the same amount of magnetic energy. However, this type of magnet also has the disadvantage of a low coercivity, which leads to an easy demagnetisation \cite{15}. 
    
   After that came the golden age of ferrites, that lasted from 1935 until 1970 \cite{16}. Ferrite magnets that mainly consist of the iron oxide $\alpha$-Fe$_2$O$_3$, are also called ceramic magnets. The greatest rise in popularity of ferrites came in 1950 with the development of hexagonal hard ferrites. The reason for this was the simplicity of the production process and the cheap raw materials used in their manufacture. This is the reason why ferrites are still the second most produced magnets in the world today \cite{17}, even though they have a lower energy product compared to the more expensive Alnico magnets. In addition, these hexagonal hard ferrites have a good corrosion resistance. The disadvantage of ferrites is that they are very fragile, as they have a pretty low mechanical strength.
    
    Towards the mid-1960s, another breakthrough led to the discovery of RE-TM magnets, the strongest magnets known today. These magnets have a high anisotropy due to the RE elements and a high Curie temperature $T_C$, as well as a high magnetisation due to the TM elements. This breakthrough was first achieved with the compound YCo$_5$ \cite{18}. Shortly afterwards, the compound SmCo$_5$ was created, whose energy product of 199~kJ/m$^3$ (25~MGOe) is significantly higher than that of the best Alnico compound with 56~kJ/m$^3$ (7~MGOe) \cite{21}. This also results again in a volume reduction of the magnetic material by a factor of 2 to obtain the same magnetic strength of SmCo$_5$ compared to the Alnico compound. The SmCo$_5$ compound was then refined with the 2:17 phase to Sm$_2$Co$_{17}$, improving the energy product to over 240~kJ/m$^3$ ($\geq$30~MGOe) (see Fig.~\ref{fig:6}). Because of their high values for $T_C$ and $M_S$, Sm$_2$Co$_{17}$ compounds are still used in high performance magnets and thus in motors. The only disadvantages of these compounds are the high costs of Sm and Co \cite{11}.  
    
    Finally, the strongest and most widely used compound to date, Nd$_2$Fe$_{14}$B, was found. This magnet was discovered simultaneously and independently in 1984 by Sagawa \textit{et al.} \cite{19} and Croat \textit{et al.} \cite{20} with a $|BH|_{max}$ of 286~kJ/m$^3$ (36~MGOe). Although this compound has a very low Curie temperature of 585~K, it is used wherever operating temperatures are not an issue. Thanks to further development, this compound has reached an energy product $|BH|_{max}$ greater than 500~kJ/m$^3$ ($>$63~MGOe). However, the low $T_C$ disqualifies this compound for high temperature applications. The addition of expensive and critical elements such as Co or Dy allows an improvement of the $T_C$ \cite{10}.

    In addition to the compounds discussed so far, Fig.~\ref{fig:6} shows as well the compound Sm-Fe-N, which has great potential due to high values for $T_C$, $M_S$ and a high coercivity. However, the Sm-Fe-N magnet is still in the research stage and therefore has not yet reached its full potential. The Sm$_{2}$Fe$_{17}$N$_{3}$ magnet is claimed to have a maximum energy product of 474~kJ/m$^3$ (60~MGOe) \cite{SmFeN1}, but usable magnets reach only up to 191~kJ/m$^3$ (24~MGOe) \cite{SmFeN}. Nevertheless, this magnet is already finding applications \textit{e.g.} in the industry. Another major area of recent and current research are ternary, quaternary and quinary compounds, like those that are going to be subject of this thesis.

\subsection{World Market and Usage of Magnets} \label{1.2}
\begin{figure}[h!]
    \centering
    \includegraphics[scale=0.56]{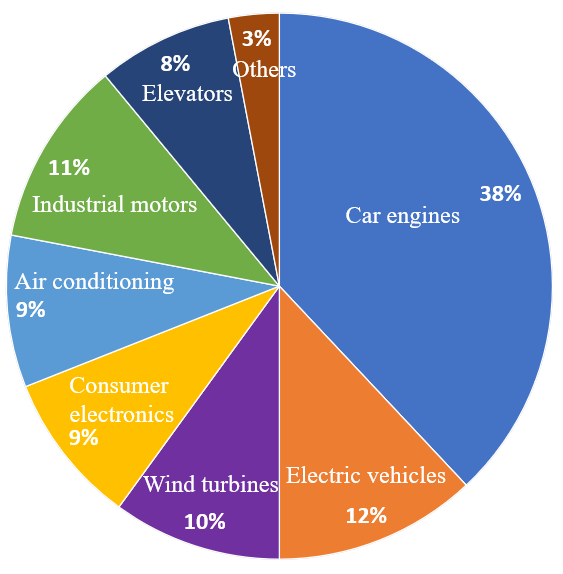}
    \caption{Application of permanent magnets by market share in 2021. The image was created from data of Rizos \textit{et al.} \cite{Fig79}.}
    \label{fig:7}
\end{figure}

    The demand for powerful magnets is increasing due to the growing number of industrial applications \cite{1}. As shown in Fig.~\ref{fig:7}, 38\% of the magnets produced are used in car engines. They are also part of household appliances such as loudspeakers (9\%), as well as part of electric vehicles (12\%). 

\begin{figure}[h]
    \centering
    \includegraphics[scale=0.65]{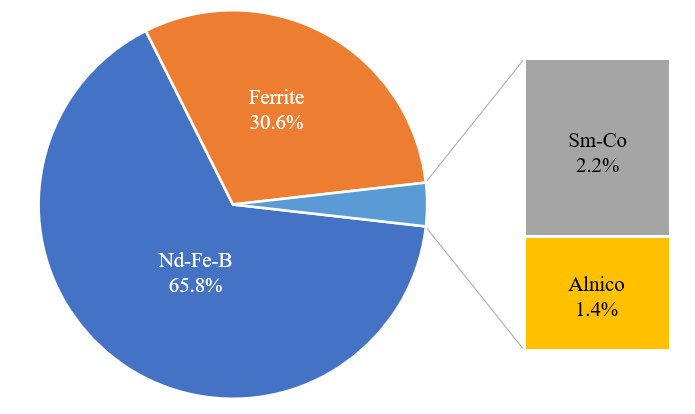}
    \caption{Percentage sales of major permanent magnet alloy types in 2022 in the global market. The image is based on data from \cite{Fig79}.}
    \label{fig:9}
\end{figure}

\begin{figure}[h!]
    \centering
    \includegraphics[scale=0.60]{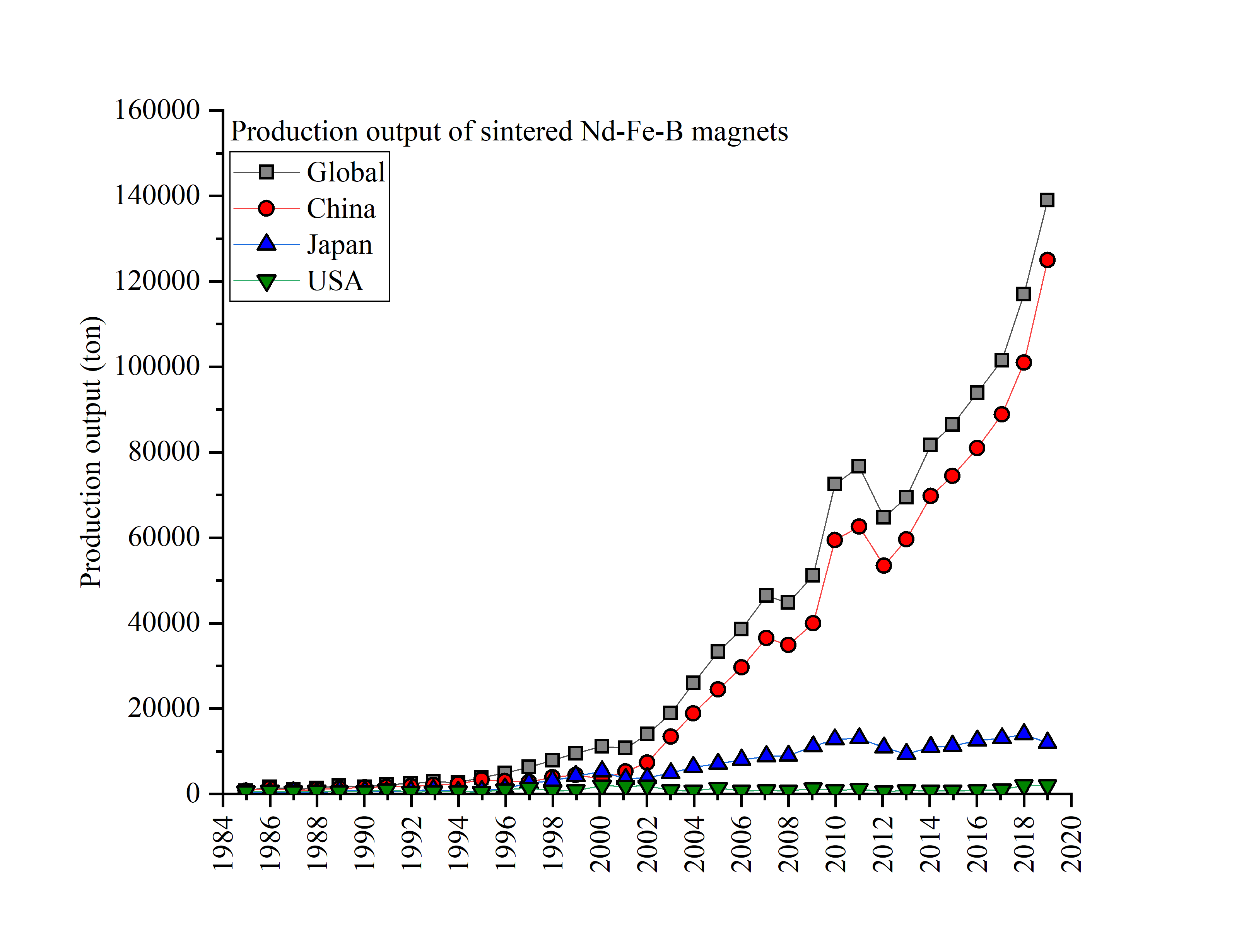}
    \caption{Worldwide production of sintered Nd-Fe-B magnets from 1984 to 2019. The picture is based on data taken from \cite{22, Fig8}.}
    \label{fig:8}
\end{figure}

    Among the used magnets, the sintered Nd-Fe-B magnet is very common with a total share of 65.8\% (see Fig.~\ref{fig:9}). The demand for this magnet increased from 20000 tonnes in the year 2000 to 120000 tonnes in 2016 (see Fig.~\ref{fig:8}). Thus, just looking at the demand for the sintered Nd-Fe-B magnet in Fig.~\ref{fig:8} from 1984 to 2019 represents an increase of more than six times. 
    
    To get a complete picture of the demand for hard magnets, Fig.~\ref{fig:9} shows the percentages of the other most commonly used magnets besides the Nd-Fe-B magnet. Ferrites have the second largest share as they are very cheap to produce. The ferrites are followed by the Sm-Co compounds, which only make up a small share due to their price. The smallest share have the Alnico compounds with only 1.4\%, as their insufficient magnetic properties make them unsuitable for widespread applications. 

    As Fig.~\ref{fig:9} shows, Nd-Fe-B magnets in particular have a large market share and are therefore used in many industrial processes. Consequently, it is necessary to have a secure and continuous supply of these hard magnets and the elements of which they consist. However, this will not be the case in the long term as almost all RE elements, such as Nd, Dy, Sm and Tb, are considered as critical resources \cite{24, 25}. Moreover, as shown in Tab.~\ref{tab.2}, the majority of RE reserves are located in China. This statement can be supported by the fact that China has produced 60\% of all RE elements measured in RE oxides in 2020. China therefore has a major influence on the price of these elements, as demonstrated in 2009 when the Chinese government imposed export quotas and other restrictions that increased the price by up to ten times. These uncertainties in the supply of RE elements have led many countries and their research to focus on RE-lean or RE-free magnets.  
    
\begin{table}[h]
\caption{Production of RE-oxides and available reserves for different countries. Data is taken from \cite{23}.}
\hspace{-0.8cm}
\label{tab.2}
\begin{tabular}{ccccccc}
\hline
Country   & \begin{tabular}[c]{@{}c@{}}Production\\ in 2017 {(}t{)}\end{tabular} & \begin{tabular}[c]{@{}c@{}}Production\\ in 2018 {(}t{)}\end{tabular} & \begin{tabular}[c]{@{}c@{}}Production\\ in 2019 {(}t{)}\end{tabular} & \begin{tabular}[c]{@{}c@{}}Production\\ in 2020 {(}t{)}\end{tabular}     & \begin{tabular}[c]{@{}c@{}}Production\\ in 2021 {(}t{)}\end{tabular}    & \begin{tabular}[c]{@{}c@{}}Reserves\\ in 2021 {(}t{)}\end{tabular} \\ \hline
China     & 105,000                                                              & 120,000                                                              & 132,000                                                              & 140,000                                                                  & 168,000                                                                 & 44,000,000                                                         \\
USA       & 0                                                                    & 18,000                                                               & 26,000                                                               & 39,000                                                                   & 43,000                                                                  & 1,800,000                                                          \\
India     & 1,500                                                                & 2,900                                                                & 3,000                                                                & 2,900                                                                    & 2,900                                                                   & 6,900,000                                                          \\
Australia & 20,000                                                               & 21,000                                                               & 21,000                                                               & 21,000                                                                   & 22,000                                                                  & 4,000,000                                                          \\
Brazil    & 2,000                                                                & 1,100                                                                & 1,000                                                                & 600                                                                      & 500                                                                     & 21,000,000                                                         \\
Thailand  & 1,600                                                                & 1,000                                                                & 1,800                                                                & 3,600                                                                    & 8,000                                                                   & n.a.                                                               \\
Vietnam   & 100                                                                  & 920                                                                  & 920                                                                  & 700                                                                      & 400                                                                     & 22,000,000                                                         \\
Myanmar   & 0                                                                    & 19,000                                                               & 25,000                                                               & 31,000                                                                   & 26,000                                                                  & n.a.                                                               \\ \hline
\end{tabular}
\end{table}

\subsection{Basic Properties of Magnets and Fundamentals} \label{1.3}
    Magnetism originates from the magnetic moments of the individual atoms in a compound. The magnetic moments of elements can be described by the classical Bohr definition. This states that the magnetic moment arises from the shift of an electric charge and therefore the movement of electrons. This orbital motion of the electrons leads to a magnetic field, which gives rise to the orbital magnetic moment. Besides the orbital magnetic moment there is the spin magnetic moment. Both are measured in Bohr magnetons \textit{\textmu}$_B$ \cite{4, 5, 6}. According to the quantum mechanical definition, the intrinsic angular momentum of an electron is \textit{s} = $\frac{1}{2}$. The total magnetic moment is composed of the orbital moment (L) and the spin moment (S), which are coupled in two different ways, LS and JJ with J=L+S. When the electron shell of an element is completely filled, the total magnetic moment is zero. This is the result of the pairing of electrons, where two electrons with opposite magnetic moments neutralise each other.
    
    From the 19$^{th}$ century onwards, a major step forward in the development of magnets was the removal of the shape dependency. Before this major step, the shape of magnets was limited by demagnetisation due to the internal magnetic field $H_d$ ($H_d$= -NM, where N is a shape-dependent demagnetisations factor and M is the magnetisation). Subsequently, U-shaped magnets were discovered, which bypassed the demagnetisation problem. In parallel, the usage of copper coils and iron led to the first magnetic motors and generators \cite{4, 6}.
    
\begin{figure}[h!]
    \centering
    \includegraphics[scale=0.60]{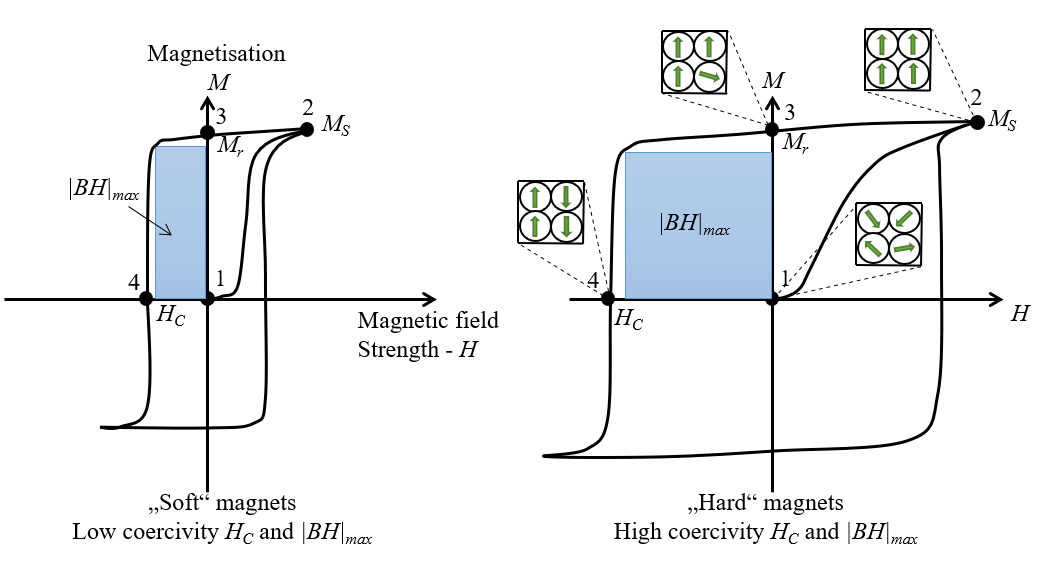}
    \caption{Hysteresis loop for soft (left) and hard (right) magnetic materials. Important magnetic properties like the magnetisation saturation $M_S$ are marked and the maximum energy product $|BH|_{max}$ is depicted as a blue square in the second quadrant. The image is based on data from \cite{metal}.}
    \label{fig:1}
\end{figure}

    However, the shape dependence of magnets was not fully resolved until the discovery of ferrimagnetic hexagonal ferrites in 1951. This discovery improved the knowledge of coercivity ($H_c$), making it possible to produce magnets with a coercivity greater than their spontaneous magnetisation. This also meant that the shape was no longer limited to needle and U shapes, which is greatly expanding the usability of magnets. High coercivity and high magnetisation are advantageous for hard magnetic materials. An explanation for these advantages can be found in the hysteresis loop (Fig. \ref{fig:1}), that is a function of the magnetisation \textit{M} against the magnetic field strength \textit{H} \cite{4, 5, 6}.

    The hysteresis loop starts at the origin (point 1) of the coordinate system. The magnetic field strength is then increased until point 2, where the maximum magnetisation of the compound and thus the magnetisation saturation ($M_s$) is reached. The magnetisation saturation can also be calculated from the density of the compound $\rho$, the Avogadro constant $N_{A}$, the molecular weight $M$ and the average magnetic moment per atom $m^{tot}_{atom}$ of the compound:
    \begin{equation}
       \mu_{0}M_{S} = m^{tot}_{atom} \cdot \mu_{B} \cdot \frac{ \rho \cdot N_{A}}{M}.
       \label{eq.37}
    \end{equation}
    
    At point 1 before the magnetic field strength is increased, the magnetic domains all align in different directions to minimise the energy. The magnetic domains are small areas which, depending on the type of magnetisation, have the same or different magnetic moments. When the external field is applied, the domain walls align themselves in parallel and with it in the direction of the applied field. Point 2 is where all domains point into the same direction. Afterwards the magnetic field is reduced to zero, which leads to an slight decrease in magnetisation. At point 3, the remanence magnetisation $M_r$ and with it the remaining magnetisation is reached. Here, most domains continue to point into one direction while a few domains point to other directions. The remanence magnetisation is an important property that should be as large as possible for an hard magnetic material, as it corresponds to the magnetisation without an external field. The next step is to reduce the magnetic field until the magnetisation is again zero. This point describes the coercive field strength $H_c$. The coercive field strength equals the field strength that needs to be applied to demagnetise the material. During demagnetisation, all domains are aligned so that they point in two opposite directions and the magnetic moment is cancelled. The rest of the hysteresis loop is a mirror image of the first two quadrants of the coordinate system. To obtain them, the field strength must be reduced and then increased to zero again. 
    
    Another important value is the efficiency of a magnet, which is quantitatively measured in terms of the maximum energy product $|BH|_{max}$. The maximum energy product corresponds to the blue square in the second quadrant. The hysteresis loops for hard and soft magnetic materials differ from each other. Soft magnetic materials have a lower coercivity and therefore a narrower hysteresis loop, a lower maximum energy product and are therefore easier to demagnetise. Hard magnetic materials have a wide hysteresis loop due to a higher coercivity, resulting in a higher resistance to demagnetisation as well as a higher maximum energy product.

\begin{figure}[h]
    \centering
    \includegraphics[scale=0.58]{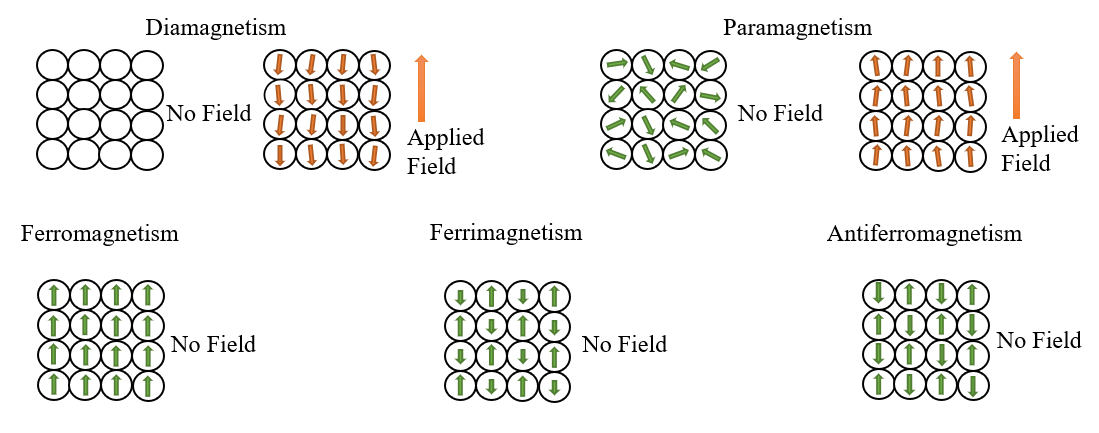}
    \caption{Different magnetic domain configurations for the different types of magnetism with and without external field application. The image is based on data from \cite{7}.}
    \label{fig:2}
\end{figure}

    There are a total of five ways in which a magnetic compound can interact with an external field $H$. The three major types of magnetism are: diamagnetism, paramagnetism and ferromagnetism. On top of these there are two subclasses of ferromagnetism, which are ferrimagnetism and antiferromagnetism \cite{5, 6}. These are shown in Fig.~\ref{fig:2}.
    
    Neither diamagnetism nor paramagnetism are noticeable without the application of an external field and both make up the majority of magnetic materials. A diamagnetic material has no magnetic moment without an external magnetic field and is only a very weak form of magnetism. When an external field is applied, a magnetic moment is induced in the material that points into the opposite direction of the applied field. This causes a weakening of the applied field. Diamagnetism is found in materials with only paired electrons. Since these electron pairs have identical but opposite magnetic moments, resulting in a total magnetic moment of zero.

\newpage

    In contrast to diamagnetism, paramagnetism has finite local magnetic moments. Without an applied external field, all domains are pointing in different directions, resulting in a total magnetic moment of zero. When the external field is applied, all the magnetic moments align in the same direction as the field, and therefore strengthening it. Paramagnetism is only found in materials containing unpaired electrons \cite{4, 5}.
    
    In ferromagnetism, all local magnetic moments are connected and point into one direction, similar to paramagnetism with an applied field. When an external field is applied, a new direction of magnetisation of the local magnetic moments can be induced, which even remains after the field is removed. This ability to induce a new direction of magnetisation is responsible for the macroscopic effect of magnetism. Ferromagnetism occurs in iron, which is where the name comes from. In addition ferromagnetism has two further sub-classes being described as ferrimagnetism and antiferromagnetism.

    One of the subclasses of ferromagnetism is ferrimagnetism, which instead of one lattice of local magnetic moments has two sublattices with alternating directions. The sublattices have magnetic moments of different strengths, resulting in an overall magnetic moment that is much weaker than in ferromagnetism. Ferrimagnetism is mainly found in multi-element compounds, such as ferrites, where each element contributes a sublattice. 
    
    The other subclass of ferromagnetism is antiferromagnetism, which likewise ferrimagnetism consists of 2 sublattices whose magnetic moments point in opposite directions. Here, the two sublattices have equally strong magnetic moments, so they oppose each other, resulting in a total magnetic moment of zero \cite{5, 6}.
    
\begin{figure}[h!]
    \centering
    \includegraphics[scale=0.65]{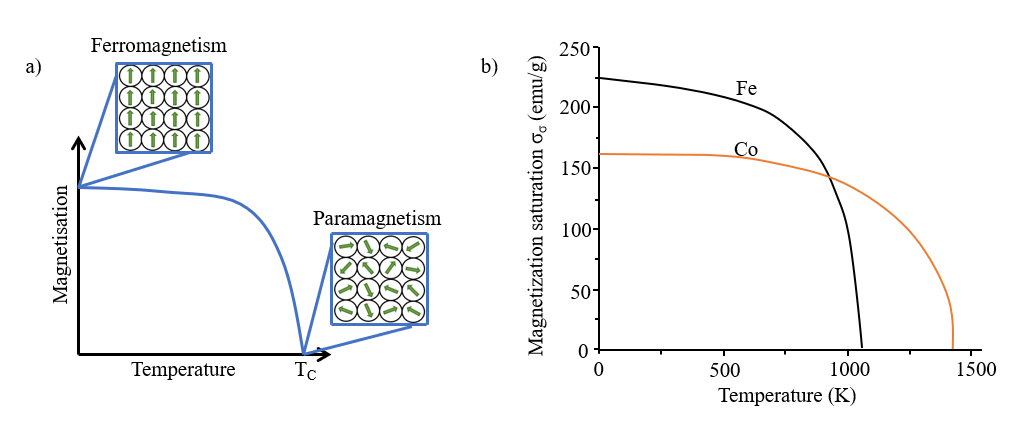}
    \caption{a) Change of the magnetic ordering from ferro- to paramagnetism with increasing temperature up to the Curie temperature $T_C$; b) Finite temperature magnetisation curves for the elements Fe and Co up to their $T_C$’s of 1044 and 1388 K. Image a) is based on data from \cite{Ste} and b) on \cite{6}.}
    \label{fig:3}
    \end{figure}

    The influence of temperature on the strength of magnetisation in ferromagnetic materials must be considered. The magnetic ordering is strongly temperature dependent as it is a thermodynamic process. At low temperature, there is a strict order of all magnetic domains, which in a ferromagnet all point into the same direction. When the temperature is slowly increased, the magnetic order is slowly disturbed as well. At a certain temperature, the type of magnetism changes from ferromagnetic to paramagnetic (see Fig.~\ref{fig:3}a). This temperature is known as Curie temperature $T_C$. Once this temperature is exceeded, the material becomes paramagnetic. Therefore, the Curie temperature is an important intrinsic magnetic property that must be taken into account. Particularly in industry with high operating temperatures, it is important that the used magnet has the highest possible Curie temperature. For the expansion of the number of application possibilities.
    
    Fig.~\ref{fig:3}b shows the magnetisation behaviour of iron and cobalt as the temperature increases. Furthermore, the finite temperature magnetisation of iron and cobalt are shown with their corresponding $T_C$'s of 1044~K and 1388~K \cite{6}.

\begin{figure}[h!]
    \centering
    \includegraphics[scale=0.79]{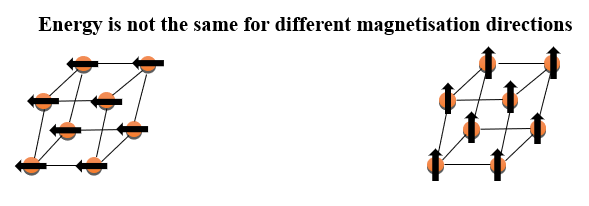}
    \caption{Unequal energies for the different magnetisation directions. Image based on \cite{Ste}.}
    \label{fig:4}
\end{figure}

    A result of deepening the knowledge of coercivity $H_c$, leads to the solution of the shape constraint and another intrinsic magnetic property was discovered. This new property was the magnetocrystalline anisotropy energy (MAE). This anisotropy states that in ferromagnets the magnetisation in one or more directions is usually energetically favourable. There is an energetic difference in the reorientation of the magnetisation in different directions (see Fig.~\ref{fig:4}). This anisotropy is referred to as uniaxial when the studied compound only has one easy direction to magnetise \cite{4, 6}. It is assumed that the ThMn$_{12}$ compounds treated in this work all exhibit unaxial anisotropy \cite{9}.
   
\newpage

\begin{figure}[h]
    \centering
    \includegraphics[scale=0.70]{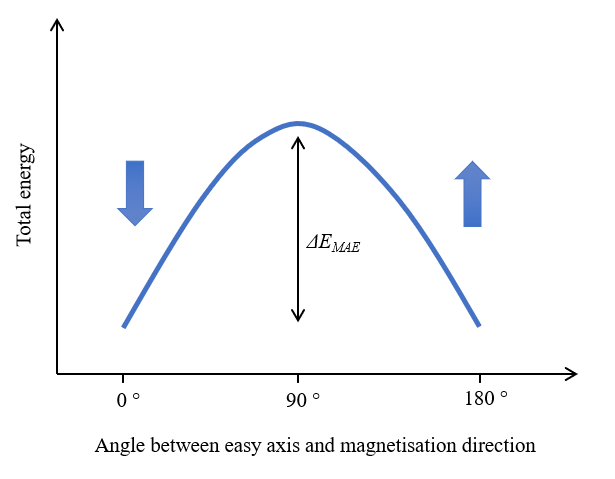}
    \caption{Change of the magnetic energy in relationship to the angle between the easy axis to magnetise and the magnetisation direction with a peak at an angle of 90°. Image based on data from \cite{Ste}.}
    \label{fig:5}
\end{figure}

    According to the Bohr definition, a charge shift and thus the movement of electrons causes magnetism. It can be concluded that this is a result of the circulating electron currents. Due to the nature of electron currents, energy is equal in opposite directions for a given magnetisation distribution (in other words $M(r)=-M(r)$). Accordingly, the energy for a given deviation of the simple magnetisation direction and the associated axis depends on the deviation angle. There is an energy peak at $\theta$ equal to 90° (see Fig.~\ref{fig:5}).

    The energy deviation ($\Delta E_{MAE}$) of the most easily magnetised axis in case of unaxial anisotropy can be calculated by the following equation \cite{10}:
\begin{equation}
       \Delta E_{MAE} = K_1 \cdot sin(\theta)^2.
       \label{eq:1}
\end{equation}
    $\theta$ is the angle of deviation from the axis corresponding to the most easily magnetised direction and $K_1$ corresponds to the anisotropy constant. All considered 1:12 compounds exhibit uniaxial anisotropy ($K_{1} > 0$), which means that the higher order anisotropy constants are negligible, as $K_{2}$ is almost two orders of magnitude smaller than $K_{1}$. $K_1$ and $\Delta E_{MAE}$ are measured in J/m$^3$, which usually reaches at least 1~kJ/m$^3$ and can be as high as 10~MJ/m$^3$ \cite{4}. 

    Another important property can be calculated from the anisotropy constant $K_1$ and the magnetisation energy. This property is called anisotropy field $H_a$, and can be calculated with the following formula \cite{10}:
\begin{equation}
       H_a = \frac{2K_1}{\mu_0M_S},
       \label{eq:2}
\end{equation}
    where $\mu_0$ is the magnetic constant with $\mu_0=4\pi \times 10^{-7}$ J/(A$^2$m).
    The upper limit of the anisotropy field $H_a$ is given by the coercive field $H_c$, with $H_a \leq H_c$. This point corresponds to the complete demagnetisation of the compound \cite{10}.

    With the properties of magnets discussed so far, the hardness factor $\kappa$ can be calculated. This value indicates whether or not the material is suitable for use as a hard magnetic material. This hardness factor can be calculated using the anisotropy constant $K_1$ and the magnetisation $M_S$ according to the next equation \ref{eq:3}:
\begin{equation}
       \kappa = \sqrt{\frac{K_1}{\mu_0M_S^2}}.
       \label{eq:3}
\end{equation}
    Depending on the resulting value, the material is suitable for certain specific applications. The material can be categorised as a hard magnet ($\kappa>1.0$), a semi-hard magnet ($0.1<\kappa<1.0$) or a soft magnet ($\kappa<0.1$). The most commonly used hard magnet, Nd$_2$Fe$_{14}$B, has a hardness factor of 1.54 (see Tab.~\ref{tab.1}).
    
\begin{table}[h]
   \caption{Curie temperature $T_C$, magnetisation $\mu_0M_S$, anisotropy constant $K_1$ and hardness factor $\kappa$ for common hard magnets. Data is taken from \cite{11, step}.}
   \centering
   \label{tab.1}
\begin{tabular}{lcccc}
 
    \hline
    Magnet           & \multicolumn{1}{l}{\textit{$T_c$} (K)} & \multicolumn{1}{l}{\textit{$\mu_0M_S$} (T)} & 
    \multicolumn{1}{l}{\textit{$K_1$} (MJ/m$^3$)} & 
    \multicolumn{1}{l}{\hspace{0.2cm}$\kappa$} \\ \hline
    Nd$_2$Fe$_{14}$B & 588                                   & 1.61                                                       & 4.9                                                         & 1.54                                      \\
    Sm$_2$Co$_{17}$  & 1190                                  & 1.21                                                       & 4.2                                                         & 1.89                                      \\
    SmCo$_{5}$      & 1020                                  & 1.05                                                       & 17.0                                                        & 4.40                                      \\
    Alnico 5         & 1210                                  & 1.40                                                       & 0.32                                                        & 0.45                                      \\
    NdFe$_{11}$Ti   & 811                                   & 1.69                                                       & 1.41                                                        & 0.88                                      \\ \hline
\end{tabular}
\end{table}    

\newpage
    In general, magnets made out of RE and TM elements have a high hardness factor as they have high anisotropy and magnetisation. The high anisotropy is due to the RE elements, while the high magnetisation is related to the TM elements such as iron or cobalt. Tab.~\ref{tab.1} lists some intrinsic magnetic properties, such as the Curie temperature $T_C$, the magnetisation $M_S$ and the anisotropy constant $K_1$, as well as the hardness factors $\kappa$ of well known hard magnetic materials.

\clearpage

\thispagestyle{plain}
\section{Case Description and Motivation of this Work}
    The demand for affordable hard magnets is steadily increasing, as they are part of technologies in the growing fields of renewable energy and electric transport \cite{1}. Currently, the strongest and most widely used magnet is Nd$_2$Fe$_{14}$B. Ongoing research on new magnets aims towards surpassing the capabilities of this Nd$_2$Fe$_{14}$B magnet, which is composed of RE and TM elements \cite{19, 20}. All RE elements, including Nd, Sm, Dy and Tb, as well as some TM elements such as Co, are considered critical for future applications \cite{24, 25}. These elements are considered critical because their reserves are slowly declining. Further, the supply monopoly of China (see Tab.~\ref{tab.2}) is a strong factor driving the development of new RE-lean or RE-free hard magnetic materials \cite{2, 3}. 
    
    Several methods have been previously documented in literature for the development of these RE-lean magnets. For instance, grain size reduction is used to achieve a higher coercive field strength and the possible omission of the critical element Dy \cite{26, 27}. The coercivity can be significantly increased by applying the grain boundary diffusion process, which leads to the reduction of critical RE elements such as Dy or Tb \cite{28, 29}. In addition, the partial replacement of the critical RE elements by more abundant and cheaper elements has been introduced (\textit{e.g.} see \cite{FU22}). To date, this process has mainly been used to produce magnets for usage at low to medium temperatures.

    The focus on the physical and magnetic properties of RE-lean permanent magnets based on the ThMn$_{12}$ prototype structure (also referred to as the 1:12 phase in this work). It should be noted that the 1:12 (RE:Fe) phase has a lower RE content than Nd$_2$Fe$_{14}$B and has promising magnetic properties \cite{32, 33, 34}. Previous research has investigated the abundant and cost-effective elements Y and Ce as potential substitutions for the critical element Nd and their impact on the magnetic properties and thermodynamic phase stabilities of the 1:12 phase \cite{35, step}. 
    
    In this thesis, Zr is considered as another promising element to replace the RE element Nd. Zr substitution can lead to more affordable and sustainable magnets as Zr is more abundant and inexpensive than Nd. The phase stability of Zr-substituted compounds with the 1:12 phase has already been reported \cite{FU22, 2016, FU21}. These studies showed that substitution of Nd with Zr has a potential to produce similar to higher saturation polarisation and an anisotropy field comparable to Nd$_2$Fe$_{14}$B \cite{FU21}. In addition, Zr substitutes on only the 2$a$ sites of the 1:12 phase, which is occupied by RE elements. However, a difficulty in the experimental studies is the substitution of Zr in these 2$a$ sites. The fact that Zr forms a solid solution with Fe, is the reason that only compounds with a Zr concentration up to 30\% have been produced so far \cite{Fan}. A possible solution to this problem is the selective laser melting method used by Neznakhin \textit{et al.} \cite{Neznakhin} as it has successfully produced the compound (Zr,Sm)Fe$_{11}$Ti (at the end of 2022) with a 1:1 ratio of Zr to RE.

    Consequently, the 30\% share of Zr from the previous experimental publications \cite{FU21, FU22, FU24} is not the highest possible concentration. Therefore, higher concentrations are taken into account here. Based on the selective laser melting method and the calculations of Sec.~\ref{5.1}, the concentration of 50\% can be considered. Nevertheless, in this work the phase stability is estimated taking into account the formation energy and the solution enthalpy at 0 K in order to find energetic favourable quaternary and quinary alloys.

    In order to obtain a systematic understanding of the intrinsic magnetic properties of Zr-substituted complex 1:12 compounds, it is started with the binaries and ternaries of M$_{2}$Fe$_{24-y}$Ti$_{y}$ alloys (where M = Zr and Nd and y: 0 $\leq$ y $\leq$ 2). Since the binary 1:12 phases are not thermodynamically stable Ti was used as a stabiliser \cite{36, 37, 38}. Then, Zr substitution was considered and Ti solubility in the quaternary compounds (Zr,Nd)Fe$_{24-y}$Ti$_{y}$ was investigated. Once the equilibrium Ti concentration was found, Co was additionally considered as quinary substitution to screen its impact on magnetic properties and stabilities.

\newpage

\thispagestyle{plain}
\section{Theoretical Background} \label{3}
    The intrinsic magnetic properties mentioned in Sec.~\ref{1.3} are calculated by using \textit{ab initio} methods. These methods are used to calculate the magnetic properties of the selected compounds in the most accurate way possible. This leads to a better atomistic understanding. The results can then be compared with the values given in the literature, if these are available.

    In order to give the reader a basic understanding of the calculations performed, their framework is provided. To begin with, density functional theory (DFT) is introduced as the basis for the calculations. Afterwards, the basic knowledge of quantum mechanics is explained, with an focus on the calculation of magnetic properties. Among these fundamentals are the Schrödinger equation, the Born-Oppenheimer approximation, the Hohenberg-Kohn equations and the Kohn-Sham equations. Following this, the exchange-correlation energy and the nuclear-electron interaction are introduced. This chapter concludes with the treatment of the $f$-electrons in the respective calculations.

\subsection{Density Functional Theory}
    Density functional theory (DFT), is used in physics and chemistry to represent the electrical structures of many-body systems and to calculate their properties. DFT is also the most widely used quantum mechanical method. The reputation of the method and the number of publications using it has grown, as better and more accurate functionals have been developed.
\begin{figure}[h]
    \centering
    \includegraphics[scale=0.54]{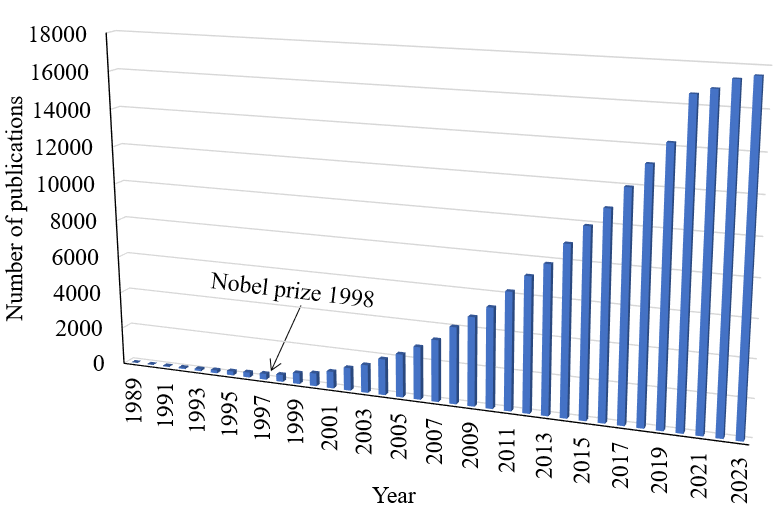}
    \caption{Increasing number of publications using DFT. The Nobel prize in chemistry is marked in accordance to the year. The values for the years 1989-2019 are taken from the source \cite{44}. The values for 2020-2023 (as of 09/10/2023) are based on the search "DFT density functional theory" in google scholar.}
    \label{fig:11}
\end{figure} 
    In fact, its reputation grew to such an extent that Walter Kohn and John Pople were awarded the Nobel Prize in Chemistry in 1998, after which the publications using DFT skyrocketed (see Fig.~\ref{fig:11}).
    
    However, the path to today's DFT methods has been long. It started with the Hohenberg-Kohn theorem, established in 1964 after its namesakes Pierre Hohenberg and Walter Kohn \cite{39}. DFT, like Hartree-Fock (HF) and post-Hartree-Fock (post-HF) methods, is a way to describe electron interactions based on quantum mechanical approaches. Quantum mechanical approaches include, \textit{e.g.} the universal principle of the Schrödinger equation. These methods (HF, post-HF, DFT) belong to the \textit{ab initio} or first-principles methods, since they do not require experimental data as input \cite{40, 41}. Nevertheless, DFT is fundamentally different from HF and post-HF methods. HF and post-HF methods use a Slater determinant, which is a determinant consisting of the wave functions of an electron. In contrast, DFT uses the electron density of the system to calculate the ground state properties of the system. The system is simplified by reducing the solution of the many-body wave function from 3\textit{N} variables to three variables. This is achieved by solving the many-body wave function (3\textit{N} variables) in pure dependence on the electron density (with dependence only 3 variables). The approach by Hohenberg and Kohn was later improved by Kohn and Sham with the Kohn-Sham equations \cite{42}. In this approach, the description of a system by the wave function is partially reintroduced. They also simplified the many-electron problem by mapping it onto a system of non-interacting quasiparticles and therefore reducing the many-electron problem to a problem of non-interacting electrons in an effective potential \cite{7}. That effective potential consists of an external potential with additional consideration of the Coulomb interaction of the electrons. The introduction of this approach corresponds to the basic of today's DFT methods \cite{43}. 
    
\subsection{Schrödinger's Equation}
    Since \textit{ab initio} methods depend on the Schrödinger equation, this fundamental principle of DFT is explained here. The Schrödinger equation is not only a fundamental principle in DFT, but also the starting point for understanding the quantum mechanical system behind a material \cite{7}. In most cases, the non-relativistic and time-independent version of the Schrödinger equation is used, which is defined as follows:
    \begin{equation}
       \hat{H}\Psi_i(x_1, x_2, ..., x_N, R_1, R_2, ..., R_M) = E_i\Psi_i(x_1, x_2, ..., x_N, R_1, R_2, ..., R_M).
       \label{eq:4}
    \end{equation}
    Here $\hat{H}$ is the Hamiltonian, which at this point is the Hamiltonian for a molecular system with \textit{M} nuclei and \textit{N}-electrons, without the influence of a magnetic or electric field. The many-body Hamiltonian is a differential operator that describes the total energy of the system. It can be described as follows:
    \begin{equation}
        \hat{H} = \hat{T}_e + \hat{T}_n + \hat{V}_{nn} + \hat{V}_{en} + \hat{V}_{ee},
       \label{eq:5}
    \end{equation}
    where $\hat{T}_e$ is the kinetic energy of the electrons and $\hat{T}_n$ is the kinetic energy of the nuclei. $\hat{V}_{nn}$ and $\hat{V}_{ee}$ define the Coulomb repulsion between identical particles (electron/electron and nucleus/nucleus repulsion), and $\hat{V}_{en}$ describes the Coulomb attraction between nuclei and electrons.
    
    Using atomic units, the Hamiltonian can be written as:
    \begin{equation}
       \hat{H} = -\frac{1}{2}\sum_{i=1}^{N}\nabla^{2}_{i} - \frac{1}{2}\sum_{A=1}^{M}\frac{1}{M_A}\nabla^{2}_{i} - \sum_{i=1}^{N}\sum_{A=1}^{M}\frac{Z_A}{r_{iA}} + \sum_{i=1}^{N}\sum_{j>i}^{N}\frac{1}{r_{ij}} + \sum_{A=1}^{M}\sum_{B>A}^{M}\frac{Z_AZ_B}{R_{AB}},
       \label{eq:6}
    \end{equation}
    with the indices \textit{i}, \textit{j} for the \textit{N}-electrons and \textit{A}, \textit{B} for the \textit{M} nuclei. On top of that, $M_A$ stands for the mass of the nucleus \textit{A} and \textit{Z} for the charge of the nucleus \cite{41}.
    
\subsection{Born-Oppenheimer Approximation}
    As can be seen from the Hamiltonian, the Schrödinger equation is often too complex to obtain an exact solution. In particular, this depends on the size and complexity of the compound being studied (systems larger than helium are already too complex). In order to achieve a solution for these impossible calculations, it is necessary to reduce the complexity by means of approximations. The best known and most commonly used approximation is the Born-Oppenheimer approximation \cite{45}. This approximation reduces complexity by separating the electronic and ionic degrees of freedom. To do this, the Born-Oppenheimer approximation states that the mass of the nuclei is much greater than the electron mass (in the case of the hydrogen atom: about 1800 times greater). In other words, the electrons are much faster and adapt to the movement of the nuclei in an instant. That allows the assumption that the nuclei can be considered as stationary, and thus the kinetic energy term ($\hat{T}_e$) in the Hamiltonian becomes zero and the nucleus/nucleus repulsion term ($\hat{V}_{nn}$) becomes a constant. This simplifies the Hamiltonian to the electronic Hamiltonian ($\hat{H}_{elec}$), which now takes the form of the following equation:
    \begin{equation}
        \hat{H}_{elec} = \hat{T}_e + \hat{V}_{en} + \hat{V}_{ee} = -\frac{1}{2}\sum_{i=1}^{N}\nabla^{2}_{i} - \sum_{i=1}^{N}\sum_{A=1}^{M}\frac{Z_A}{r_{iA}} + \sum_{i=1}^{N}\sum_{j>i}^{N}\frac{1}{r_{ij}}.
       \label{eq:7}
    \end{equation}
    
    After solving the Schrödinger equation with the electronic Hamiltonian ($\hat{H}_{elec}$) it results in the electronic Schrödinger equation ($\Psi_{elec}$) and the electronic energy ($E_{elec}$). Here $\Psi_{elec}$ depends only on the coordinates of the electrons and the coordinates of the nuclei are parametric and not explicitly included. The total energy ($E_{tot}$) is then obtained by adding the electric energy $E_{elec}$ to the nuclear energy $E_{nuc}$, which consists of the nuclear repulsion term in the form $E_{nuc} = \sum_{A=1}^{M}\sum_{B>A}^{M}\frac{Z_AZ_B}{R_{AB}}$ \cite{41}.

    From the calculated total ground state energy of the electronic system according to Eq.~\ref{eq:7}, the ground state properties can be obtained. The ground state energy $E_0=E~[N,~V_{ext}]$ and the ground state wave function $\Psi_{0}$ can be calculated using the variational principle as well as the number of \textit{N}-electrons and a known nuclear potential $V_{ext}$. To obtain the true ground state, a full energy minimisation $E[\Psi]$ must be performed, using the following equation:
    \begin{equation}
        E_0 = \min_{\Psi\to N} E[\Psi] = \min_{\Psi\to N}\langle\Psi|\hat{T}_e + \hat{V}_{en} + \hat{V}_{ee}|\Psi\rangle.
       \label{eq:8}
    \end{equation}
    
\subsection{The Hohenberg-Kohn Equations}
    Often the Born-Oppenheimer approximation is not sufficient enough, as in case of the materials investigated in this study. Although the Born-Oppenheimer approximation simplifies the Hamiltonian significantly, it is still too complicated for complex compounds. In particular, the electron/electron coulomb repulsion term $\hat{V}_{ee}$ with 3\textit{N} dimensionality is still difficult to calculate. A further step towards simplification was postulated by Hohenberg and Kohn in 1964 in the form of the Hohenberg-Kohn theorem \cite{40, 41}. This was the cornerstone of density functional theory, postulating that the many-body wave function can be reduced to the electron density $\rho{(r)}$. 

    Hohenberg and Kohn's first theorem contains the proof that the electron density determines the number of electrons in a system. That leads to the conclusion that the electron density with $\rho{(r)}$ can also be used to calculate the ground state wave function $\Psi$ and thus any other electronic property. This possibility is shown by the fact that only the minimum energy principle is applied to the ground state.

    The first theorem can be proven by a \textit{reductio ad absurdum}, which means to assume the opposite and prove that it is false \cite{Molecular}. At the beginning of this proof, the electron density is given by $\rho{(r)}$ for the non-degenerate ground state of a \textit{N} electron system, where \textit{N} is defined by quadrature. The electron density also defines the external potential $\nu(r)$ and all other properties. First, two different potentials $\nu$ and $\nu_0$, differing by more than one constant and each having the same density $\rho$ in its ground state, are considered. Then two different Hamiltonian $\hat{H}$ and $\hat{H}_0$ with the same ground state densities but with different wave functions $\Psi$ and $\Psi_0$ are obtained. By solving the $\hat{H}$ problem with the trial function $\Psi_0$ or vice versa $\hat{H}_0$ with the trial function $\Psi$, the following two expressions are received:
    \begin{equation}
        E_0 < \langle\Psi'|\hat{H}|\Psi'\rangle = \langle\Psi'|\hat{H}'|\Psi'\rangle + \langle\Psi'|\hat{H} - \hat{H}'|\Psi'\rangle = E_0' + \int\rho(r)[\nu(r) - \nu'(r)] dr,
       \label{eq:9}
    \end{equation}
    and:
    \begin{equation}
        E_0' < \langle\Psi|\hat{H}'|\Psi\rangle = \langle\Psi|\hat{H}|\Psi\rangle + \langle\Psi|\hat{H}' - \hat{H}|\Psi\rangle = E_0 - \int\rho(r)[\nu(r) - \nu'(r)] dr.
       \label{eq:10}
    \end{equation}

    In these expressions, $E_0$ and $E'_0$ correspond to the ground state energies of the Hamiltonian $\hat{H}$ and $\hat{H}'$. The contradiction $E_0+E'_0<E'_0+E_0$ is obtained by adding these expressions. Since this implies that there cannot be two different $\nu$ with the same electron density for their ground states, this is a contradiction. With this pursuit of \textit{reductio ad absurdum} it can be proven that the electron densities determine \textit{N} and $\nu$ and all other ground state properties such as the kinetic energy $T(\rho)$, the potential energy $V(\rho)$ and the total energy $E(\rho)$ \cite{40, 41, Molecular}.

    The second theorem of Hohenberg and Kohn introduces the variational principle in the form:
    \begin{equation}
        E_0 \leq E(\tilde{\rho}) = T(\tilde{\rho}) + V_{en}(\tilde{\rho}) + V_{ee}(\tilde{\rho}),
       \label{eq:11}
    \end{equation}
    with $\tilde{\rho}(r)$ as the experimental density, satisfying two boundary conditions $\tilde{\rho}(r)~\geq~0$ and $\int \tilde{\rho}(r) dr = N$. These conditions ensure that only the correct ground state energy leads to the energy minimum, and any other experimental density would result in a higher energy. The lowest energy can subsequently be obtained by varying the experimental density $\tilde{\rho}(r)$ \cite{40, 41}. These assumptions provide the following equation for obtaining the energy minimum:
    \begin{equation}
        \langle\tilde{\Psi}|\hat{H}|\tilde{\Psi}\rangle = T[\tilde{\rho}] + V_{ee}[\tilde{\rho}] + \int\tilde{\rho}(r)\nu_{ext}dr = E[\tilde{\rho}] \geq E_0[\tilde{\rho}_0] = \langle\Psi_0|\hat{H}|\Psi_0\rangle.
       \label{eq:12}
    \end{equation}

    Furthermore, the kinetic energy $T (\rho)$ and the electron/electron Coulomb repulsion term $V_{ee} (\rho)$ can be summarised in the Hohenberg-Kohn functional $F_{HK}(\rho)$, taking the form of:
    \begin{equation}
        F_{HK}(\rho) = T(\rho) + V_{ee}(\rho).
       \label{eq:13}
    \end{equation}
    The electron/electron coulomb repulsion term $V_{ee}(\rho)$ has two additional terms, which are corresponding to the classical repulsion term $J(\rho)$ and the non-classical repulsion term $K_{ex}(\rho)$.
    \begin{equation}
        V_{ee}(\rho) = J(\rho) + K_{ex}(\rho).
       \label{eq:14}
    \end{equation}

\subsection{The Kohn-Sham Equations}
    The Hohenberg-Kohn equations with Eq.~\ref{eq:12} do not give the exact energy of the system. Therefore, a new scheme is required. This new scheme was introduced by Kohn and Sham in 1965 \cite{42}, where the complete interacting electronic system is projected onto a fictitious system of non-interacting quasi particles moving in an effective potential. Therefore, the Kohn-Sham (KS) equation can be formulated as follows:
    \begin{equation}
        \hat{H}_{KS}\psi_i = \epsilon_i\psi_i,
       \label{eq:15}
    \end{equation}
    where $\epsilon_i$ reflects the orbital energy of the corresponding KS orbital $\psi_i$. The associated KS Hamiltonian operator can then be written as:
    \begin{equation}
        \hat{H}_{KS} = [-\frac{1}{2}\nabla^2 + V_{eff} (r)],
       \label{eq:16}
    \end{equation}
    in which the term $V_{eff}(r)$ in the KS Hamiltonian operator corresponds to the value of the effective potential. 

    Having introduced this scheme, the many-body Schrödinger equation can be solved by solving the single-body equations. According to the HK theorem, the KS Hamiltonian represents a functional of an electron at point $r$, so that the electrons can be determined by the following equation:
    \begin{equation}
        \rho (r) = \sum_{i=1}^{N}|\psi_i(r)|^2.
       \label{eq:17}
    \end{equation}
    The kinetic energy term and the classical Coulomb interaction take the following forms:
    \begin{equation}
        T_e = -\frac{1}{2}\sum_{i=1}^{N}\int d^3r |\nabla\psi_i(r)|^2,
       \label{eq:18}
    \end{equation}
    and:
    \begin{equation}
        J(\rho) = \frac{1}{2}\int d^3rd^3r'
        \frac{\rho(r)\rho(r')}{|r - r'|}.
       \label{eq:19}
    \end{equation}
    This leads to the conclusion, that the total energy of the system according to the KS-approach can be calculated as follows:
    \begin{equation}
        E_{KS} = \sum_{i}^{N}\epsilon_i - J(\rho) + E_{xc} - \int\frac{\delta E_{xc}}{\delta\rho(r)}.
       \label{eq:20}
    \end{equation}
    The values for $\epsilon_i$ result from the KS equations and correspond to the one-electron energies. However, the physical significance of these one-electron energies is limited. $E_{xc}$ corresponds to the exchange correlation term of all many-body interactions of the exchange. Consequently, this term is a part of the electron interaction which has to be considered besides the Hartree term \cite{7}. The term $E_{xc}$ can be calculated by the following equation:
    \begin{equation}
        E_{xc} = T_{exact}(\rho) - T_S(\rho) + V_{ee}(\rho) - J(\rho) = T_{exact}(\rho) - T_S(\rho) + K_{ex}(\rho).
       \label{eq:21}
    \end{equation}
    $E_{xc}$ emerges from the difference of the exact kinetic energy of the interacting system ($T_{exact}(\rho)$), the kinetic energy of the non-interacting system ($T_S(\rho)$) and the non-classical exchange part ($K_{ex}(\rho)$). Here the $T_S(\rho)$ term is determined by the KS approach and the $K_{ex}(\rho)$ term results from the Coulomb interaction term of the electrons ($V_{ee}(\rho)$). The difficulty with this scheme is that the exchange correlation energy cannot be determined exactly because $T_{exact}(\rho)$ and $K(\rho)$ are not known \cite{40, 41}.
    
\subsection{Exchange-Correlation Energy}
\begin{figure}[h]
    \centering
    \includegraphics[scale=0.85]{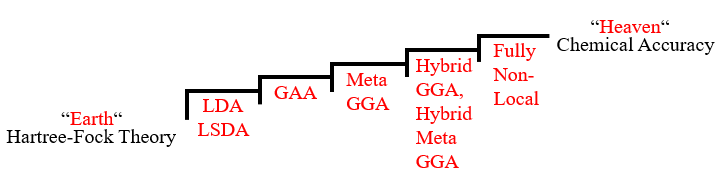}
    \caption{Jacob’s ladder showing the different functionals (LDA, LSDA, GGA, Meta GGA, Hybrid GGA, Hybrid Meta GGA and Fully Non-Local) leading to increases of the accuracy of calculations. The image is adapted from \cite{47}.}
    \label{fig:12}
\end{figure}

    With the methods described up to this point, the exact energy cannot be calculated and only approximations are possible. For these approximations, several methods have already been invented (see Fig.~\ref{fig:12}).
    The methods shown in Fig.~\ref{fig:12} and the quality of the resulting values vary considerably depending on the system. It is therefore required to find the most appropriate method for the system considered. It should be noted that the more accurate a calculation is, the greater is the impact on the computing time. It is therefore important to find the best balance between computation time and accuracy. The categorisation of exchange correlation energy calculation methods shown in Fig.~\ref{fig:12} is also called the Jacob´s ladder, and was introduced by Perdew \textit{et al.} \cite{46} in 2001.

    In Jacob's ladder, methods are ranked according to their accuracy. The ladder starts at the 'Earth' level, which corresponds to the HF method. The HF method is followed by the local density approximation (LDA) \cite{48}, and then the generalised gradient approximation (GGA) \cite{49}. The Jacob's ladder is completed when the "Heaven" level is reached, which corresponds to the exact exchange correlation energy. In this work the focus lies on the GGA level, but the LDA level is partly used as well. 
    
\subsection{Local Density Approximation}
    The first step above the "earth" in the Jacob´s ladder corresponds to the local density approximation (LDA). This approximation is the basis of all other exchange correlation functionals, although it is notoriously inaccurate. It is based on the model of a uniform electron gas where the electrons move on a positive background charge distribution to represent a neutral system. In this model, the number of electrons $N$ and the volume of the electron gas $V$ are assumed to be infinite. The electron density ($\frac{N}{V}$), on the other hand, is assumed to be always finite and constant everywhere. The exchange correlation energy according to the LDA can be set up as follows:
    \begin{equation}
        E_{xc}^{LDA} = \int\rho(r)\epsilon_{xc}[\rho(r)]dr.
       \label{eq:22}
    \end{equation}
    In the LDA exchange correlation functional, $\epsilon_{xc}[\rho(r)]$ corresponds to the exchange correlation energy per particle in the uniform electron gas with density $\rho (r)$. This gets weighted by the probability $\rho (r)$ to ensure that the electron is really at that location in space. 
    
    Physically, this model corresponds to an idealised metal with a perfect crystal of valence electrons and a distorted positive core. This model can therefore be applied relatively well to systems with small fluctuations in density, such as sodium. However, in most real cases the density changes rapidly, so the results of this model are usually far from reality. The functional of this approximation is the only one in DFT whose form is known with high accuracy \cite{40, 41}.
    
\subsection{Local Spin Density Approximation}
    An unconstrained version of the LDA is the local spin density approximation (LSDA), which uses the spin densities $\rho_{\alpha}(r)$ and $\rho_{\beta}(r)$ (with $\rho_{\alpha}(r) + \rho_{\beta}(r) = \rho (r)$) instead of the electron density $\rho(r)$. This splitting of one density term into two spin density terms provides the functional with more flexibility. It increases the accuracy of the calculation in all cases where the electron numbers for $\alpha$ and $\beta$ are different. In consequence, the LDA functional can be transformed into that of LSDA \cite{40, 41}:
    \begin{equation}
        E_{xc}^{LSDA}[\rho_{\alpha}\rho_{\beta}] = \int\rho(r)\epsilon_{xc}[\rho_{\alpha}(r), \rho_{\beta}(r)]dr.
       \label{eq:23}
    \end{equation}
    In spin-compensated situations (like $\rho_{\alpha}(r) = \rho_{\beta}(r) = \frac{1}{2}\rho(r)$) and in spin-polarised situations (such as $\rho_{\alpha}(r) \neq \rho_{\beta}(r)$), LSDA can be used to calculate exchange and correlation energies.

    Like LDA, LSDA is on the first level of the Jacob´s ladder, which suggests disadvantages of the LSDA functional. These first level functionals tend to underestimate the ground state energy and ionisation of the system and overestimate the binding energy. LSDA finds their most valid applications in high spin systems \cite{41}.
    
\subsection{Generalised Gradient Approximation}
    Chemistry usually requires methods that can produce higher quality results than LDA or LSDA calculations, which is why they are rarely used in computational chemistry. A higher quality method is the generalised gradient approximation (GGA), which is one step above LDA in the Jacob´s ladder. As the name of the method suggests, the functional of GGA uses the density gradient $\nabla\rho(r)$ instead of the charge density at a given point $r$ \cite{40, 41}. 
    \begin{equation}
        E_{xc}^{GGA}[\rho] = \int\rho(r)\epsilon_{xc}(\rho(r), \nabla\rho)dr.
       \label{eq:24}
    \end{equation}

    GGA takes into account a slow electron density variation and thus provides data that is lost in LDA and LSDA. This is also the reason why GGA usually provides a better result than LDA or LSDA, with a few exceptions. The most commonly used GGA functionals are those of Perdew and Wang (PW91) \cite{50, 51} and Perdew, Burke and Ernzerhof (PBE) \cite{49}.

    Based on the results obtained by Sözen \textit{et al.} \cite{7, 35}, the lattice parameters and magnetic properties can be correctly calculated with the PBE functional. Furthermore, Erdmann \textit{et al.} \cite{step} showed that GGA gives partly inaccurate results in the calculation of the anisotropy, which is why the LSDA functional is also used in this work.
    
\subsection{Ultra-soft Pseudopotentials and the Projector - Augmented Wave Method}
    Another interaction to be considered in DFT besides the exchange correlation energy $E_{xc}$ is the nuclear-electron interaction. The two best-known and most effective methods accounted in this interaction are the projector-augmented-wave (PAW) method of Blöchl \cite{54, 55} and the ultra-soft pseudo-potential (USPP) method of Vanderbilt \cite{56}. In these methods, the motion effect of the nuclei is replaced by an effective potential. 

    In the USPP method, the radius of the nuclei is increased by soft and knotless pseudowave functions, which results in a reduction of the required basis set. An important prerequisite in the selection of the pseudowave functions is that they must have the same norm as the original wave function within a certain radius of the nucleus. Furthermore, the pseudowave functions outside this radius must have the same norm as the original wave function, too. This method has two serious disadvantages when calculating elements with highly localised electrons. Firstly, the calculation is expensive because it requires a very large base data set. Secondly, the pseudowave functions mentioned above are very difficult to generate, as many parameters have to be chosen. 

\begin{figure}[h]
    \centering
    \includegraphics[scale=0.89]{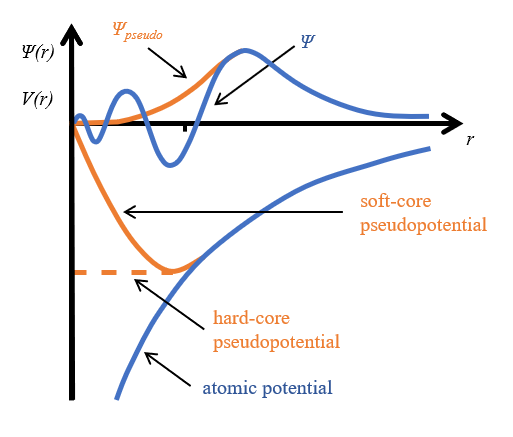}
    \caption{Radial dependence of typical pseudopotentials compared to the atomic Coulomb potential (blue). In the lower part a soft-core pseudopotential and a hard-core pseudopotential are shown with the solid and dashed orange curves, respectively. In the upper part the wave functions and the pseudo wave function resulting from the atomic (blue) and the pseudopotential (orange) are shown. Note that both wave functions have the same values above a certain point r. The picture is based on data by Böer \textit{et al.} \cite{57}.}
    \label{fig:13}
\end{figure} 

    The disadvantages of the USPP method are not found in the PAW method, since a linear transformation of the pseudowave functions to the many-particle wave function is used. In the PAW method the total energy is calculated by applying this linear transformation to the KS functionals. In addition, the lattice of augmentation charges is divided into a radial support lattice and a regular lattice, allowing to apply this method directly onto the many-electron wave function and the corresponding many-electron potentials \cite{7}.

\subsection{Magnetism in Density Functional Theory}
    Since in this thesis spin-polarised cases of the investigated hard magnetic materials are considered, the theorems of Hohenberg and Kohn have to be complemented. In particular, the theorems of Hohenberg and Kohn are only applicable to spinless cases and are therefore insufficient for spin-polarised cases. For spin-polarised cases, Barth and Hedin \cite{58} extended these theorems by replacing the scalar density by a hermitian 2x2 matrix, defined as follows:
    \begin{equation}
        n_{\alpha\beta}(r) = \sum_{i=1}^{N}\psi_i^{*\alpha}(r) \psi_i^{\beta}(r),
       \label{eq:25}
    \end{equation}
    where $\alpha$, $\beta$ = (+), (-). This matrix can be decomposed into a scalar and a vector part corresponding to the charge and magnetisation densities, respectively. Furthermore, the matrix with the Pauli matrices can be reformulated as follows:
    \begin{equation}
        n(r) = \frac{1}{2} (n(r)I+\sigma \cdot m(r)) = \frac{1}{2} 
    \left(\begin{array}{rr} 
        n(r)+m_z(r) & M_x(r)-im_y(r)\\
        m_x(r)+im_y(r) & n(r)-m_z(r)
    \end{array}\right).
       \label{eq:26}
    \end{equation}

    In the same way, the Schrödinger equation can be written as:
    \begin{equation}
        [(-\frac{\hbar^2}{2m}\nabla^2 + \sum_{\alpha}\int\frac{n_{\alpha\alpha (r')}}{|r - r'|}dr')I + \nu(r) + \frac{\delta E_{xc}}{\delta n(r)}]
    \left(\begin{array}{r} 
        \psi_i^{(+)}r\\
        \psi_i^{(-)}
    \end{array}\right) = \epsilon_i
    \left(\begin{array}{r} 
        \psi_i^{(+)}r\\
        \psi_i^{(-)}
    \end{array}\right).
       \label{eq:27}
    \end{equation}
    Here $I$ and the exchange correlation matrices correspond to 2x2 matrices, where $I$ corresponds to the unit matrix \cite{59}.

    The potential matrix can also take the following forms:
    \begin{equation}
        \nu(r) = \nu(r)I + \mu_B\sigma \cdot B(r),
       \label{eq:28}
    \end{equation}
    or
    \begin{equation}
        \nu_{XC}(r) = \nu_{XC}(r)I + \mu_B\sigma \cdot B_{xc}(r).
       \label{eq:29}
    \end{equation}
    In the equations \ref{eq:28} and \ref{eq:29} $B (r)$ corresponds to the magnetic field and $\mu_B$ with $\mu_B=\frac{e\hbar}{2mc}$ corresponding to the Bohr magneton. It is assumed that the magnetic structure is collinear and that the potential matrix is diagonal. As a result, the magnetic and exchange fields point in the direction of $z$. This leads to the fact that the already reformulated Schrödinger equation is decoupled and has to be rewritten as:
    \begin{equation}
        (-\frac{\hbar^2}{2m}\nabla^2 + \nu_{Coul}(r) + \nu (r) + B_z(r) + \nu_{xc}^{(+)}(r))\psi_i^{(+)}(r) = \epsilon_i^{(+)}\psi_i^{(+)}(r),
       \label{eq:30}
    \end{equation}
    \begin{equation}
        (-\frac{\hbar^2}{2m}\nabla^2 + \nu_{Coul}(r) + \nu (r) - B_z(r) + \nu_{xc}^{(-)}(r))\psi_i^{(-)}(r) = \epsilon_i^{(-)}\psi_i^{(-)}(r).
       \label{eq:31}
    \end{equation}
    Here $\nu_{Coul}$ corresponds to the classical Coulomb potential and $\nu_{xc}^{+,-}$ to the exchange correlation potential, with the two cases of spin up and spin down of the diagonal density matrix. The density matrices with spin up or spin down have to be solved separately \cite{59}. In case of collinear magnetic structures with the solution of these equations all kinds of magnetic structures like ferromagnetism, antiferromagnetism or ferrimagnetism can be obtained.

    The spin density and the spin moment are given by the following two equations:
    \begin{equation}
        m(r) = -\mu_B \sum_{\alpha,\beta}\psi_{\alpha}^{(+)}(r)\sigma_{\alpha\beta} \psi_{\beta}(r),
       \label{eq:32}
    \end{equation}
    and 
    \begin{equation}
        M_{spin} = \int m (r) dr = \int (n^{(+)}(r) - n^{(-)}(r))dr.
       \label{eq:33}
    \end{equation}
    
    The quality of the results then depend only on the functional used, which usually corresponds to that of GGA. Rarely the LSDA functional gives better result than that of GGA \cite{7}.
    
\subsection{Treatment of \textit{f}-electrons}
    Since $f$-electrons have a high risk of self-interaction, strongly correlating $f$-electrons cannot be taken into account in DFT. Electrical self-interaction is the phenomenon of the electron seeing itself and thus causing an artificial delocalisation of the electrons. Even the GGA and LSDA terms cannot completely eliminate this phenomenon in the Hartree term. That is why partially occupied $f$-states of the RE elements of Pr-Eu and Tb-Yb are particularly problematic. Exceptions in the series of RE elements are Gd with a half-filled $f$-shell and Ce in whose $\alpha$-Ce phase the electrons are delocalised (in $\gamma$-Ce strong localisation). 

    Thus a method for the treatment of $f$-electrons is needed. Some examples of such methods that allow $f$-electrons to be considered in DFT are the DFT+$U$ method \cite{60, 61} or the dynamical mean-field theory (DMFT) \cite{62}. In this thesis the DFT+$U$ method is used.

    The DFT+$U$ method adds a corrective energy term called the Hubbard $U$ repulsion. This energy term originates from the Hubbard model, where the Hamiltonian $\hat{H}$ describes the motion of particles on a lattice. In addition, Hubbard adds two further terms to the Hamiltonian. The first addition is a jump term which describes the probability that a particle will jump from one point to another. Secondly, Hubbard adds a penalty term corresponding to $U$, which represents the repulsion at the point that should cancel out the artificial delocalisation. The resulting energy functional of the DFT+$U$ method has the form:
    \begin{equation}
        D^{DFT+U}[n] = E^{DFT}[n] - E^{dc}[n_i] + E_{U}[n_i].
       \label{eq:34}
    \end{equation}
     
    In the energy functional, $E^U$ is the repulsion term given by~ $E^U[n_i]=\frac{1}{2}U \sum_{i\neq j}n_in_j$ added to the population of $n_i$. $E^{dc}$ corresponds to the correction term representing the double counting of the correlation effects.

    The DFT+$U$ method is usually used in the form of the Liechtenstein \cite{63} or the Dudarev \cite{61} interpretation. In this work, only the second interpretation is used. The energy functional of the Dudarev method of GGA+$U$ is as follows:
    \begin{equation}
        E_{Dudarev}^{DFT+U} = E^{DFT} + \frac{U_{eff}}{2} \sum_{I,\sigma}\sum_i n_i^{I,\sigma}(1 - n_i^{I,\sigma}),
       \label{eq:35}
    \end{equation}
    where $U_{eff}$ is the effective $U$ parameter, describing the Coulomb interaction at the site and playing a central role in this approach. Density of states (DOS) calculations can be performed to define this $U$ value. According to Lang \textit{et al.} \cite{74}, these calculations yield an experimentally determined value of 4.65 eV below the Fermi energy for $U_{eff}$ in the case of NdFe$_{11}$Ti for the $f$-peak. For 1:12 Nd-based compounds, the literature gives a Hubbard correction of $U$ = 5-6 eV \cite{step, 35, p60}.
    
\newpage

\thispagestyle{plain}
\section{Computational Details} \label{4}
    All theoretical calculations were performed by spin-polarised density functional theory (DFT) using the Vienna \textit{ab-initio} Simulation Package (VASP) \cite{64, 65}. The projector-augmented wave (PAW) method was applied, as implemented in VASP. The exchange correlation effects were treated within the Perdew-Burke-Ernzerhof (PBE) \cite{49} generalised gradient approximation (GGA). For magnetocrystalline anisotropy energy (MAE), the local spin density approximation (LSDA) \cite{48} was considered in addition to GGA. In these MAE calculations, optimised structures from GGA were used. 

    To scan the Brillouin zone, a $\Gamma$-centered $k$-point grid of 12$\times$12$\times$10 meshes was used for the 26-atom supercells employed. The cut-off energy of the plane wave basis was set to 500 eV and the smearing parameter was set to 0.1 eV. The convergence criteria within the self-consistent field scheme was set to $10^{-5}$ eV for all calculations except the anisotropy calculations, where the criteria was set to $10^{-7}$~eV. The choice of input parameters yields an energy convergence within an error of less than 1~meV/atom. 

    The $4f$-electrons of Nd were treated carefully, although this poses a challenge with DFT. The $f$-electrons of Nd have been considered in the in-core state, which provides good mechanical and elastic properties. However, the calculation of the magnetic properties fails due to the absence of $f$-electrons. Therefore, it is necessary to go beyond DFT, and therefore a DFT+$U$ treatment by the Dudarev method \cite{61} with a Hubbard correction of \textit{U} = 6 eV was performed. This Hubbard correction term has been used previously for 1:12 compounds on Nd basis in the literature \cite{step, 35, p60}, and good results have been obtained.

    Treating the 4$f$-electrons in the in-core state, hybridisation with other orbitals is completely neglected and instead atomic physics are applied to the Nd-4$f$ states. It has to be noted that only the spin magnetic moment is calculated from the self-consistent DFT calculations. To obtain the total magnetic moment, the orbital magnetic moment must therefore be added. The orbital magnetic moment composes of $g_{j} \cdot J$ where $g_{j}$ is the Landé $g$-factor ($g_{Nd}$ = $\frac{8}{11}$ \cite{72}) and $J$ denotes the total angular momentum of $\frac{9}{2}$. The total angular momentum results from Hund's first rule and the full spin polarisation ($S = \frac{3}{2}$) of the three $4f$-electrons of the element Nd. The orbital magnetic moment of Nd can be estimated to be 3.273~$\mu_B$, utilizing the equation $J = |L - S|$ (for elements with less than half a full $f$-shell) with $L = 6$. In case of the GGA+$U$ calculations, the calculated spin part of the Nd-$4f$-electrons and the value of $g_{j} \cdot J$ was added to avoid double counting for the spin amount of the $4f$-electrons.  

    In order to calculate the magnetocrystalline anisotropy, a collinear self-consistent calculation was performed. The resulting relaxed structures from GGA calculations were then used as input for the non-collinear calculations, which take the spin-orbit coupling into account. For each non-collinear run, the magnetic moment was aligned with one of the crystallographic directions [001], [010] and [100]. To show the progression of the magnetisation, three intermediate calculations were made between the easiest and the hardest of the three original directions.

    For Curie temperature ($T_C$), the Liechtenstein method \cite{28} was employed to determine the exchange interaction energies $J_{ij}$ for the ferromagnetic (FM) and local moment disorder (LMD) states of the 1:12 phase. This was achieved by applying DFT through the Korringa-Kohn-Rostoker (KKR) \cite{66, 67} Green's function method, as implemented in the AkaiKKR \cite{68} code, also known as MACHIKANEYAMA. This method implements the atomic sphere approximation (ASA) incorporating the coherent potential approximation (CPA) \cite{69, 70}. Continuous concentration changes of both the RE and TM sublattices were considered based on the CPA. Moruzzi, Janak and Williams (MJW) \cite{71} parameterised the local density approximation (LDA) \cite{34, 35}, which serves as the foundation for all KKR calculations. In this study, the scattering of the system is considered up to $d$-scattering ($l_{max}=2$), \textit{i.e.} the $f$-electrons are placed into the valence state on basis of the open core approximation \cite{72, 73}. The respective results of the relaxation calculations of GGA-PBE are used as structural input for the $T_C$ calculations.

\newpage
\thispagestyle{plain}
\section{Results and Analysis}
    Now that the intrinsic magnetic properties (see Sec.~\ref{1.3}) as well as the theoretical background (see Chapter \ref{3}), and the detailed computational parameters (see Chapter \ref{4}) have been explained, the results can be presented. Shown are the physical and magnetic properties of the M$_{2}$Fe$_{24-y}$Ti$_{y}$ compounds (where M$_{2}$ = Zr$_{2}$, ZrNd and Nd$_{2}$ and y: 0 $\leq$ y $\leq$ 2) investigated in this work.
    
    In the first section of the following chapter, the phase stability of Zr-containing compounds are considered in terms of formation energy and solution enthalpy. Furthermore, the physical properties such as lattice constants and cell volume are discussed. In the following two sections the magnetisation is examined with the total magnetic moment and the saturation magnetisation, followed by the maximum energy product. Beyond that, the Curie temperature, then the anisotropy and finally the hardness factor are discussed.
    
\subsection{Stability of Ti, Co and Zr substituted 1:12 compounds} \label{5.1}
\begin{figure}[h]
    \centering
    \includegraphics[scale=0.54]{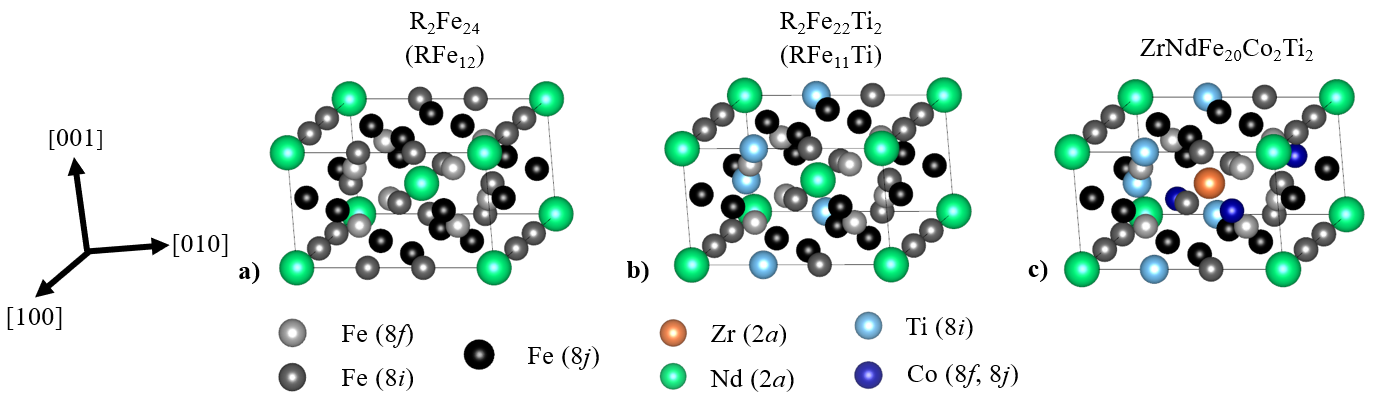}
    \caption{Schematic representation of the 2 formula unit (f.u.) (26 atoms) body-center-tetragonal ThMn$_{12}$ (I4/$mmm$, space group number 139) 1:12 phases. (a) Unstable RFe$_{12}$, (b) 2 Ti atom substituted (7.7~at.\%) R$_{2}$Fe$_{22}$Ti$_{2}$ (RFe$_{11}$Ti) and (c) 2 Ti and 2 Co substituted (each 7.7~at.\%) supercell RFe$_{10}$CoTi (R$_{2}$Fe$_{20}$Co$_{2}$Ti$_{2}$). Ti atoms are substituted in energetically favorable 8$i$ sites and Co atoms are substituted in energetically favorable 8$f$ and 8$j$ sites.}
    \label{fig.16}
\end{figure}

    Besides the well-known Nd$_{2}$Fe$_{14}$B magnet \cite{p39, p40}, another class of typical compounds fulfilling the technological requirements for hard magnetic applications is in the form of R-Fe$_{12-y}$Ti$_{y}$. Fig.~\ref{fig.16}(a) shows a schematic representation of the RFe$_{12}$ compound. 

    There is a well-known issue that REFe$_{12}$ compounds are considered as thermodynamically unstable \cite{36, Halil2019} and partial substitution of Fe atoms is needed for the stabilisation of a bulk system with the ThMn$_{12}$ structure. Therefore, Ti, which is a commonly used and typical stabilising element \cite{33, p41} is used. The effect of Co substitution is investigated on both the stabilisation and on the magnetic properties. To these TM-site substitutions, the Zr substitution on RE-site was examined to make Nd concentration lean. 
 
    Primarily, the phase formation energies of the ternary phases with an X substitution (X = Ti and Co) are considered:
 
\begin{equation} \label{eq:36}
\begin{split}
       E^{f}(\textrm{Nd}_2\textrm{Fe}_{24-y}\textrm{X}_y) &= E(\textrm{Nd}_2\textrm{Fe}_{24-y}\textrm{X}_y) - 2\mu_{Nd}-(24-y)\mu_{Fe}-y\mu_{X} \\
      & = 2E^f(\textrm{NdFe}_{12})+E^{sol}(y)
\end{split}
\end{equation}

    $E(\textrm{Nd}_2\textrm{Fe}_{24-y}\textrm{X}_y)$ is the total energy of X substituted 1:12 phase. X can be Ti or Co at the TM sublattice and for a given Ti and Co concentration all possible configurations have been investigated and the energetically most favorable configuration is taken into account in this formula. Note, that a similar formulation can be written for Zr substitution on RE-site as well. $\mu$ stands for the ground state of the considered elements. In case of $\mu_{Fe}$ this is the body-centered cubic (BCC) and ferromagnetic (FM) structure. For the elements Co ($\mu_{Co}$), Ti ($\mu_{Ti}$), Nd ($\mu_{Nd}$) and Zr ($\mu_{Zr}$) the hexagonal structure (HCP) was considered. The elements Ti, Nd, and Zr are treated as non-magnetic (NM) and Co in FM state. 

    The phases are thermodynamically stable (unstable), for negative (positive) formation energies. Although pure unaries will be considered as reference, the results are expressed in terms of the energy of formation $E^f(\textrm{NdFe}_{12})$ of the unstable phase NdFe$_{12}$. Here, instead of the base compound, different formulations can be taken into account. However, due to the missing quaternary phase diagrams, natural reference was considered, which are again base compounds and unaries:
 
\begin{equation} \label{eq:365}
    E^{sol}(y) = E(\textrm{Nd}_2\textrm{Fe}_{24-y}\textrm{X}_y)-2E(\textrm{NdFe}_{12})+y(\mu_{Fe}-\mu_{Ti}).
\end{equation}
    
    Since the solution enthalpy calculations include reference elements, which are critical for the resulting energies, an accurate modeling is necessary. Table~\ref{tab.3} compares the physical and magnetic properties of the pure elements with experimental data. 

    According to the calculations the Nd-based 1:12 compound, NdFe$_{12}$, is thermodynamically unstable with a formation energy of 0.94~eV, which agrees to previous reports \cite{36, 37, 38}. Nevertheless, ZrFe$_{12}$ was calculated to be stable with a formation energy of -0.76~eV, which agrees to the other theoretical work \cite{appl}. It is not the case for the 50\% replacement of Nd with Zr, (ZrNd)Fe$_{12}$ compound, as it has a formation energy of 0.16~eV.
    
\begin{table}[h]
\tiny
\caption{Calculated physical and magnetic properties of the pure elements that are included in solution enthalpy formalism in Eq.~\ref{eq:365}. \textit{a} and \textit{c} are the lattice constants, $B_0$ is the bulk modulus and $m_{tot}$ is the magnetic moment (where NM stands for nonmagnetic).}
\centering
\label{tab.3}
\hspace*{\fill}\begin{tabular}{cccccccccc}
\hline
Alloy & \multicolumn{3}{c}{\begin{tabular}[c]{@{}c@{}}Lattice constants\\ (\AA)\end{tabular}}                & \multicolumn{2}{c}{\begin{tabular}[c]{@{}c@{}}Lattice constants\\ experimental (\AA)\end{tabular}}                & \begin{tabular}[c]{@{}c@{}}Bulk modulus\\ $B_{0}$\end{tabular} & \begin{tabular}[c]{@{}c@{}}Bulk modulus\\ experimental\end{tabular} & $m_{tot}$        & \begin{tabular}[c]{@{}c@{}}$m_{tot}$ \\ experimental\end{tabular} \\
      & \textit{a}                      & \textit{b}                      & \textit{c}                       & \textit{a}                                              & \textit{c}                                              & (GPa)                                                          & (GPa)                                                               & ($\mu_{B}$/atom) & ($\mu_{B}$/atom)                                                  \\ \hline
Fe    & 2.832                           & 2.832                           & 2.832                            & 2.668$^{h}$ \cite{Fe}                                   & 2.668$^{h}$ \cite{Fe}                                   & 185.47                                                         & 159-173$^{d}$ \cite{Fe}                                             & 2.21             & 2.22$^{a}$ \cite{mag}                                             \\
Co    & 2.489                           & 2.489                           & 4.036                            & 2.5$^{a}$ \cite{Co}                                     & 4.1$^{a}$ \cite{Co}                                     & 212.15                                                         & 199 $\pm$ 6$^{a}$ \cite{Co}                                             & 1.63             & 1.53$^{e}$ \cite{mag}                                             \\
Ti    & 4.553                           & 4.553                           & 2.817                            & 4.60$^{a}$ \cite{Ti}                                    & 2.83$^{a}$ \cite{Ti}                                    & 128.93                                                         & 102-119$^{a}$ \cite{Ti}                                             & NM               & NM$^{f}$ \cite{magti}                                             \\
Nd    & 3.704                           & 3.704                           & 11.935                           & 3.658$^{d}$ \cite{Nd}                                   & 11.839$^{d}$ \cite{Nd}                                  & 33.84                                                          & 25.4-34.9$^{d}$ \cite{Nd}                                           & NM               & NM$^{a}$ \cite{magnd}                                             \\
Zr    & 3.235                           & 3.235                           & 5.166                            & 3.233$^{b}$ \cite{Zr}                                   & 5.146$^{b}$ \cite{Zr}                                   & 93.57                                                          & 92-102$^{b}$ \cite{Zr}                                              & NM               & NM$^{g}$ \cite{magzr}                                             \\ \hline
\end{tabular}\hspace*{\fill}
\begin{tablenotes}[flushleft]
{\scriptsize \setstretch{0.25}
    \item $^a$ X-ray diffraction (XRD) analysis at 300 K.\newline
    \item $^b$ High-pressure (ultrasonic) measurements combined with X-ray techniques.\newline
    \item $^c$ X-ray diffraction (XRD) analysis at $>$573 K. \newline
    \item $^d$ With Vibrating Sample Magnetometer at 300 K. \newline
    \item $^e$ Neutron Diffraction at 300~K.\newline
    \item $^f$ By measuring the flux change.
    }
\end{tablenotes}
\end{table}
    
    As 50\% substitution of Nd by Zr is apparently not sufficient to stabilise the ternary phase, additional substitution of Fe with Ti is required. For the compounds containing one Ti atom, NdFe$_{11.5}$Ti$_{0.5}$ and (ZrNd)Fe$_{11.5}$Ti$_{0.5}$, the formation energies were calculated to be -0.03~eV and -0.79~eV, respectively. Comparing these values, it can be seen that in the case of the Zr-substituted alloy, the substitution of one Ti atom is sufficient to stabilise the compound. In case of the ternary compound without Zr, the formation energy is just below 0~eV. Since Zr has the potential to reduce Ti concentration, the magnetic properties are enhanced.

\begin{figure}[h]
    \centering
    \includegraphics[scale=0.43]{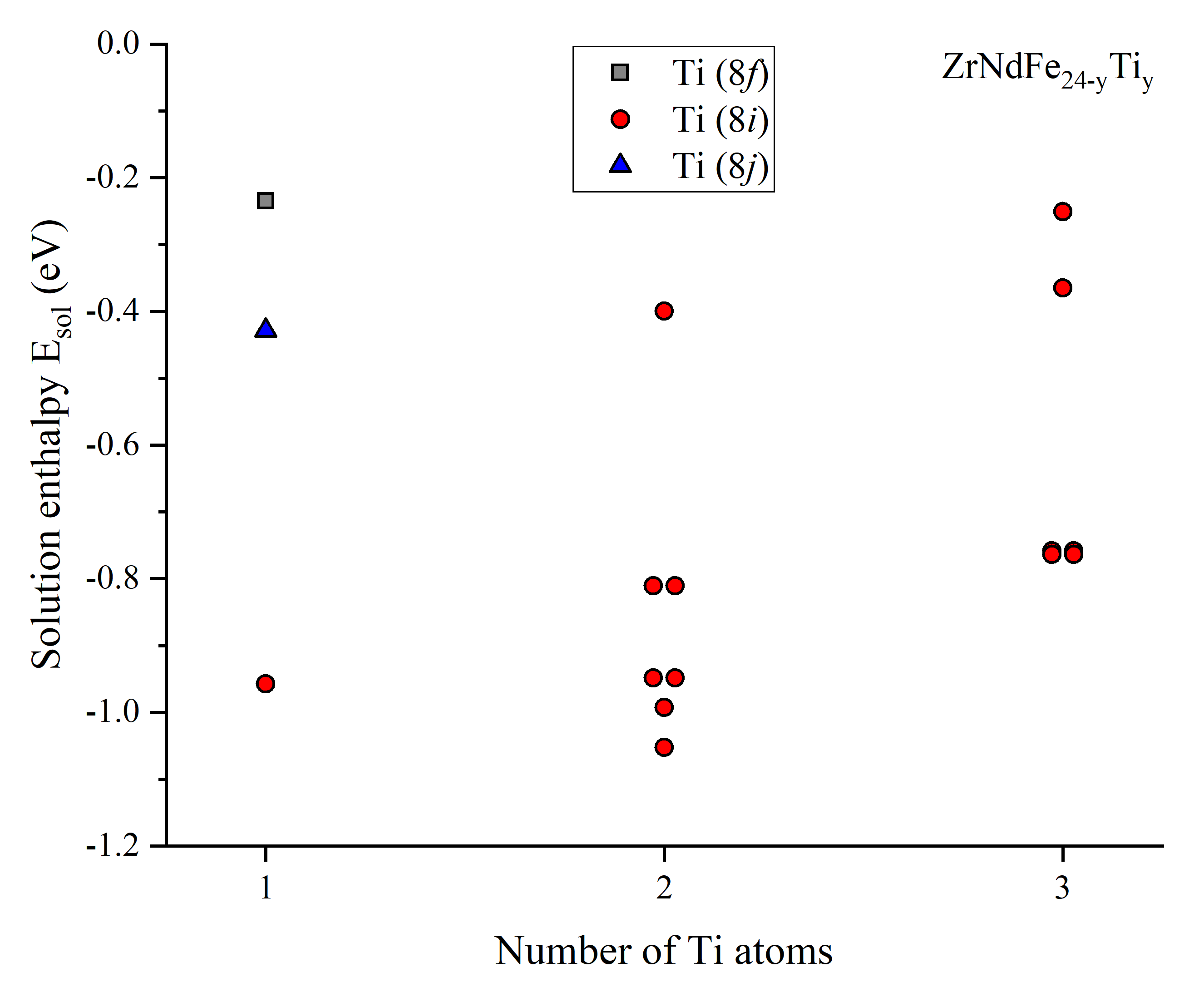}
    \caption{Calculated Ti solution enthalpies for (ZrNd)Fe$_{24-y}$Ti$_y$ according to Eq.~\ref{eq:365}. The first Ti atom is checked in 8$i$, 8$j$ and 8$f$ sites one by one. Then the remaining Ti atoms are substituted on 8$i$ site due to the lower solution enthalpy and all possible configurations are considered.}
    \label{fig:14}
\end{figure}

    The effect of Ti stabilisation has been investigated for the ZrNdFe$_{24-y}$Ti$_y$ compounds (see Fig.~\ref{fig:14}). It is started with the substitution of the first Ti atom in each TM sites, \textit{i.e.} 8$i$, 8$j$ and 8$f$, in the supercell consisting of 26 atoms (2 formula units (f.u.)). Note that a single Ti atom corresponds to 3.8~at.\%. The calculated solution enthalpy values are shown in Fig.~\ref{fig:14}. As in previous works for Ce and Nd-based 1:12 compounds \cite{Halil2019, 35}, the 8$i$ site is calculated to be the energetically most favorable site. The solution enthalpy for the 8$j$ and 8$f$ sites are higher by 0.53 and 0.72~eV, respectively. Thus only the 8$i$ sublattice was considered for the substitution for the remaining Ti atoms.
 
    The exclusion of the 8$j$ and 8$f$ sublattices is also supported by the experimental works \cite{9, p45}. The origin of the 8$i$ preference is due to the larger Wigner-Seitz radius of this site compared to the 8$f$ and 8$j$ sites \cite{p47}. In case of the substitution of the second and third Ti atom all possible configurations on 8$i$ sites were considered. The Ti solubility investigations on (ZrNd)Fe$_{24-y}$Ti$_y$ yield a equilibrium Ti concentration of 7.7 at.\%, which is the same as in the case of NdFe$_{12-x}$Ti$_x$ and CeFe$_{12-x}$Ti$_x$ \cite{35, step, Halil2019}. The corresponding structure is given in Fig.~\ref{fig.16}(b).
    
\newpage

    Although Ti stabilises the Zr substituted 1:12 compounds, it reduces the total magnetic moment due to its nonmagnetic (NM) nature. For instance (ZrNd)Fe$_{22}$Ti$_2$ has 4.06 $\mu_{B}$/f.u. less total magnetic moment than the Ti-free 1:12 compound. Therefore, ferromagnetic (FM) Co was considered as an another substitutional element for (ZrNd)Fe$_{24-x}$Co$_x$ and (ZrNd)Fe$_{22-x}$Ti$_2$Co$_x$, to improve magnetic properties and understand its impact on the stability. 

\begin{figure}[t]
        \centering
        \includegraphics[scale=0.47]{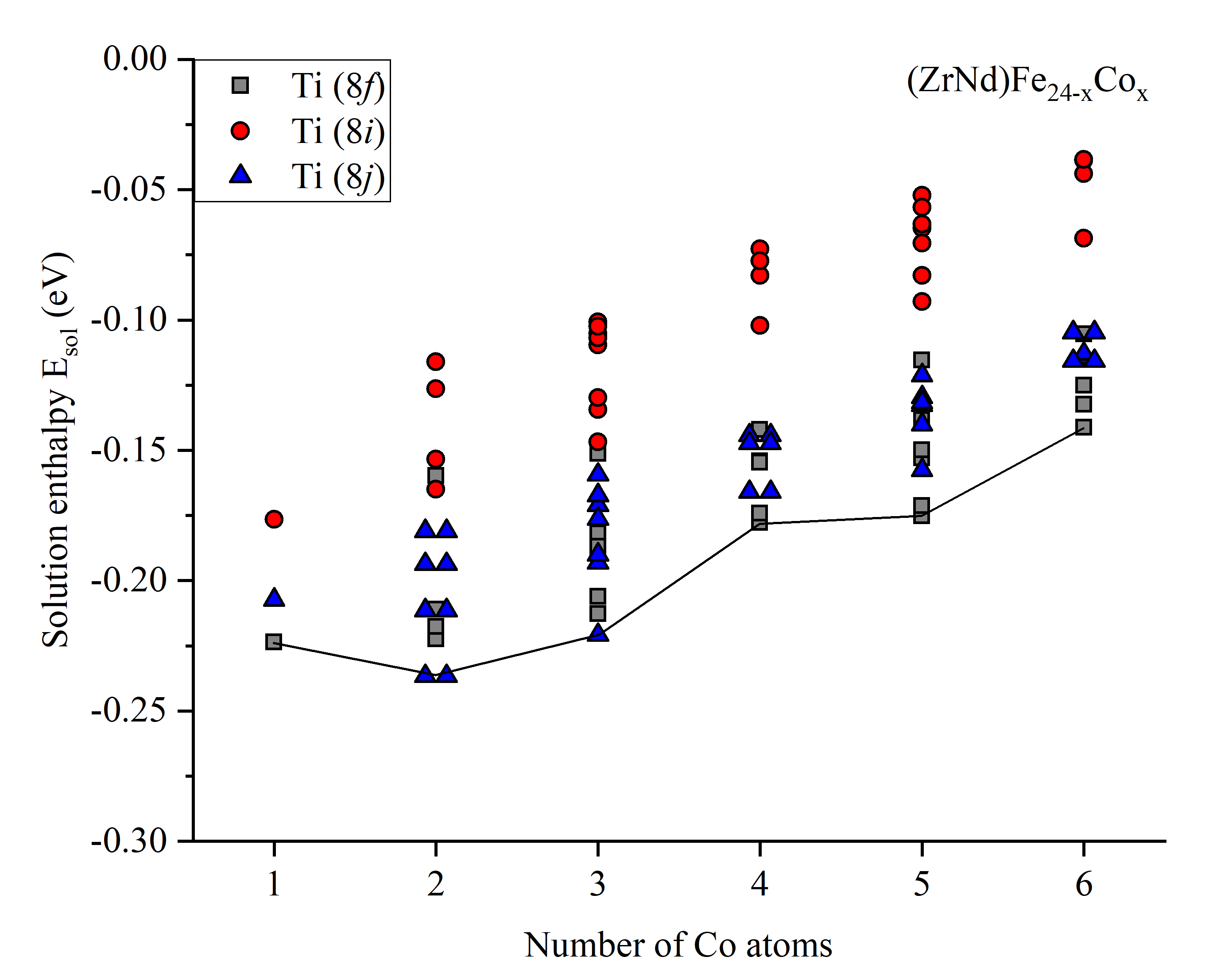}
        \caption{Calculated Co solution enthalpies for chemical compositions with the formula (ZrNd)Fe$_{24-x}$Co$_x$ according to Eq.~\ref{eq:365}.}
        \label{fig.15a}
    \end{figure}

\begin{figure}[h]
        \centering
        \includegraphics[scale=0.47]{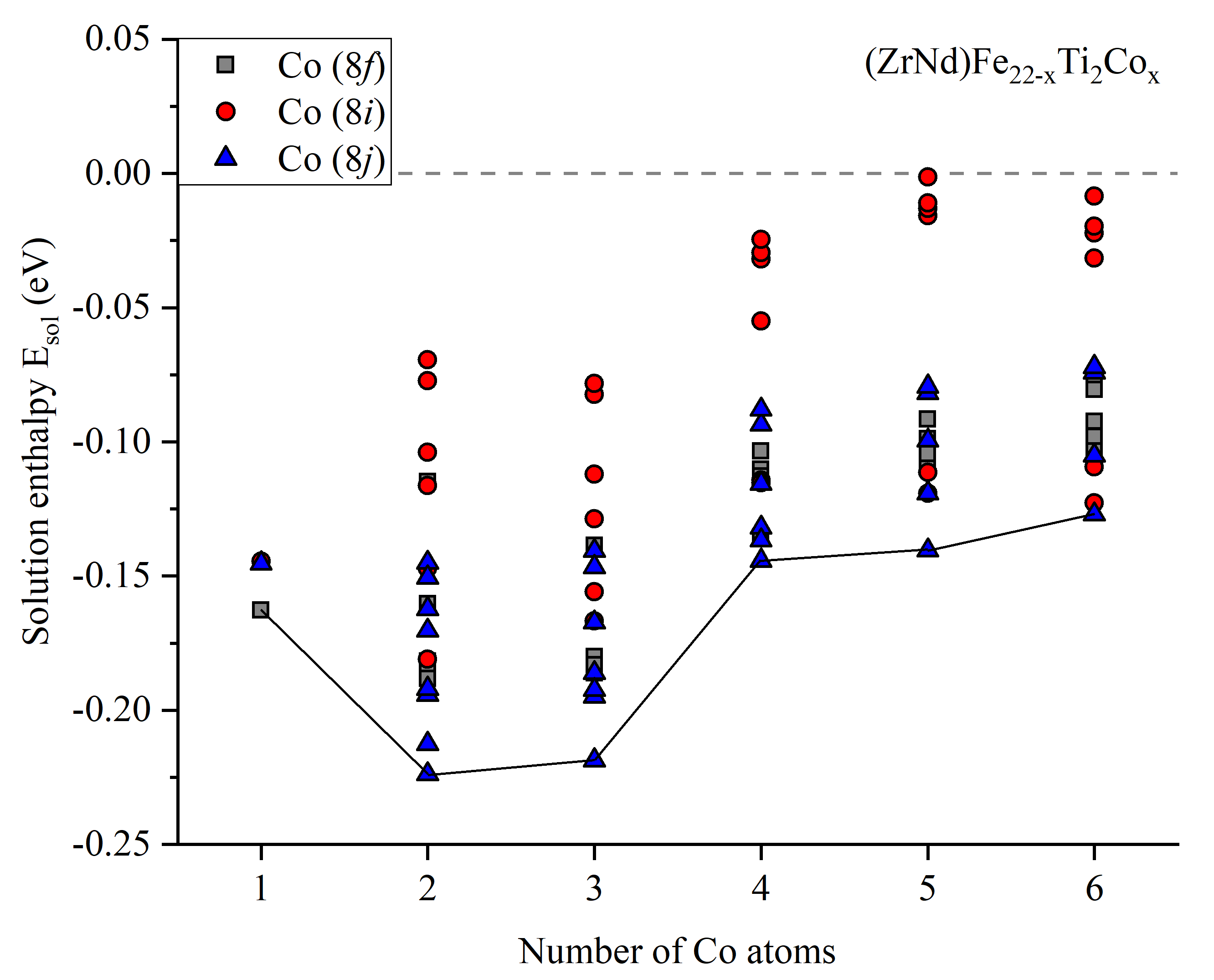}
        \caption{Calculated Co solution enthalpies for chemical compositions with the formula (ZrNd)Fe$_{22-x}$Ti$_2$Co$_x$ according to Eq.~\ref{eq:365}.}
        \label{fig.15b}
    \end{figure}

    Figures~\ref{fig.15a} and \ref{fig.15b} show the site preference of Co substitutions and equilibrium concentrations for the considered 1:12 compounds. As shown, Co prefers to substitute on 8$f$ and 8$j$ sites for both (ZrNd)Fe$_{24-x}$Co$_x$ and (ZrNd)Fe$_{22-x}$Ti$_2$Co$_x$ cases. In case of the (ZrNd)Fe$_{24-x}$Co$_x$ compounds, it can be seen in Fig.~\ref{fig.15a} that the first Co atom is substituted into the 8$f$ side with -0.22~eV, followed by the 8$j$ and 8$i$ with -0.21~eV and -0.18~eV, respectively. After keeping the 1$^{st}$ Co atom in the 8$f$ position and replacing the second Co atom, the lowest values of the 8$j$, 8$f$ and 8$i$ sides are obtained with -0.24~eV, -0.22~eV and -0.17~eV, respectively. As the equilibrium concentration is at 2 Co atoms, additional Co atoms lead to higher values and therefore less stabilisation. The 8$j$ site gives the most negative value for the third Co atom. From the 4$^{th}$ to the 6$^{th}$ Co atom the 8$f$ side gives the lowest energies. 
    
\newpage
    
    As can be seen in Fig.~\ref{fig.15b} in case of the (ZrNd)Fe$_{22-x}$Ti$_2$Co$_x$ compounds, again the substitution of Co into the 8$f$ side with -0.16~eV, followed by 8$j$ and 8$i$ with about -0.14~eV, leads to the most stabilised compound. For the 2$^{nd}$ Co atom the 8$j$ side with -0.22~eV again gives the most negative value, followed by 8$f$ with -0.19~eV \\ and 8$i$ with -0.18~eV. For the 3$^{rd}$ to 6$^{th}$ Co atoms, the 8$j$ side gives the lowest energies. The resulting preference of Co in the 8$f$ and 8$j$ sublattices is likely due to the fact that the 8$i$ sublattice has a larger Wigner-Seitz radius and Co has a smaller atomic radius than Fe. The theoretically calculated site preference is in agreement with experimental reports \cite{FU22, FU24, 2016}, which found that Co prefers the 8$f$ and 8$j$ sites. In addition, for both series of compounds, a trend is observed that the 8$i$ site generally has higher solution enthalpies and the 8$j$ and 8$f$ sites are more stable. A Co equilibrium concentration of 7.7 at\% was calculated for both compounds. The schematic representation of the most stable quinary phase is shown in Fig.~\ref{fig.16}(c).

    The physical and magnetic properties of the lowest energy configurations of considered Co concentrations for (ZrNd)Fe$_{24-x}$Co$_x$ and (ZrNd)Fe$_{22-x}$Ti$_2$Co$_x$, are given in Tab.~\ref{tab.4}. In case of Co substitution in (ZrNd)Fe$_{24-x}$Co$_x$ a slight increase in magnetic properties is observed. For example, the total magnetic moment increases by 0.44 $\mu_{B}$ from 0 Co atoms to 6 Co atoms. In the case of substitution in (ZrNd)Fe$_{22-x}$Ti$_2$Co$_x$, a decrease of the total magnetic moment of -0.90 $\mu_{B}$ is observed with the substitution of 6 Co atoms. Note, that this increase (decrease) becomes visible only with the substitution of the 4$^{th}$ Co atom. Thus, Co in its equilibrium concentration has only a minor effect on the magnetic properties. The substitution of a Ti or Co atom into (ZrNd)Fe$_{24}$ leads to a stabilisation of -0.96~eV or -0.22~eV, respectively. Therefore, Co has a smaller effect on the stability of the compound than Ti. It can be seen that in case without Ti, the cell volume constantly increases (from 165.46~\AA$^{3}$ (0 Co) to 165.91~\AA$^{3}$ (6 Co)) with Co, whereas with Ti a decrease (from 167.93~\AA$^{3}$ (0 Co) to 166.84~\AA$^{3}$ (6 Co)) can be seen.
    
\begin{table}[h]
\tiny
   \caption{DFT calculated physical and magnetic properties of the most stable Co containing compounds (see Figs.~\ref{fig.15a} and \ref{fig.15b}) starting from ZrNdFe$_{24}$ and ZrNdFe$_{22}$Ti$_{2}$ with increasing Co content.}
   \centering
   \label{tab.4}
\begin{tabular}{ccccccccc}
\hline
\multirow{3}{*}{Alloy}        & \multicolumn{3}{c}{Lattice constants {(}\AA{)}}              & \multirow{3}{*}{Cell volume}      & \multirow{3}{*}{$m^{2a}_{tot}$ }                    & \multirow{3}{*}{$m_{tot}$}                 & \multirow{3}{*}{$\mu_{0}M_{S}$}               & \multirow{3}{*}{$|BH|_{max}$}                   \\
                              & \multirow{2}{*}{$a$} & \multirow{2}{*}{$b$} & \multirow{2}{*}{$c$} & \multirow{3}{*}{{(}\AA$^{3}${)}}  & \multirow{3}{*}{{(}$\mu_{B}$/f.u.{)}}               & \multirow{3}{*}{{(}$\mu_{B}$/f.u.{)}}      & \multirow{3}{*}{{(}T{)}}                      & \multirow{3}{*}{{(}kJ/m$^{3}${)}}               \\
                              &                    &                    &                    &                                   &                                                     &                                            &                                               &                                                 \\ \hline
ZrNdFe$_{24}$                 & 8.422              & 8.422              & 4.666              & 165.46                            & -0.50 (m$^{2a}_{Zr}$)  -0.26 (m$^{2a}_{Nd}$)        & 27.58                                      & 1.94                                          & 750.70                                          \\
ZrNdFe$_{23}$Co               & 8.419              & 8.419              & 4.659              & 165.15                            & -0.50 (m$^{2a}_{Zr}$)  -0.26 (m$^{2a}_{Nd}$)        & 27.34                                      & 1.93                                          & 740.30                                          \\
ZrNdFe$_{22}$Co$_{2}$         & 8.407              & 8.423              & 4.664              & 165.13                            & -0.50 (m$^{2a}_{Zr}$)  -0.26 (m$^{2a}_{Nd}$)        & 27.34                                      & 1.93                                          & 740.97                                          \\
ZrNdFe$_{21}$Co$_{3}$         & 8.417              & 8.417              & 4.665              & 165.24                            & -0.50 (m$^{2a}_{Zr}$)  -0.26 (m$^{2a}_{Nd}$)        & 27.46                                      & 1.94                                          & 745.95                                          \\
ZrNdFe$_{20}$Co$_{4}$         & 8.419              & 8.417              & 4.672              & 165.55                            & -0.50 (m$^{2a}_{Zr}$)  -0.26 (m$^{2a}_{Nd}$)        & 27.70                                      & 1.95                                          & 756.67                                          \\
ZrNdFe$_{19}$Co$_{5}$         & 8.417              & 8.417              & 4.682              & 165.83                            & -0.51 (m$^{2a}_{Zr}$)  -0.26 (m$^{2a}_{Nd}$)        & 27.95                                      & 1.96                                          & 767.69                                          \\
ZrNdFe$_{18}$Co$_{6}$         & 8.410              & 8.410              & 4.691              & 165.91                            & -0.50 (m$^{2a}_{Zr}$)  -0.26 (m$^{2a}_{Nd}$)        & 28.02                                      & 1.97                                          & 770.45                                          \\ \hline
ZrNdFe$_{22}$Ti$_{2}$         & 8.408              & 8.493              & 4.703              & 167.93                            & -0.49 (m$^{2a}_{Zr}$)  -0.27 (m$^{2a}_{Nd}$)        & 23.07                                      & 1.60                                          & 509.90                                          \\
ZrNdFe$_{21}$CoTi$_{2}$       & 8.408              & 8.495              & 4.708              & 168.13                            & -0.49 (m$^{2a}_{Zr}$)  -0.26 (m$^{2a}_{Nd}$)        & 23.19                                      & 1.61                                          & 513.86                                          \\
ZrNdFe$_{20}$Co$_{2}$Ti$_{2}$ & 8.401              & 8.490              & 4.712              & 168.03                            & -0.49 (m$^{2a}_{Zr}$)  -0.26 (m$^{2a}_{Nd}$)        & 23.07                                      & 1.60                                          & 509.24                                          \\
ZrNdFe$_{19}$Co$_{3}$Ti$_{2}$ & 8.407              & 8.481              & 4.715              & 168.11                            & -0.49 (m$^{2a}_{Zr}$)  -0.26 (m$^{2a}_{Nd}$)        & 23.14                                      & 1.60                                          & 511.78                                          \\
ZrNdFe$_{18}$Co$_{4}$Ti$_{2}$ & 8.399              & 8.469              & 4.719              & 167.84                            & -0.48 (m$^{2a}_{Zr}$)  -0.26 (m$^{2a}_{Nd}$)        & 22.95                                      & 1.59                                          & 505.11                                          \\
ZrNdFe$_{17}$Co$_{5}$Ti$_{2}$ & 8.402              & 8.459              & 4.707              & 167.28                            & -0.47 (m$^{2a}_{Zr}$)  -0.25 (m$^{2a}_{Nd}$)        & 22.54                                      & 1.57                                          & 490.70                                          \\
ZrNdFe$_{16}$Co$_{6}$Ti$_{2}$ & 8.407              & 8.449              & 4.698              & 166.84                            & -0.47 (m$^{2a}_{Zr}$)  -0.25 (m$^{2a}_{Nd}$)        & 22.17                                      & 1.55                                          & 477.03                                          \\ \hline
\end{tabular}
\end{table}

\newpage
    The calculated physical properties of the compounds considered, including lattice constants and bulk moduli, are given in Tab.~\ref{tab.5}. A trend of decreasing lattice constants and cell volumes with Nd$_{2}$Fe$_{24-y}$Ti$_{y}$ $>$ ZrNdFe$_{24-y}$Ti$_{y}$ $>$ Zr$_{2}$Fe$_{24-y}$Ti$_{y}$ with $0 \leq y \leq 2$ can be demonstrated. This trend is in good agreement with existing theoretical and experimental values \cite{p45, FU21, step}. This behaviour has also been previously documented by Sakuma \textit{et al.} \cite{FU22}. Furthermore, an increasing volume with increasing Ti concentration is observed, which is consistent with the work of Erdmann \textit{et al.} \cite{step}. The substitution of Co into the quaternary compound ZrNdFe$_{22}$Ti$_{2}$ shows a minimal increase in cell volume (by 0.03 \AA$^3$) and the $c$ (by 0.015~\AA) lattice constant, while the $a$ (by 0.020~\AA) and $b$ (by 0.005~\AA) lattice constants are reduced slightly. As mentioned above, a DFT+$U$ treatment was also included for Nd containing compounds (see values in parenthesis in Tab.~\ref{tab.5}), showing the same trends. Looking at the bulk moduli, the same softening effect of Ti can be seen as in the work of Erdmann \textit{et al.} \cite{step}. This effect occurs again not only in the Nd$_{2}$Fe$_{24-y}$Ti$_{y}$ but also in the Zr$_{2}$Fe$_{24-y}$Ti$_{y}$ compounds as well as in the compounds containing both Nd and Zr with ZrNdFe$_{24-y}$Ti$_{y}$ (y: $0 \leq y \leq 2$) (see Tab.~\ref{tab.5}).

\begin{figure}[t!]
    \centering
    \includegraphics[scale=0.50]{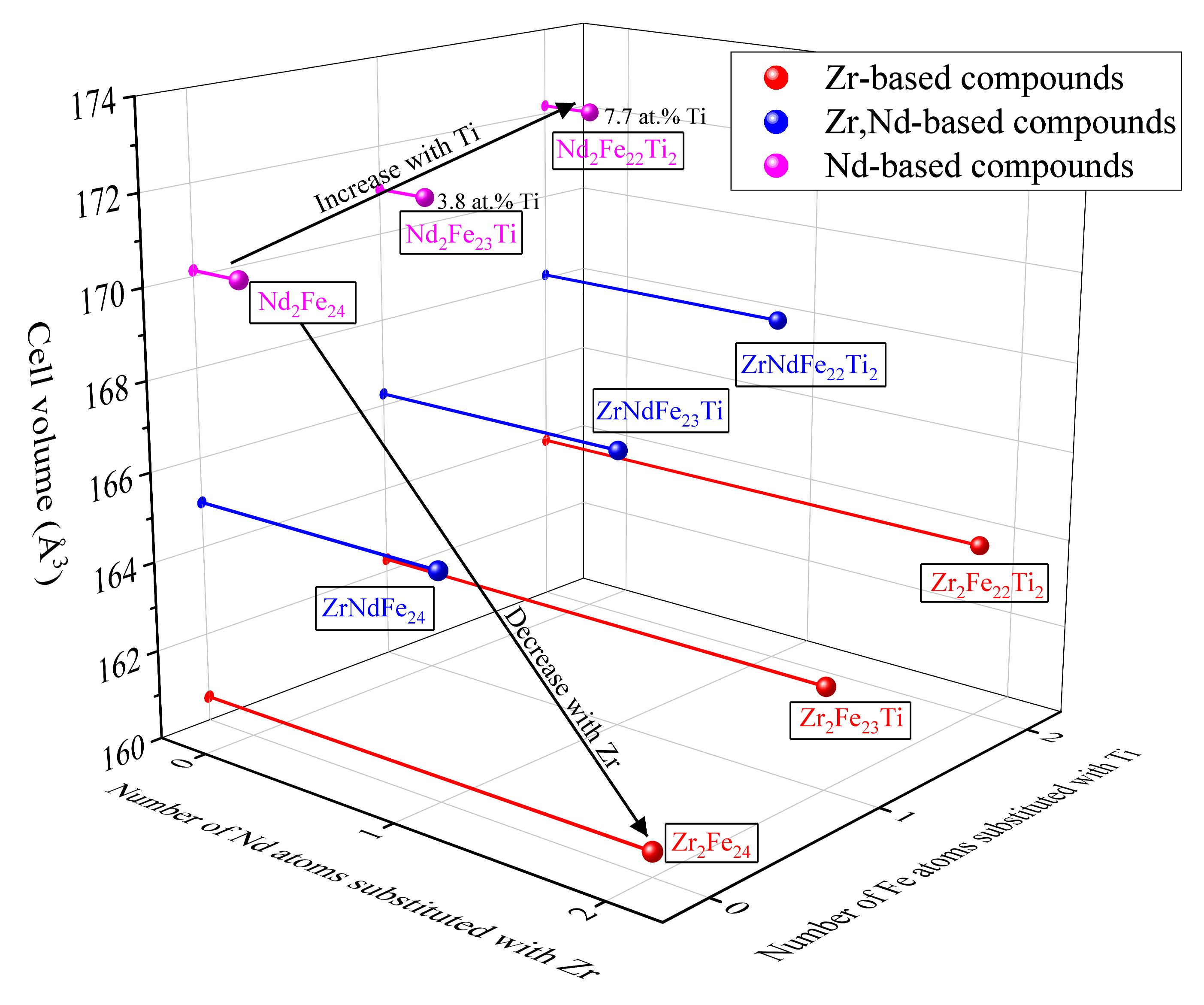}
    \caption{Schematic representation of cell volume evolution for two different substitutions, namely substitution of Ti into the Fe sublattice and substitution of Zr into the Nd sublattice, using GGA calculations. Pure Nd-containing compounds are shown in pink while compounds containing only Zr are shown in red. The quaternary compounds containing Zr and Nd are shown in blue.}
    \label{fig.17}
\end{figure}

\newpage

\begin{table}[h!]
\tiny
   \caption{Calculated lattice constant, cell volume and bulk modulus of the considered alloys using DFT. In case of Nd containing stable compounds DFT+$U$ has been considered for Nd 4$f$-electrons with $U$ = 6 eV and is given in parenthesis.}
   \centering
   \hspace{-1.5cm}
   \label{tab.5}
\begin{tabular}{cccccccccc}
\hline
\multirow{2}{*}{Alloy}                                              & \multicolumn{3}{c}{Lattice constants {(}\AA{)}}  & \multicolumn{2}{c}{Lattice constants }                         & \multirow{2}{*}{Cell}                  & \multirow{2}{*}{Cell volume}                      & \multirow{2}{*}{Bulk}                   & \multirow{2}{*}{Bulk modulus}                    \\
                                                                    &              &              &                    & \multicolumn{2}{c}{experimental {(}\AA{)}}                     & \multirow{2}{*}{volume}                & \multirow{2}{*}{experimental}                     & \multirow{2}{*}{modulus}                & \multirow{2}{*}{Literature}                      \\ 
                                                                    & $a$          & $b$          & $c$                & $a$                       & $c$                                & \multirow{2}{*}{{(}\AA$^{3}${)}}       & \multirow{2}{*}{{(}\AA$^{3}${)}}                  & \multirow{2}{*}{{(}GPa{)}}              & \multirow{2}{*}{{(}GPa{)}}                       \\ 
                                                                    &              &              &                    &                           &                                    &                                        &                                                   &                                         &                                                  \\ \hline
Nd$_{2}$Fe$_{24}$                                                   & 8.535        & 8.535        & 4.671              & 8.574* \cite{55}          & 4.907* \cite{55}                   & 170.13                                 & 180.37* \cite{55}                                 & 129.45                                  & 127.96* \cite{step}                              \\
(NdFe$_{12}$)                                                       &              &              &                    &                           &                                    &                                        &                                                   &                                         &                                                  \\
Nd$_{2}$Fe$_{23}$Ti                                                 & 8.568        & 8.530        & 4.680              & 8.552* \cite{step}        & 4.686* \cite{step}                 & 171.02                                 & 171.73* \cite{step}                               & 128.74                                  & 127.17* \cite{step}                              \\
(NdFe$_{11.5}$Ti$_{0.5}$)                                           & (8.558)      & (8.533)      & (4.675)            & (8.561)* \cite{step}      & (4.692)* \cite{step}               & (171.06)                               & (171.81)* \cite{step}                             &                                         &                                                  \\
Nd$_{2}$Fe$_{22}$Ti$_{2}$                                           & 8.593        & 8.535        & 4.696              & 8.574$^{a}$ \cite{54}     & 4.907$^{a}$ \cite{54}              & 172.19                                 & 176.2$^{a}$ \cite{54}                             & 128.89                                  & 127.20* \cite{step}                              \\
(NdFe$_{11}$Ti)                                                     & (8.584)      & (8.532)      & (4.703)            & 8.574$^{a}$ \cite{56}     & 4.794$^{a}$ \cite{56}              & (172.21)                               & 180.35$^{a}$ \cite{56}                            &                                         &                                                  \\ \hline
Zr$_{2}$Fe$_{24}$                                                   & 8.312        & 8.312        & 4.647              &                           &                                    & 160.54                                 &                                                   & 144.46                                  &                                                  \\
(ZrFe$_{12}$)                                                       &              &              &                    &                           &                                    &                                        &                                                   &                                         &                                                  \\
Zr$_{2}$Fe$_{23}$Ti                                                 & 8.285        & 8.377        & 4.677              &                           &                                    & 162.29                                 &                                                   & 139.03                                  &                                                  \\
(ZrFe$_{11.5}$Ti$_{0.5}$)                                           &              &              &                    &                           &                                    &                                        &                                                   &                                         &                                                  \\
Zr$_{2}$Fe$_{22}$Ti$_{2}$                                           & 8.258        & 8.433        & 4.708              & 8.358* \cite{ndzr}        & 4.715* \cite{ndzr}                 & 163.93                                 &                                                   & 139.58                                  &                                                  \\
(ZrFe$_{11}$Ti)                                                     &              &              &                    &                           &                                    &                                        &                                                   &                                         &                                                  \\ \hline
ZrNdFe$_{24}$                                                       & 8.419        & 8.419        & 4.656              &                           &                                    & 165.01                                 &                                                   & 132.98                                  &                                                  \\
((Zr$_{0.5}$Nd$_{0.5}$)Fe$_{12}$)                                   &              &              &                    &                           &                                    &                                        &                                                   &                                         &                                                  \\
ZrNdFe$_{23}$Ti                                                     & 8.439        & 8.439        & 4.669              &                           &                                    & 166.27                                 &                                                   & 130.23                                  &                                                  \\
((Zr$_{0.5}$Nd$_{0.5}$)Fe$_{11.5}$Ti$_{0.5}$)                       & (8.439)      & (8.440)      & (4.666)            &                           &                                    & (166.20)                               &                                                   &                                         &                                                  \\
ZrNdFe$_{22}$Ti$_{2}$                                               & 8.411        & 8.497        & 4.704              &                           &                                    & 168.08                                 &                                                   & 130.59                                  &                                                  \\
((Zr$_{0.5}$Nd$_{0.5}$)Fe$_{11}$Ti)                                 & (8.352)      & (8.508)      & (4.703)            &                           &                                    & (167.10)                               &                                                   &                                         &                                                  \\ \hline
ZrNdFe$_{20}$Co$_{2}$Ti$_{2}$                                       & 8.391        & 8.492        & 4.719              &                           &                                    & 168.11                                 &                                                   & 133.30                                  &                                                  \\
                                                                    & (8.399)      & (8.483)      & (4.721)            &                           &                                    & (168.17)                               &                                                   &                                         &                                                  \\
(Nd$_{0.7}$Zr$_{0.3}$)$_{2}$(Fe$_{0.75}$Co$_{0.25}$)$_{23}$Ti       &              &              &                    & 8.536$^{a}$ \cite{FU21}   & 4.770$^{a}$ \cite{FU21}            &                                        & 173.8$^{a}$ \cite{FU21}                           &                                         &                                                  \\ \hline
\end{tabular}
\begin{tablenotes}[flushleft] 
{\footnotesize{\scriptsize \setstretch{0.25}
    \item * Theoretical references are represented by *. \newline
    \item $^a$ X-ray diffraction (XRD) analysis at 300 K.
    }}
\end{tablenotes}
\end{table}

    As a first conclusion, the initial calculations show that Ti-enriched compounds (\textit{i.e.} ZrNdFe$_{24-y}$Ti$_{y}$) can be regarded as promising candidates for future investigations. Note that Co-enrichment appears to be inferior to Ti, as it has less effect on stabilisation. Moreover, the first calculations predict only a small effect of Co on the magnetic properties (see Tab.~\ref{tab.4}).

\subsection{Total Magnetic Moment and Magnetisation}
    The theoretically calculated values of the intrinsic magnetic properties, such as the total magnetic moment $m_{tot}$, the spin magnetic moment of the atoms in 2$a$ position $m^{2a}_{spin}$ and the saturation magnetisation $\mu_{0}M_{S}$, are presented in Tab.~\ref{tab.6}. These values are displayed together with previous theoretical and experimental values of the same or similar compounds. It is known that the used PBE functional tends to overestimate the magnetic moments \cite{p57}, which means that the calculated values are in the expected range. This overestimation can be evaluated in case of NdFe$_{11}$Ti using literature values. In Zr$_{2}$Fe$_{24-y}$Ti$_{y}$ (y: $0 \leq y \leq 2$) compounds it is not possible, since neither theoretical nor experimental values have been published so far. It should also be noted that even the relaxed structure does not guarantee that the magnetic properties are correctly reproduced. For the quaternary ZrNdFe$_{24-y}$Ti$_{y}$ (y: $0 \leq y \leq 2$) alloys, it can be estimated that the values are conceivable, since both experimental and theoretical values for similar compounds are available and in the expected range (see Tab.~\ref{tab.6}). For the Zr$_{2}$Fe$_{24-y}$Ti$_{y}$ (y: $0 \leq y \leq 2$) compounds it can at least be assumed that the magnetic moments of Fe and Ti are quite similar to those of Nd$_{2}$Fe$_{24-y}$Ti$_{y}$ (y: $0 \leq y \leq 2$) compounds, since only the $2a$ sites differ with Zr at -0.47~$\mu_{B}$ and Nd at -0.27~$\mu_{B}$. 
    
\begin{figure}[h]
    \centering
    \includegraphics[scale=0.52]{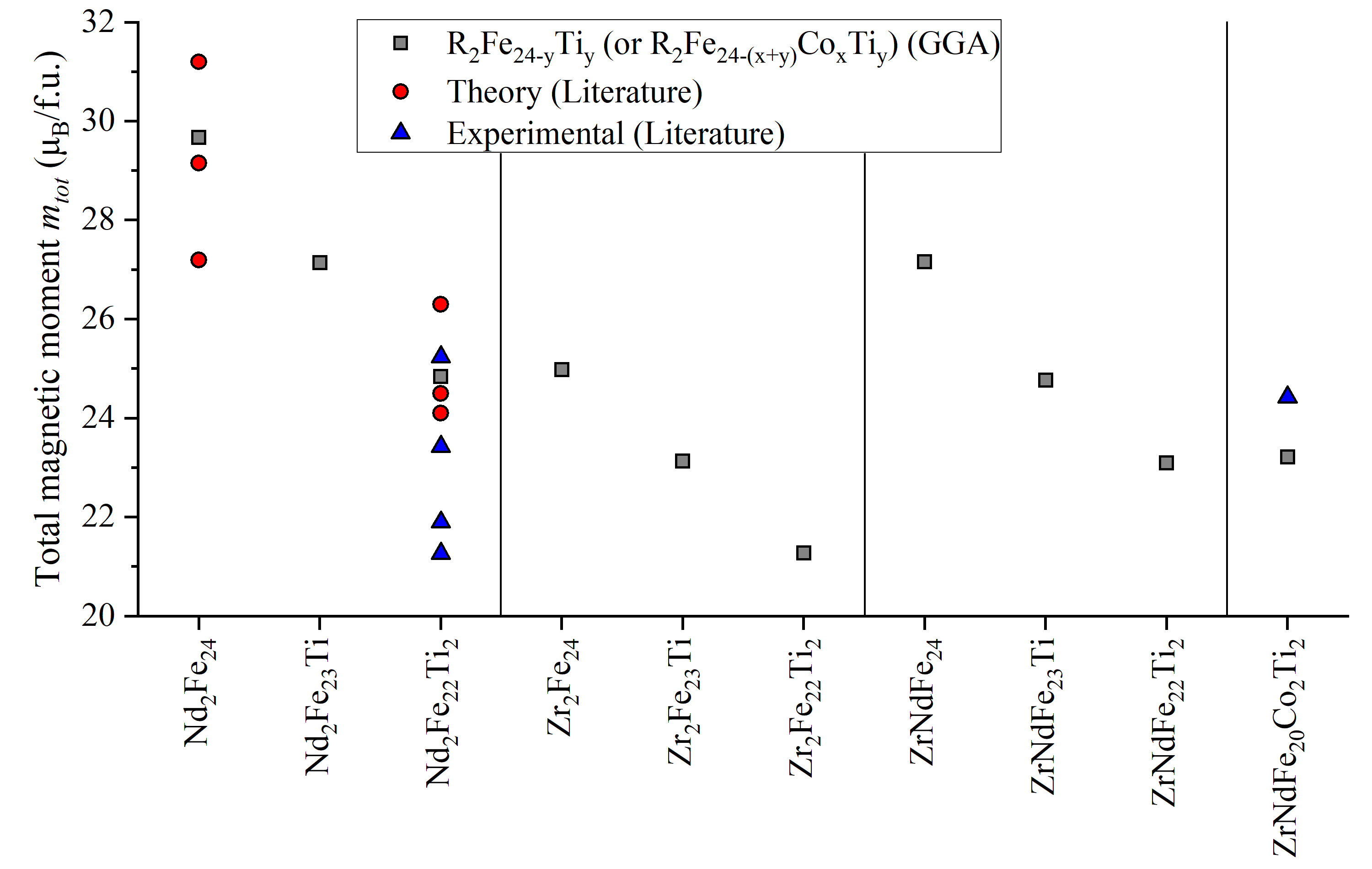}
    \caption{Total magnetic moments of R$_{2}$Fe$_{24}$\_$_{y}$Ti$_{y}$ (R: Zr and Nd; y: $0 \leq y \leq 2$). Since GGA+$U$ calculations do not change the results significantly, only GGA results are given. The red circles show theoretical data \cite{p40, p55, p63, p64} for the alloys from others and the blue triangles show experimental data \cite{p45, p54, p56, p60, FU22} for the alloys. Here all literature values correspond to the same compounds. Only the experimental value at ZrNdFe$_{20}$Co$_{2}$Ti$_{2}$ corresponds to the (Zr$_{0.3}$Nd$_{0.7}$)$_{2}$(Fe$_{0.75}$Co$_{0.25}$)$_{23}$Ti compound.}
    \label{fig.18}
\end{figure}    
    
    The DFT calculations treating the $f$-electrons by means of the in-core treatment, give reasonable $m_{tot}$ values for the Nd$_{2}$Fe$_{22-y}$Ti$_{y}$ compounds. The $m_{tot}$ value of NdFe$_{11}$Ti is 24.84~$\mu_{B}$ and the experimentally determined value of Herper \textit{et al.} \cite{p60} is 23.43~$\mu_{B}$. However, the calculated value including the in-core approximation neglects the $f$-electrons. Including the $f$-electrons by considering them as fully localized and applying the DFT+$U$ treatment, a value for $m_{tot}$ of 25.04~$\mu_{B}$ (including the orbital magnetic moment) is calculated. Further DFT+$U$ treatment also yields a value of 1.69 T for the saturation magnetisation, which is in the range of the experimentally determined values of 1.38 T to 1.70~T. In case of DFT+$U$ approximation, a reduced spin magnetic moment is obtained with a magnitude of the Nd ion being -3.30 $\mu_{B}$. The reduction of the value of about -3.0~$\mu_{B}$ is due to the $f$-electrons, since the value in DFT is about -0.27~$\mu_{B}$. Nevertheless, the results for both treatments are acceptable compared to experimental values. The calculated results are based on the addition of the orbital magnetic moment of the Nd-4$f$-electrons with $g_{j} \cdot J = |L - S| = 3.273~\mu_{B}$. 
    
\newpage

\begin{table}[h]
\tiny
   \caption{Calculated spin magnetic moments at rare-earth site $m^{2a}_{spin}$, total magnetic moments $m_{tot}$ ($\mu_{B}$/f.u.), maximum energy product $|BH|_{max}$ (kJ/m$^{3}$) and saturation magnetisation $\mu_{0}M_{S}$ (T) with comparison against available experimental data. In case of Nd containing compounds DFT+$U$ with $U = 6$ eV is applied to Nd 4$f$-electrons only and results are given in parenthesis.}
   \centering
   \hspace{-1cm}
   \label{tab.6}
\begin{tabular}{cccccccc}
\hline
Alloy                                                         & \begin{tabular}[c]{@{}c@{}}$m^{2a}_{spin}$\\   {(}$\mu_{B}$/f.u.{)}\end{tabular}           & \begin{tabular}[c]{@{}c@{}}$m_{tot}$ \\ {(}$\mu_{B}$/f.u.{)}\end{tabular}          & \begin{tabular}[c]{@{}c@{}}$m_{tot}$ \\experimental\\ {(}$\mu_{B}$/f.u.{)}\end{tabular} & \begin{tabular}[c]{@{}c@{}}$\mu_{0}M_{S}$\\  {(}T{)}\end{tabular}         & \begin{tabular}[c]{@{}c@{}}$\mu_{0}M_{S}$ \\experimental \\ {(}T{)}\end{tabular}                                                              & \begin{tabular}[c]{@{}c@{}}$|BH|_{max}$\\  {(}kJ/m$^{3}${)}\end{tabular}         & \begin{tabular}[c]{@{}c@{}}$|BH|_{max}$ \\experimental \\ {(}kJ/m$^{3}${)}\end{tabular}                                                                        \\ \hline
Nd$_{2}$Fe$_{24}$                                             & -0.27                                                                                      & 29.67                                                                              & 27.2* \cite{p63}                                                                        & 2.03                                                                      & 1.73* \cite{p63}                                                                                                                              & 821.55                                                                           & 818* \cite{step}                                                                                                                                             \\
(NdFe$_{12}$)                                                 &                                                                                            &                                                                                    & 29.15* \cite{p40}                                                                       &                                                                           & 1.99* \cite{p40}                                                                                                                              &                                                                                  &                                                                                                                                                              \\
                                                              &                                                                                            &                                                                                    & 31.20* \cite{p55}                                                                       &                                                                           & 2.01* \cite{p55}                                                                                                                              &                                                                                  &                                                                                                                                                              \\
Nd$_{2}$Fe$_{23}$Ti                                           & -0.27                                                                                      & 27.14                                                                              &                                                                                         & 1.85                                                                      &                                                                                                                                               & 680.30                                                                           & 683* \cite{step}                                                                                                                                             \\
(NdFe$_{11.5}$Ti$_{0.5}$)                                     & (-3.29)                                                                                    & (27.31)                                                                            &                                                                                         & (1.86)                                                                    &                                                                                                                                               & (688.49)                                                                         & 701* \cite{step}                                                                                                                                             \\
Nd$_{2}$Fe$_{22}$Ti$_{2}$                                     & -0.28                                                                                      & 24.84                                                                              & 21.27$^{a}$ \cite{p54}                                                                  & 1.68                                                                      & 1.38$^{a}$ \cite{p54}                                                                                                                         & 562.30                                                                           & 438* \cite{p40}                                                                                                                                               \\
(NdFe$_{11}$Ti)                                               & (-3.30)                                                                                    & (25.04)                                                                            & 21.90$^{b}$ \cite{p56}                                                                  & (1.69)                                                                    & 1.48$^{b}$ \cite{p56}                                                                                                                         & (571.39)                                                                         & 569* \cite{step}                                                                                                                                             \\
                                                              &                                                                                            &                                                                                    & 23.43$^{c}$ \cite{p60}                                                                  &                                                                           & 1.58$^{c}$ \cite{p60}                                                                                                                         &                                                                                  & 575* \cite{step}                                                                                                                                             \\
                                                              &                                                                                            &                                                                                    & 24.10* \cite{p64}                                                                       &                                                                           & 1.63* \cite{p64}                                                                                                                              &                                                                                  &                                                                                                                                                              \\
                                                              &                                                                                            &                                                                                    & 24.50* \cite{p40}                                                                       &                                                                           & 1.65* \cite{p40}                                                                                                                              &                                                                                  &                                                                                                                                                              \\
                                                              &                                                                                            &                                                                                    & 25.24$^{d}$ \cite{p45}                                                                  &                                                                           & 1.70$^{d}$ \cite{p45}                                                                                                                         &                                                                                  &                                                                                                                                                              \\
                                                              &                                                                                            &                                                                                    & 26.30* \cite{p55}                                                                       &                                                                           & 1.70* \cite{p55}                                                                                                                              &                                                                                  &                                                                                                                                                              \\ \hline
Zr$_{2}$Fe$_{24}$                                             & -0.47                                                                                      & 24.98                                                                              &                                                                                         & 1.81                                                                      &                                                                                                                                               & 654.10                                                                           &                                                                                                                                                              \\
(ZrFe$_{12}$)                                                 &                                                                                            &                                                                                    &                                                                                         &                                                                           &                                                                                                                                               &                                                                                  &                                                                                                                                                              \\
Zr$_{2}$Fe$_{23}$Ti                                           & -0.46                                                                                      & 23.13                                                                              &                                                                                         & 1.66                                                                      &                                                                                                                                               & 548.87                                                                           &                                                                                                                                                              \\
(ZrFe$_{11.5}$Ti$_{0.5}$)                                     &                                                                                            &                                                                                    &                                                                                         &                                                                           &                                                                                                                                               &                                                                                  &                                                                                                                                                              \\
Zr$_{2}$Fe$_{22}$Ti$_{2}$                                     & -0.45                                                                                      & 21.28                                                                              &                                                                                         & 1.51                                                                      &                                                                                                                                               & 455.19                                                                           &                                                                                                                                                              \\
(ZrFe$_{11}$Ti)                                               &                                                                                            &                                                                                    &                                                                                         &                                                                           &                                                                                                                                               &                                                                                  &                                                                                                                                                              \\ \hline
ZrNdFe$_{24}$                                                 & \begin{tabular}[c]{@{}c@{}}-0.50 ($m^{2a}_{Zr}$)\\ -0.26 ($m^{2a}_{Nd}$)\end{tabular}      & 27.16                                                                              &                                                                                         & 1.92                                                                      & \begin{tabular}[c]{@{}c@{}}1.93* (Zr$_{0.1}$) \cite{ndzr}\\ 1.91* (Zr$_{0.2}$) \cite{ndzr}\\ 1.89* (Zr$_{0.3}$) \cite{ndzr}\end{tabular}      & 732.20                                                                           & \begin{tabular}[c]{@{}c@{}}741* (Zr$_{0.1}$) \cite{ndzr}\\ 726* (Zr$_{0.2}$) \cite{ndzr}\\ 711* (Zr$_{0.3}$) \cite{ndzr}\end{tabular}   \\
ZrNdFe$_{23}$Ti                                               & \begin{tabular}[c]{@{}c@{}}-0.50 ($m^{2a}_{Zr}$)\\ -0.26 ($m^{2a}_{Nd}$)\end{tabular}      & 24.77                                                                              &                                                                                         & 1.74                                                                      &                                                                                                                                               & 599.63                                                                           &                                                                                                                                                              \\
                                                              & \begin{tabular}[c]{@{}c@{}}((-0.50 ($m^{2a}_{Zr}$))\\ (-3.27 ($m^{2a}_{Nd}$))\end{tabular} & (24.82)                                                                            &                                                                                         & (1.74)                                                                    &                                                                                                                                               & (602.83)                                                                         &                                                                                                                                                              \\
ZrNdFe$_{22}$Ti$_{2}$                                         & \begin{tabular}[c]{@{}c@{}}-0.49 ($m^{2a}_{Zr}$)\\ -0.27 ($m^{2a}_{Nd}$)\end{tabular}      & 23.10                                                                              &                                                                                         & 1.60                                                                      &                                                                                                                                               & 510.17                                                                           &                                                                                                                                                              \\
                                                              & \begin{tabular}[c]{@{}c@{}}((-0.48 ($m^{2a}_{Zr}$))\\ (-3.31 ($m^{2a}_{Nd}$))\end{tabular} & (23.28)                                                                            &                                                                                         & (1.62)                                                                    &                                                                                                                                               & (524.51)                                                                         &                                                                                                                                                              \\ \hline
ZrNdFe$_{20}$Co$_{2}$Ti$_{2}$                                 & \begin{tabular}[c]{@{}c@{}}-0.49 ($m^{2a}_{Zr}$)\\ -0.26 ($m^{2a}_{Nd}$)\end{tabular}      & 23.22                                                                              &                                                                                         & 1.61                                                                      &                                                                                                                                               & 515.62                                                                           &                                                                                                                                                              \\
                                                              & \begin{tabular}[c]{@{}c@{}}((-0.49 ($m^{2a}_{Zr}$))\\ (-3.28 ($m^{2a}_{Nd}$))\end{tabular} & (23.28)                                                                            &                                                                                         & (1.61)                                                                    &                                                                                                                                               & (517.70)                                                                         &                                                                                                                                                              \\ \hline
Nd$_{2}$(Fe$_{0.75}$Co$_{0.25}$)$_{23}$Ti                     &                                                                                            &                                                                                    & 25.41$^{d}$ \cite{FU24}                                                                 &                                                                           & 1.63$^{d}$ \cite{FU24}                                                                                                                        &                                                                                  & 529$^{d}$ \cite{FU24}                                                                                                                                      \\
(Nd$_{0.7}$Zr$_{0.3}$)$_{2}$(Fe$_{0.75}$Co$_{0.25}$)$_{23}$Ti &                                                                                            &                                                                                    & 24.43$^{d}$ \cite{FU24}                                                                 &                                                                           & \begin{tabular}[c]{@{}c@{}}1.66$^{d}$ \cite{FU21}\\ 1.63$^{d}$ \cite{FU24}\end{tabular}                                                       &                                                                                  & \begin{tabular}[c]{@{}c@{}}548$^{d}$ \cite{FU21}\\ 529$^{d}$ \cite{FU24}\end{tabular} \\ \hline                                                                                                              
\end{tabular}
\begin{tablenotes}[flushleft] 
{\footnotesize{\scriptsize \setstretch{0.25}
    \item * Theoretical references are represented by *. \newline
    \item $^a$ Extraction sample magnetometer analysis at 1.5~K. \newline
    \item $^{b}$ Mössbauer measurement at 4.2~K. \newline
    \item $^{c}$ Single crystal, physical property measurement system (PPMS) measured at 10 K. \newline
    \item $^{d}$ Vibrating sample magnetometer (VSM) analysis at 4.2~K.
    }}
\end{tablenotes}
\end{table}
    
    The substitution of Zr atoms into the 1:12 phase reduces the total magnetic moment in average by 2.25~$\mu_{B}$/f.u. (ZrNdFe$_{24-y}$Ti$_{y}$) up to 4.21~$\mu_{B}$/f.u. (Zr$_{2}$Fe$_{24-y}$Ti$_{y}$). In DFT, this reduction originated from the lower magnetic moment of Zr (-0.47~$\mu_{B}$) compared to Nd (-0.27~$\mu_{B}$). The opposite is the case with the DFT+$U$ treatment, since the inclusion of $f$-electrons lowers the magnetic moment of Nd to -3.3~$\mu_{B}$. 
    The effect of Ti substitution on the total magnetic moments of the considered compounds are shown in Fig.~\ref{fig.18}. In GGA and GGA+$U$ the total magnetisation of Ti atoms is about -1.1~$\mu_{B}$. This is due to the antiferromagnetic orientation of Ti atoms relative to Fe atoms. It can be seen that the total magnetic moment decreases with more Ti in a compound. The trend of the values from the Nd$_{2}$Fe$_{24-y}$Ti$_{y}$ (y: $0 \leq y \leq 2$) compounds shows the same tendency as in the work of Erdmann \textit{et al.} \cite{step}. The development of the values of all considered compounds show a similar decrease of the total magnetic moment of about 2-3~$\mu_{B}$/f.u. per Ti atom.
    
    The influence of Co on the total magnetic moment is only a minor one as shown in Tab.~\ref{tab.4} and Tab.~\ref{tab.6}. When comparing ZrNdFe$_{22}$Ti$_{2}$ and ZrNdFe$_{20}$Co$_{2}$Ti$_{2}$, only a minimal difference (increase of 0.12~$\mu_{B}/f.u.$) can be seen in the DFT values. The same values were obtained in the DFT+$U$ approach. 
    
    As a result from the theoretical observations and calculations in this section quaternary Zr-substituted compounds (such as ZrNdFe$_{22}$Ti$_{2}$) are favourable for future experimental investigations, in order to find promising new permanent hard magnets, particularly in comparison to the Co-enriched quinary compound (ZrNdFe$_{20}$Co$_{2}$Ti$_{2}$). The latter one is not going to be a viable option, mainly due to its higher costs due to Co without providing additional benefits.

\subsection{Maximum Energy Product}
\begin{figure}[h]
    \centering
    \includegraphics[scale=0.38]{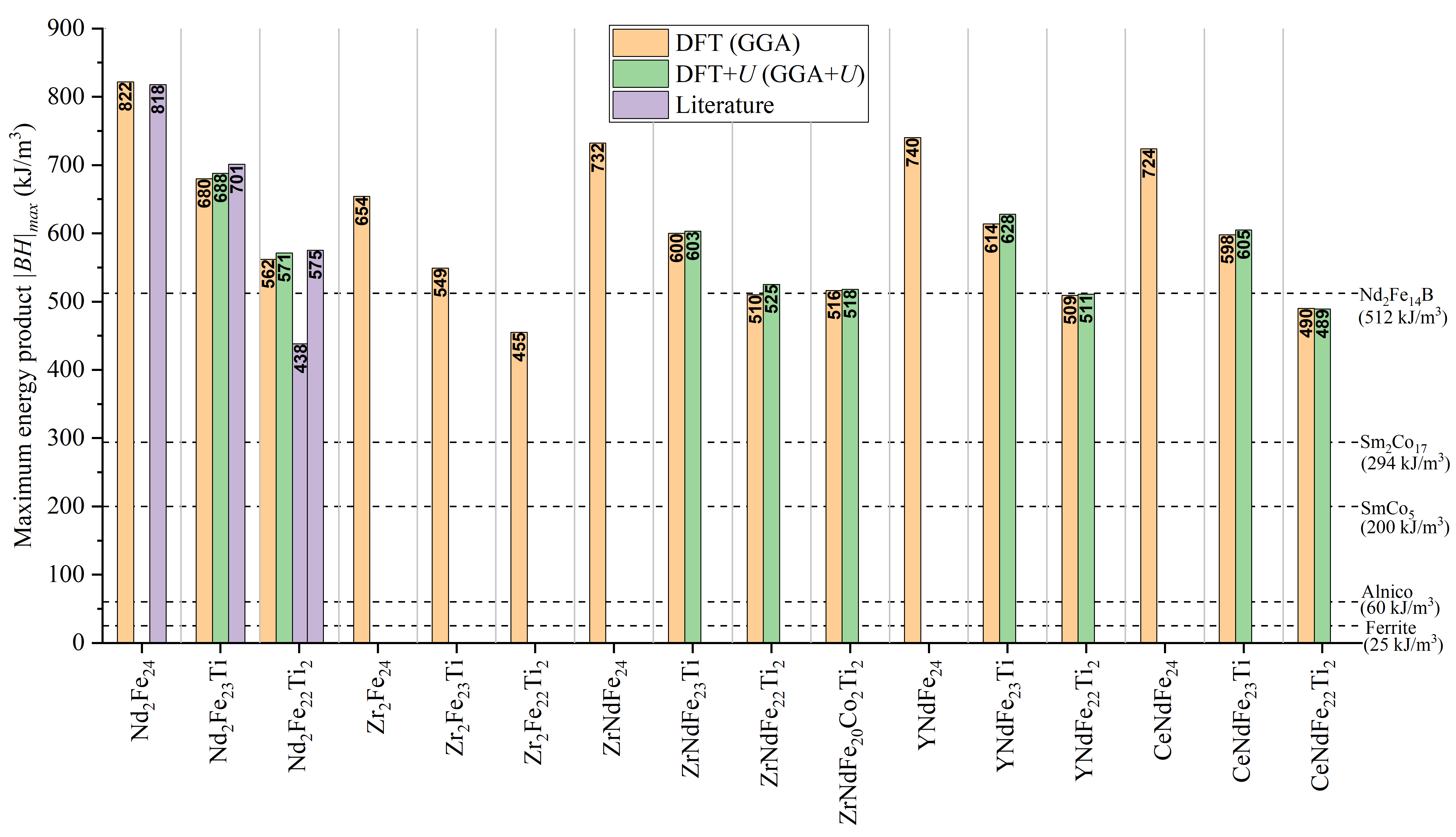}
    \caption{Theoretical maximum energy product $|BH|_{max}$ values for the compounds considered and the literature for comparison \cite{step, p40}. In addition to conventional DFT, DFT+$U$ with $U$ = 6 eV has been considered for stable Nd containing alloys. Experimental values \cite{10, p67} of the most common hard magnets are given as horizontal dotted lines.}
    \label{fig.19}
\end{figure}

    The maximum energy product $|BH|_{max}$ is one of the performance measures of a permanent magnet. This important parameter describes the strength of a permanent magnetic material and how the magnet should be tuned so that its operating point is close to the $|BH|_{max}$ for the most efficient use. The $|BH|_{max}$ also reflects the upper limit of magnetic energy that can be stored in free space by a unit volume of an permanent magnet \cite{p65}. Through the relation of $|BH|_{max}$ to the saturation magnetisation $M_{S}$, $|BH|_{max}$ can be calculated for an ideal quadratic Hysteresis loop as done by Coey \cite{10}:
    \begin{equation}
       |BH|_{max} = \frac{\mu_{0}M_{S}^{2}}{4},
       \label{eq.3}
    \end{equation}
        where $\mu_{0}$ is the vacuum permeability ($\mu_{0} = 4\pi \cdot 10^{-7}$ NA$^{-2}$). The calculated $|BH|_{max}$ values of the selected compounds are presented in Tab.~\ref{tab.6} and Fig.~\ref{fig.19}.
        
    For a hard magnetic compound to be considered as promising, a $|BH|_{max}$ $>$ 400 kJ/m$^{3}$ is important. Higher $|BH|_{max}$ values are received by Ti-free RFe$_{12}$ (R: Y, Zr and Ce) compounds. Substituting Fe against one Ti atom (3.8~at.\% in the 2 f.u. supercell) reduces the value by about 120~kJ/m$^{3}$ in average. A second Ti atom reduces the value by another 100~kJ/m$^{3}$ (7.7~at.\% in the 2 f.u. supercell).
    
    Due to the lack of experimental $|BH|_{max}$ data, no direct comparison with the theoretical values can be made. However, there are other theoretical and experimental results for similar compounds against which the values can be compared. Körner \textit{et al.} \cite{p40} calculated the $|BH|_{max}$ value of various 1:12 phases using the tight-binding linear-muffin-tin-orbital atomic-sphere approximation (TB-LMTO-ASA). Among these compounds they calculated NdFe$_{11}$Ti with a value of 438~kJ/m$^{3}$, which is a smaller value then the calculated one of 562~kJ/m$^{3}$ in this thesis. Furthermore, the calculated value for $|BH|_{max}$ is in the same range of the work of Erdmann \textit{et al.} \cite{step} with 569~kJ/m$^{3}$. An even better agreement with the work of Erdmann \textit{et al.} \cite{step} is obtained for the DFT+$U$ treatment with $U$ = 6 eV. In this work the value of 571~kJ/m$^{3}$ has been calculated, whereas a value of 575~kJ/m$^{3}$ \cite{step} has been found before.

    In case of Zr$_{2}$Fe$_{24-y}$Ti$_{y}$ (y: $0 \leq y \leq 2$) compounds, the results obtained are $\geq 400$~kJ/m$^{3}$. Nevertheless, the $|BH|_{max}$ values of the ternary Zr$_{2}$Fe$_{24-y}$Ti$_{y}$ compounds are about 130~kJ/m$^{3}$ lower than those of the Nd$_{2}$Fe$_{24-y}$Ti$_{y}$ (y: $0 \leq y \leq 2$) compounds. As an illustration, the ratio of ZrFe$_{11}$Ti to NdFe$_{11}$Ti is given with 455~kJ/m$^{3}$ and 562~kJ/m$^{3}$, respectively. The quaternary ZrNdFe$_{24-y}$Ti$_{y}$ compounds lie, as expected, between the values of the R$_{2}$Fe$_{24-y}$Ti$_{y}$ (R: Nd and Zr, y: $0 \leq y \leq 2$) compounds, making them promising candidates as well. The comparison of the calculated $|BH|_{max}$ value of ZrNdFe$_{23}$Ti (600~kJ/m$^{3}$) with the experimentally determined value of (Zr$_{0.3}$Nd$_{0. 7}$)$_{2}$(Fe$_{0.75}$Co$_{0.25}$)$_{23}$Ti (548~kJ/m$^{3}$) \cite{FU21}, shows good agreement, keeping in mind that they are only similar to each other.

    In Fig.~\ref{fig.19}, the $|BH|_{max}$ values of well known hard magnets are listed as reference. When the value of the well known hard magnet Nd$_{2}$Fe$_{14}$B of 512~kJ/m$^{3}$ is considered and compared with the quaternary alloys, it can be seen that the values are just above. For comparison, the quaternary compounds with Y and Ce calculated from the work of Erdmann \textit{et al.} \cite{step} are also shown in Fig.~\ref{fig.19}. It can be concluded that in case of the (R,Nd)Fe$_{11.5}$Ti$_{0.5}$ (R: Y, Zr and Ce) alloys, the Y containing alloy has the highest $|BH|_{max}$ value, followed by Ce and Zr containing alloys. The case is different for (R,Nd)Fe$_{11}$Ti (R:~Y, Zr and Ce) compounds. The Zr-containing compound has the highest value, followed by the Y and Ce compounds, respectively. As a first intermediate result, it can be concluded, that from the viewpoint of the $|BH|_{max}$ values the quaternary Zr-containing compounds could join the ranks of promising hard magnets.

\subsection{Curie Temperature}
    In addition to the already mentioned intrinsic magnetic properties, a high Curie temperature ($T_{C}$) is important and required for stable magnets at application related temperatures. The well known and widely used Nd$_{2}$Fe$_{14}$B magnet has a relatively low $T_{C}$ of 588~K \cite{10}, while traditional RE based permanent magnets have higher $T_{C}$ values (\textit{e.g.} SmCo$_{5}$ with 1020~K \cite{10}). In consequence, the search for new magnets is complex, because the RE-lean or RE-free magnets need to have a better balance between saturation magnetisation and Curie temperature. 

    To calculate $T_{C}$, an effective spin Heisenberg model is selected, which is solved in Mean Field Approximation (MFA). In this Heisenberg model, the Hamiltonian of the spin-spin interaction is given by $\mathcal{H} = -\Sigma_{i \neq j} J_{ij} \overrightarrow{S}_{i} \overrightarrow{S}_{j}$, where $\overrightarrow{S}_{i}$ and $\overrightarrow{S}_{j}$ correspond to the spin of place $i$ and $j$, respectively. $J_{ij}$ corresponds to the exchange coupling constant between the two places $i$ and $j$ and was calculated with AkaiKKR \cite{68}. The absolute value of the magnetic ordering temperature $T_{C}$ resulting from MFA is $k_{B}T_{C} = \frac{3}{2} J_{0}$. Here, $k_{B}$ corresponds to the Boltzmann constant. $J_{0}$ results from the sum of $J_{ij}$ and is therefore the total effective exchange of a given lattice site connected to all others.

    In addition to MFA, where the systematic overestimation of $T_{C}$ is a well-known problem, the Local Moment Disorder (LMD) approach is considered along with the FM states. Note that the margin of error for the FM-based calculations of the ternary compounds is about 45\%. LMD is also called disorder local moment (DLM) \cite{p70} and preserves the magnetic moments which do not vanish. In the KKR-CPA method, the concept of an LMD state corresponds to the description of a paramagnetic state of a ferromagnet above its $T_{C}$ \cite{p69}. There is an advantage in using the Green's function theory based KKR, as it allows both the exchange interaction energy situations to be calculated. Furthermore, not only the Curie temperature of the FM states, but also those of the related LMD states can be determined.

\begin{figure}[h]
    \centering
    \includegraphics[scale=0.44]{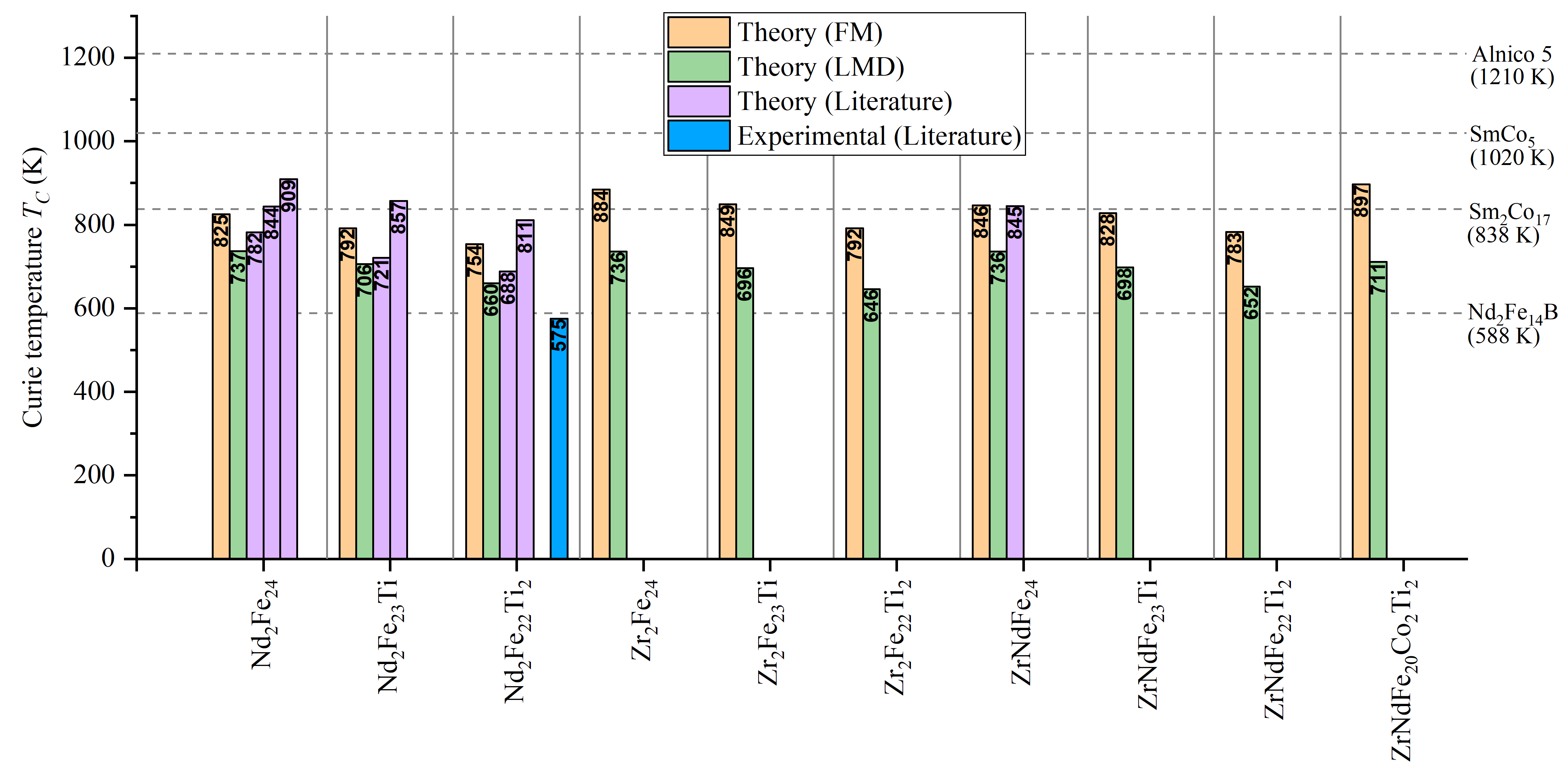}
    \caption{Calculated Curie temperatures $T_{C}$ for all considered 1:12 phases. The exchange interaction energies have been calculated for both ordered ferromagnetic (FM) and disordered local momentum (LMD) states, and the lattice information comes from the theoretically relaxed calculations (see Tab.~\ref{tab.5}). The experimental data is taken from \cite{p54, step, ndzr}.}
    \label{fig.20}
\end{figure}

    The calculated Curie temperatures are shown in Fig.~\ref{fig.20} with available theoretical and experimental literature values. The $T_{C}$ of known permanent magnets are shown as horizontal dashed lines for comparison. As expected, the FM ground state approximation overestimates the Curie temperature, which can be deduced from the significantly higher values of $T_{C}$ in contrast to those of the LMD states. This overestimation is on one hand due to the problematic description of the delocalised electronic state in the magnetism of intermetallics on basis of the localised degrees of freedom. On the other hand, the exchange coupling between the local moments is calculated for the ground state, and this assumption does not have to be valid at high temperatures close to $T_{C}$.

    In the work of Erdmann \textit{et al.} \cite{step}, a good agreement with Miyake \textit{et al.} \cite{p71} was found for the Nd$_{2}$Fe$_{22}$Ti$_{2}$ compound. It should be noted, that the FM states overestimate the value of $T_{C}$. As can be seen, this work is in good agreement with the earlier work in case of the Nd$_{2}$Fe$_{24-y}$Ti$_{y}$ (y: 0 $\leq y \leq 2$) compounds. Miyake \textit{et al.} \cite{p71} reported that a better agreement of LMD calculations strongly depends on the crystal structure, and in some cases the LMD values are more accurate than the FM values. Furthermore, a complex dependence of the performance of LMD on the crystalline structure and the transition metals has been reported \cite{p71}. This dependence could be due to a difference in spin fluctuation near a magnetic transition, but this is not clearly understood and does require further research. In case of the Zr$_{2}$Fe$_{24-y}$Ti$_{y}$ (y: 0 $\leq y \leq 2$) compounds there is no literature data available yet and therefore the calculated values can only be seen as indicators for the $T_{C}$ values. In case of (Zr$_{0.3}$Nd$_{0.7}$)$_{2}$(Fe$_{0.75}$Co$_{0.25}$)$_{23}$Ti, Suzuki \textit{et al.} \cite{FU24} estimates the value of $T_{C}$ to be $>$840~K, which suggests that the FM values of the ZrNdFe$_{24-y}$Ti$_{y}$ (y: $0 \leq y \leq 2$) and the ZrNdFe$_{20}$Co$_{2}$Ti$_{2}$ compounds are more likely than the LMD results. 

    Similar to $m_{tot}$ and $|BH|_{max}$, an increasing Ti concentration decreases the Curie temperature, as can be seen in Fig.~\ref{fig.20}). The value of $T_{C}$ decreases by about 45~K per Ti atom (3.8~at.\%). This also corresponds to the average decrease found in the work of Erdmann \textit{et al.} \cite{step}. The theoretical FM (LMD) results give the following trend with the calculated $T_{C}$ values of 792 (646), 783 (652) and 754 (660)~K for R$_{2}$Fe$_{22}$Ti$_{2}$ (R$_{2}$: Zr$_{2}$, ZrNd, Nd$_{2}$, respectively). Due to the lack of literature data, the trend of the calculated compounds can only be assumed by literature of the compounds (Zr$_{0.3}$Nd$_{0. 7}$)$_{2}$(Fe$_{0. 75}$Co$_{0. 25}$)$_{23}$Ti \cite{FU24} and Nd$_{2}$Fe$_{24-y}$Ti$_{y}$ (y: 0 $\leq y \leq 2$) \cite{step, p71} to be true. In addition, the $T_{C}$ can be increased by substitution of Fe with Co (by 7.7 at.\% of Co, see Fig.~\ref{fig.20}) by at least 85~K.
    
    The substitution of Zr into the 1:12 phase starting from Nd$_{2}$Fe$_{24}$, can be assumed to result in an increase of the Curie temperature. Comparing the mother phase Nd$_{2}$Fe$_{22}$Ti$_{2}$ and the quaternary compound ZrNdFe$_{22}$Ti$_{2}$, only minor changes are observed. On average, a slight increase can be assumed with the substitution of Zr. Comparing the value of Nd$_{2}$Fe$_{14}$B magnets of 588~K \cite{10}, both the LMD value of 652~K and the FM value of 783~K of ZrNdFe$_{22}$Ti$_{2}$ are higher. The FM value of 838~K of Sm$_{2}$Co$_{17}$ can even be exceeded by Co substitution. 
    
    In general, it can be postulated that all considered compounds have $T_{C}$ values around or slightly below the critical temperature of Nd$_{2}$Fe$_{14}$B (588~K) because of the expected overestimation. The expected $T_{C}$ values for the Zr-enriched quaternary hard magnet candidates focused on in this work will require experimental confirmation \textit{i.e.} via sintering and measuring all required parameters of these compounds. However, as the theoretical calculations already show promising $T_{C}$ values around 600 K, an investigation of these substances is justified.

\subsection{Magnetocrystalline Anisotropy Energy}
    The next magnetic property studied in this section is the magnetocrystalline anisotropy energy (MAE). The MAE can be determined from the magnetic force theorem \cite{p72}, or by calculating the energy difference of the magnetisation directions. In this thesis the MAE is calculated using the latter approach. The equation used to calculate the MAE in this work is as follows:
    \begin{equation}
       \Delta E_{MAE} = E_{[010]} - E_{[001]},
       \label{eq.39}
    \end{equation}
    where $E_{[001]}$ and $E_{[010]}$ correspond to the total energies of the magnetisation directions [001] and [010] of the supercell, respectively.

\begin{figure}[h]
    \centering
    \includegraphics[scale=0.58]{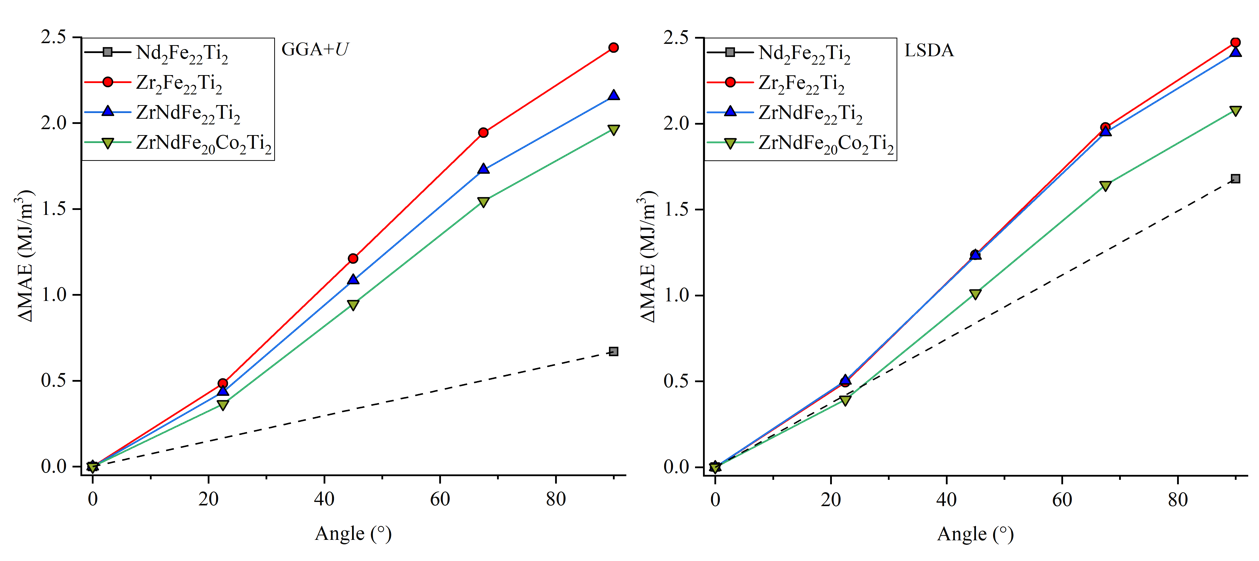}
    \caption{Calculated energy differences of magnetocrystalline energy for the shift of the crystallographic direction from [001] to [010] for all compounds considered in MAE calculation. LSDA results are shown on the left and DFT+$U$ results on the right. For the change from [001] to [010], angles 0, 22.5, 45, 67.5 and 90° are taken into consideration. For NdFe$_{11}$Ti only the values of the angles 0 and 90° are shown.}
    \label{fig.22}
\end{figure}

    In Magnetocrystalline Anisotropy Energy (MAE) calculations the orientation of magnetisation along the crystallographic directions plays an important role. In Fig.~\ref{fig.22} the resulting energy differences based on DFT+$U$ and LSDA calculations are shown for changes from the easiest magnetisation axis [001] to the hardest axis [010]. The directional changes from [001] to [010] are shown for the different compounds. LSDA is used because Erdmann et al. \cite{step} showed that GGA+$U$ sometimes gives inaccurate results when calculating anisotropy.

    From the total energies and the resulting energy differences the anisotropy constant $K_{1}$ and the anisotropy field $H_{a}$ can be calculated. The anisotropy energy can be written as:
    \begin{equation}
       \Delta E_{MAE} = K_{1} \cdot sin(\phi)^{2},
       \label{eq.40}
    \end{equation}
    where $\phi$ is the angle between the direction of magnetisation and the axis that is easiest to magnetise. In this formulation only the leading term containing $K_{1}$ is shown, being sufficiently accurate in this case. All considered 1:12 phases were found to exhibit uniaxial anisotropy ($K_{1} > 0$). The higher order anisotropy constants are negligible because $K_{2}$ is almost two orders of magnitude smaller than $K_{1}$. Thus makes $K_{1}$ sufficient to describe the magnetocrystalline anisotropy (MCA) (for further details see \cite{p50}). A resulting positive (or negative) value for $K_{1}$ corresponds to uniaxial (or planar) anisotropy.

    The anisotropy field $H_{a}$ corresponds to the upper limit of the coercivity and can be calculated using Eq.~\ref{eq:2}. $H_{a}$ reflects an important property of an permanent magnet and should be as high as possible. At this point it is important to mention that the values calculated here for $K_{1}$ and $M_{S}$ are valid for T~=~0~K. This excludes the temperature dependence of $H_{a}$. The calculated anisotropy constant $K_{1}$ and the anisotropy field $H_{a}$, calculated by GGA+$U$ and LSDA, are shown in Tab.~\ref{tab.7}. 
    
\begin{table}[h]
   \caption{Calculated theoretical magnetocrystalline anisotropy constant $K_{1}$ according to Eq.~\ref{eq.40}, anisotropy field $H_{a}$ according to Eq.~\ref{eq:2} and magnetic hardness factor $\kappa$ from Eq.~\ref{eq:3} for stable compounds with available experimental data. The values are given for both GGA and LSDA treatment for the exchange–correlation potential. DFT+$U$ ($U$~=~6~eV) results are reported in parenthesis for Nd containing compounds. The experimental value of $K_{1}$ and $\kappa$ of well known hard magnets are also given for comparison.}
   \tiny
   \centering
   \hspace{-1.2cm}
   \label{tab.7}
\begin{tabular}{cccccccccc}
\hline
Alloy                                                         & \begin{tabular}[c]{@{}c@{}}$K_{1}$ GGA \end{tabular}         & \begin{tabular}[c]{@{}c@{}}$K_{1}$ LSDA \end{tabular}         & \begin{tabular}[c]{@{}c@{}}$K_{1}$ experi- \end{tabular}      & \begin{tabular}[c]{@{}c@{}}$H_{a}$ GGA \end{tabular}    & \begin{tabular}[c]{@{}c@{}}$H_{a}$ LSDA \end{tabular}   & \begin{tabular}[c]{@{}c@{}}$H_{a}$ experi- \end{tabular}      & \begin{tabular}[c]{@{}c@{}}$\kappa$ GGA \end{tabular}         & \begin{tabular}[c]{@{}c@{}}$\kappa$ LSDA \end{tabular}         & \begin{tabular}[c]{@{}c@{}}$\kappa$ experi- \end{tabular}        \\ 
                                                              & {(}MJ/m$^{3}${)}                                             & {(}MJ/m$^{3}${)}                                              & mental                                                        & {(}T{)}                                                 & {(}T{)}                                                 & mental                                                        &                                                               &                                                                & mental                                              \\ 
                                                              &                                                              &                                                               & {(}MJ/m$^{3}${)}                                              &                                                         &                                                         & {(}T{)}                                                       &                                                               &                                                                &                                                     \\ \hline
Nd$_{2}$Fe$_{22}$Ti$_{2}$                                     & 0.67                                                         & 1.68                                                          & 0.61* \cite{step}                                             & 0.80                                                    & 2.00                                                    & 0.72* \cite{step}                                             & 0.49                                                          & 0.77                                                           & 0.52* \cite{step}                                        \\
(NdFe$_{11}$Ti)                                               &                                                              &                                                               & 1.41´ \cite{step}                                             &                                                         &                                                         & 1.86´ \cite{step}                                             &                                                               &                                                                & 0.88´ \cite{step}                                        \\
                                                              &                                                              &                                                               & 1.78$^{b}$ \cite{p60}                                         &                                                         &                                                         & 1.9$^{a}$ \cite{p77}                                          &                                                               &                                                                &                                                          \\
                                                              &                                                              &                                                               &                                                               &                                                         &                                                         & 2.0$^{a}$ \cite{p77}                                          &                                                               &                                                                &                                                          \\ \hline
Zr$_{2}$Fe$_{22}$Ti$_{2}$                                     & 2.44                                                         & 2.47                                                          &                                                               & 3.22                                                    & 3.27                                                    &                                                               & 1.03                                                          & 1.04                                                           &                                                          \\
(ZrFe$_{11}$Ti)                                               &                                                              &                                                               &                                                               &                                                         &                                                         &                                                               &                                                               &                                                                &                                                          \\ \hline
ZrNdFe$_{22}$Ti$_{2}$                                         & 2.16                                                         & 2.41                                                          &                                                               & 2.69                                                    & 3.01                                                    &                                                               & 0.92                                                          & 0.97                                                           &                                                          \\
((Zr,Nd)Fe$_{11}$Ti)                                          &                                                              &                                                               &                                                               &                                                         &                                                         &                                                               &                                                               &                                                                &                                                          \\
ZrNdFe$_{20}$Co$_{2}$Ti$_{2}$                                 & 1.97                                                         & 2.08                                                          &                                                               & 2.44                                                    & 2.58                                                    &                                                               & 0.87                                                          & 0.90                                                           &                                                          \\
((Zr,Nd)Fe$_{10}$CoTi)                                        &                                                              &                                                               &                                                               &                                                         &                                                         &                                                               &                                                               &                                                                &                                                          \\
(Zr$_{0.3}$Nd$_{0.7}$)$_{2}$(Fe$_{0.75}$Co$_{0.25}$)$_{23}$Ti &                                                              &                                                               & 1.41$^{c}$ \cite{FU21}                                        &                                                         &                                                         & 1.7$^{c,d}$ \cite{FU21}                                       &                                                               &                                                                & 0.71$^{c}$ \cite{FU21}                                   \\ \hline
                                                              &                                                              &                                                               &                                                               &                                                         &                                                         &                                                               &                                                               &                                                                &                                                          \\
Nd$_{2}$Fe$_{14}$B                                            &                                                              &                                                               & 4.9$^{e}$ \cite{11}                                           &                                                         &                                                         &                                                               &                                                               &                                                                & 1.54$^{e}$ \cite{11}                                     \\
Sm$_{2}$Co$_{17}$                                             &                                                              &                                                               & 4.2$^{e}$ \cite{11}                                           &                                                         &                                                         &                                                               &                                                               &                                                                & 1.89$^{e}$ \cite{11}                                     \\
SmCo$_{5}$                                                    &                                                              &                                                               & 17.0$^{e}$ \cite{11}                                          &                                                         &                                                         &                                                               &                                                               &                                                                & 4.40$^{e}$ \cite{11}                                     \\
Alnico 5                                                      &                                                              &                                                               & 0.32$^{e}$ \cite{11}                                          &                                                         &                                                         &                                                               &                                                               &                                                                & 0.45$^{e}$ \cite{11}                                     \\ \hline
\end{tabular} 
\begin{tablenotes}[flushleft] 
{\footnotesize{\scriptsize \setstretch{0.25}
    \item * Theoretical GGA+$U$ references are represented by *. \newline
    \item ´ Theoretical LSDA references are represented by ´. \newline
    \item $^{a}$ Derived from magnetisation curve at 300 K. \newline
    \item $^{b}$ Single Crystal, Physical Property Measurement System (PPMS) measured at 10 K. \newline
    \item $^{c}$ Calculated from source data. \newline
    \item $^{d}$ Calculated from experimental $\mu_{0}M_{S}$ using the law of approaching saturation. \newline
    \item $^{e}$ Properties measured at 300 K.
    }}
\end{tablenotes}
\end{table}

    In the work of Erdmann \textit{et al.} \cite{step}, an increase in the anisotropy constant with increasing Ti concentration was observed. Based on this, the MAE of the likely stable compounds RFe$_{11}$Ti (R: Zr, (ZrNd), Nd) and (ZrNd)Fe$_{10}$CoTi are calculated.
    
    In case of the NdFe$_{11}$Ti compound, the DFT (GGA+$U$) calculation yields an anisotropy constant of 0.67~MJ/m$^{3}$. This is a significant underestimation of the experimental value of Herper \textit{et al.} \cite{p60} with 1.78~MJ/m$^{3}$. Therefore, the local spin density approximation (LSDA) was used to study the MCA. With LSDA, a significant increase of $K_{1}$ is obtained with 1.68~MJ/m$^{3}$. This improvement is in accordance with the work of Erdmann \textit{et al.} \cite{step}, providing values of 0.61 (GGA+$U$) and 1.41~MJ/m$^{3}$ (LSDA).

    According to the calculations, the substitution of Nd with Zr improves the MAE compared to NdFe$_{11}$Ti. The quaternary 1:12 compound (Zr,Nd)Fe$_{11}$Ti exhibits $K_{1}$ values of 2.16 (GGA+$U$) and 2.41~MJ/m$^{3}$ (LSDA), which are both larger than the experimental value of 1.78~MJ/m$^{3}$ of NdFe$_{11}$Ti. The complete Zr-based 1:12 phase with ZrFe$_{11}$Ti leads to another small increase in the $K_{1}$ value to 2.44 (GGA+$U$) and 2.47~MJ/m$^{3}$ (LSDA). Due to the lack of literature, an estimation of the Zr-containing phase is rather difficult to assess. In comparison the NdFe$_{11}$Ti phase shows a larger difference between the GGA+$U$ and LSDA values for $K_{1}$ than the ZrFe$_{11}$Ti phase. It should be noted that the difference between the LSDA value of (Zr,Nd)Fe$_{11}$Ti of 2.41~MJ/m$^{3}$ and the values of ZrFe$_{11}$Ti of 2.44 (GGA+$U$) and 2.47 (LSDA)~MJ/m$^{3}$ is very small. It appears that the GGA+$U$ method becomes more accurate as the Zr content increases. In case of the (Zr,Nd)Fe$_{11}$Ti compound, the GGA+$U$ value can be estimated to be more precise.
    
    In case of the effect of Co on $K_{1}$, it can be seen that the values decrease. This decrease can be seen by comparing the values of (Zr,Nd)Fe$_{11}$Ti with 2.16 (GGA+$U$) and 2.41 (LSDA) MJ/m$^{3}$ with those of (Zr,Nd)Fe$_{10}$CoTi with 1.97 (GGA+$U$) and 2.08 (LSDA) MJ/m$^{3}$. A rather small lowering effect of Co on $H_{a}$ and on $K_{1}$ was found by Suzuki \textit{et al.} \cite{FU24} as well.
    
\begin{figure}[h]
    \centering
    \includegraphics[scale=0.50]{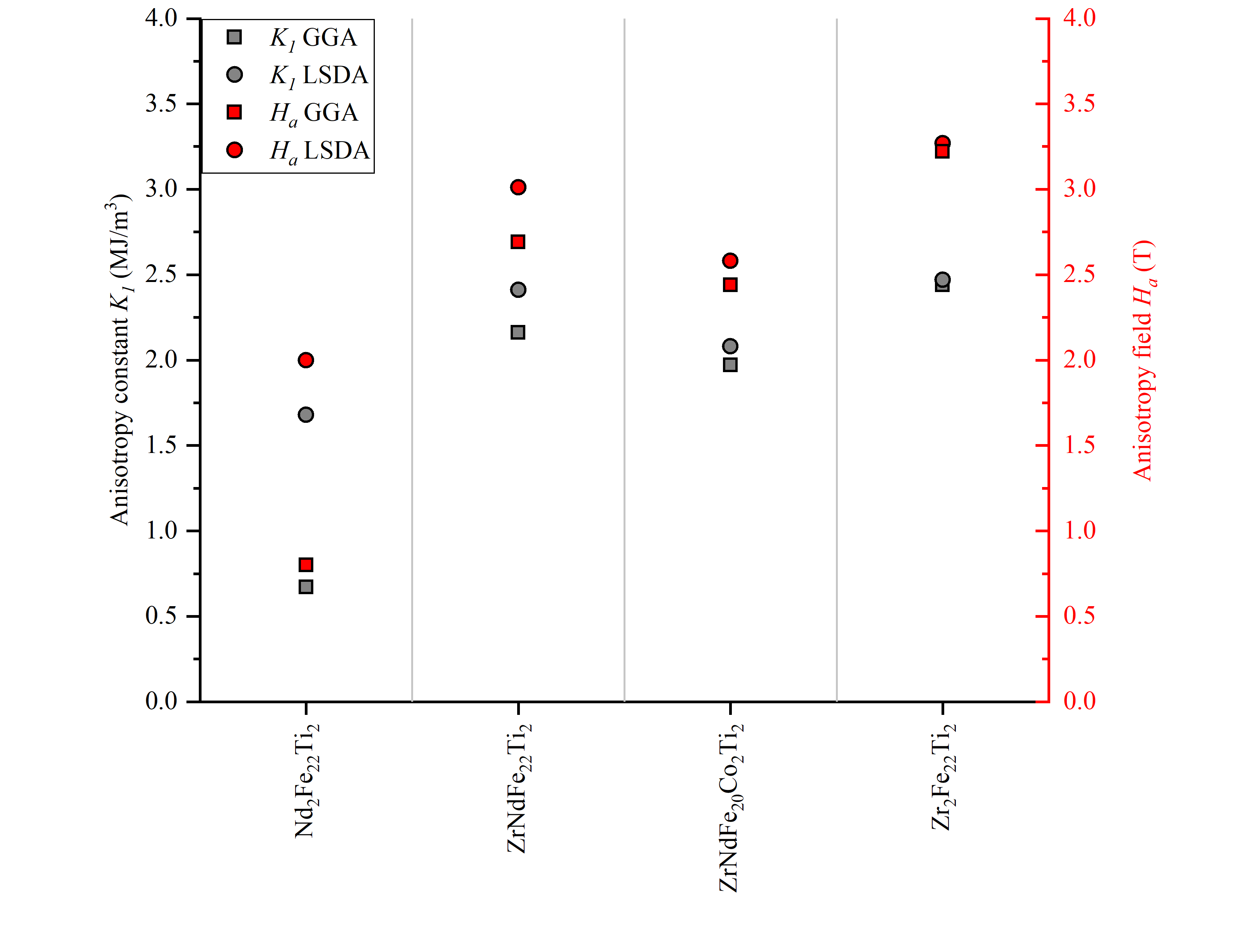}
    \caption{The values of the anisotropy constant $K_{1}$ and the anisotropy field $H_{a}$ are shown in black and red, respectively. The values calculated with the GGA+$U$ functional are shown as circles, while those using the LSDA functional are shown as squares.}
    \label{fig.23}
\end{figure}

    Furthermore, if the $K_{1}$ values of RFe$_{11}$Ti (R: Y, Zr, Ce and Nd) are compared, the following trend of the GGA+$U$ values can be observed with Nd (0.67~MJ/m$^{3}$) $<$ Y (0.92~MJ/m$^{3}$ \cite{step}) $<$ Ce (2.08~MJ/m$^{3}$ \cite{step}) $<$ Zr (2.16~MJ/m$^{3}$). From the LSDA values the same trend can be observed with Nd (1.68~MJ/m$^{3}$) $<$ Y (1.73~MJ/m$^{3}$ \cite{step}) $<$ Ce (2.12~MJ/m$^{3}$ \cite{step}) $<$ Zr (2.41~MJ/m$^{3}$). From these trends, Zr causes the greatest increase in the value of $K_{1}$. Moreover, the value obtained for the Zr-containing compounds can be considered rather reasonable, since it is not too far from that of the mentioned Ce compound. All considered quaternary compounds from this work and for comparison those from Erdmann \textit{et al.} \cite{step}, show the following trend of GGA+$U$ (and LSDA) treatments with YNd with 0.91 (1.65) \cite{step} $<$ CeNd with 1.35 (1.61) \cite{step} $<$ ZrNd with 2.16 (2.41)~MJ/m$^{3}$.

    As can be seen in Eq.~\ref{eq:2}, the anisotropy field $H_{a}$ is calculated from the anisotropy constant $K_{1}$ and the saturation magnetisation $M_{S}$. As in case of $K_{1}$, there is no literature for comparison of Zr-containing compounds. For NdFe$_{11}$Ti, the values can be compared with the experimental values obtained by Bouzidi \textit{et al.} \cite{p77} with 1.9 T and Akayama \textit{et al.} \cite{p45} with 2.0~T. It shows that the calculated value obtained from GGA+$U$ treatment of 0.80~T is well below literature. The LSDA treatment leads to a value of 2.00~T and is therefore in good agreement. For ZrFe$_{11}$Ti, (Zr,Nd)Fe$_{11}$Ti and (Zr,Nd)Fe$_{10}$CoTi, the GGA+$U$ (and LSDA) treatments yielded values of 3.22 (3.27), 2.69 (3.01) and 2.44 (2.58)~T, respectively. These values follow the same mentioned trends as for $K_{1}$. Similar trends can be seen when comparing the Y and Ce compounds from Erdmann \textit{et al.} \cite{step}. For the quaternary compounds, the value increases with GGA+$U$ (LSDA) treatment as follows: YNd with 1.19 (2.33) \cite{step} $<$ CeNd with 1.73 (2.30) \cite{step} $<$ ZrNd with 2.69 (3.01)~T.
    
    This section concludes that the MCA including $K_1$ and $H_a$ of the ZrNdFe$_{22}$Ti$_{2}$ compound is promising. According to calculations, the values are even higher than for quaternary Y and Ce compounds. This needs future experimental verification of this compound.

\subsection{Hardness Factor}
    With the anisotropy constant $K_{1}$ and the saturation magnetisation $M_{S}$, the magnetic hardness factor $\kappa$ can be calculated. This is done using the following equation:
\begin{equation}
    \kappa = \sqrt{\frac{K_{1}}{\mu_{0}M_{S}^{2}}}. \tag{\ref{eq:3}}
\end{equation}
    The resulting $\kappa$ value can be used to estimate the potential of a material as a permanent magnet. A magnetic compound can be classified as a soft ($\kappa < 0.1$), semi-hard ($0.1 < \kappa < 1$) or hard ($\kappa > 1$) magnet. With $\kappa > 1$ the material is suitable as a permanent magnet regardless of its shape, whereas in the semi-hard range ($0.1 < \kappa < 1$) a shape/dimension limitation as with Alnico magnets \cite{p72} has to be considered. 
    
    The $\kappa$ values calculated according to Eq.~\ref{eq:3} are shown in Tab.~\ref{tab.7}. Comparing the values from the earlier work of Erdmann \textit{et al.} \cite{step} of NdFe$_{11}$Ti (0.52 (GGA+$U$), 0.88 (LSDA)) with the newly calculated values (0.49 (GGA+$U$), 0.77 (LSDA)), a good theoretical agreement can be seen. Looking at the values of RFe$_{11}$Ti (R: Y, Zr, Ce and Nd), it is shown that Zr with 1.04 increases the $\kappa$ value less in case of LSDA than Y or Ce with 1.18 \cite{step} and 1.31 \cite{step}, respectively. Comparing the Nd (0.77) and Zr (1.04) compounds, the $\kappa$ value of Zr is higher. In case of GGA+$U$ treatment the value of Zr with 1.03 is higher than that of Nd with 0.49 and Y with 0.71 \cite{step}, but lower than that of Ce with 1.12 \cite{step}. Thus, Zr improves the hardness factor in comparison to the Nd compound, but increases the value less than Y or Ce.

    The calculated $\kappa$-values of the quaternary compound (Zr,Nd)Fe$_{11}$Ti are 0.92 (GGA +$U$) and 0.97 (LSDA). As there is no direct literature on the Zr compound, it has to be classified as a semi-hard magnet for the time being, as it gives values $\kappa < 1.00$. In comparison to the quaternary compounds (Y,Nd)Fe$_{11}$Ti (0.67 (GGA+U) and 1.02 (LSDA)) and (Ce,Nd)Fe$_{11}$Ti (0.83 (GGA+U) and 1.02 (LSDA)) \cite{step}, the Zr compound has a comparable or slightly lower hardness factor. It can be seen that GGA+$U$ gives better values in case of Zr than for Y or Ce. This is indicated by the smaller difference between the GGA+$U$ and LSDA values in case of Zr.
    
\begin{figure}[h]
    \centering
    \includegraphics[scale=0.50]{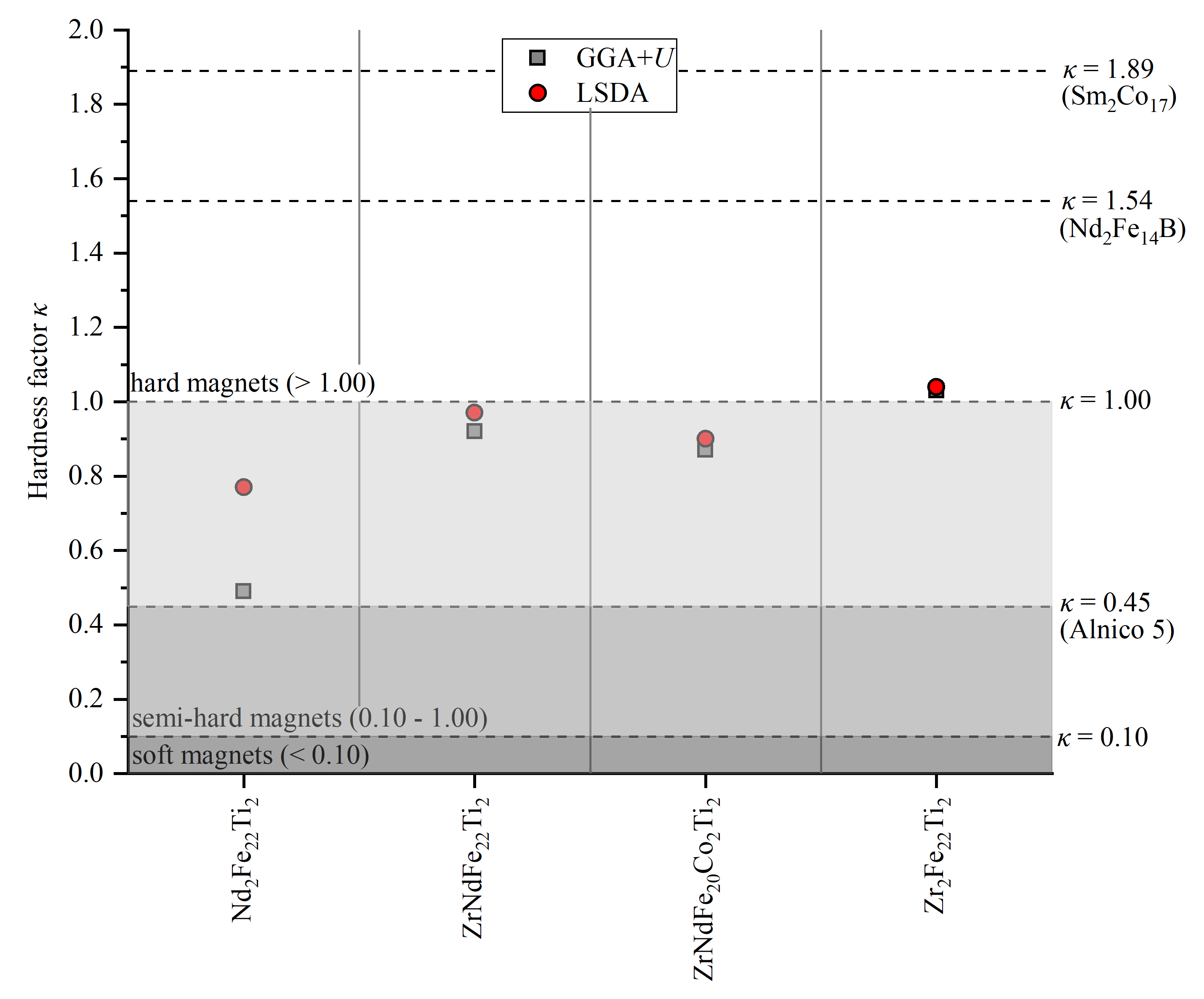}
    \caption{Calculated hardness factors $\kappa$ for GGA+$U$ (squares) and LSDA (circles) functionals. The magnetic hardness factors of well known magnets are shown by horizontal dashed lines.}
    \label{fig.24}
\end{figure}

    As in the case of $K_{1}$ and $H_{a}$ values, the Co substitution lowers the $\kappa$ value. Since the $\kappa$ values of (Zr,Nd)Fe$_{10}$CoTi are 0.87 (GGA+$U$) and 0.90 (LSDA), both are lower than those of the (Zr,Nd)Fe$_{11}$Ti compound with 0.92 (GGA+$U$) and 0.97 (LSDA). As there is no comparable literature for $K_{1}$, $H_{a}$ and $\kappa$ for (Zr,Nd)Fe$_{11}$Ti or (Zr,Nd)Fe$_{10}$CoTi, these values can only be roughly compared with the similar  (Zr$_{0.3}$Nd$_{0.7}$)(Fe$_{0.75}$Co$_{0.25}$)$_{11.5}$Ti$_{0.5}$ compound. The literature sample reaches 1.41 MJ/m$^{3}$ for $K_{1}$, 1.7 T for $H_{a}$ and a $\kappa$ value of 0.71 \cite{FU21}. Therefore, the values of the quaternary and quinary compounds of this work are significantly higher than those of the mentioned literature compound. The deviations can be explained by a higher Zr content as well as a higher Ti and lower Co concentration of the chosen compounds. Thus, a good accordance with the limited available literature can be assumed.
    
    In summary, the hardness factor of the ZrNdFe$_{22}$Ti$_{2}$ compound is lower than the values of the quaternary compounds containing Y and Ce. This is because the calculations yield values for $\kappa$ that are just below 1. However, the computed values of 0.92 (GGA+$U$) and 0.97 (LSDA) are very close to $\kappa = 1$, requiring experimental verification.

\newpage
\thispagestyle{plain}
\section{Conclusion and Outlook}
    In summary, starting from the promising findings on the phase stability and magnetic properties of Nd-lean permanent magnets such as (X,Nd)Fe$_{12-y}$Ti$_{y}$ (X: Y and Ce; y: $0 \leq y \leq 1$) \cite{step, 35}, systematic investigations on the intrinsic magnetic properties of (Zr,Nd)Fe$_{12-y}$Ti$_{y}$ compounds were performed. Zr has been chosen as an alternative to Nd, being an abundant, less expensive and less critical material to be procured. Since there was no available literature on the chosen Zr-containing phases, only the NdFe$_{12-y}$Ti$_{y}$ compounds were compared with literature directly. However, the Zr-containing phases were compared with the similar (Zr$_{0.3}$Nd$_{0.7}$)(Fe$_{0.75}$Co$_{0.25}$)$_{11.5}$Ti$_{0.5}$ compound. As Co is substituted together with Zr in the literature \cite{2016, FU21, FU22, FU24}, this is also discussed by considering the compound (Zr,Nd)Fe$_{10}$CoTi. The resulting good agreement of the literature with the calculated values gives great confidence in the presented magnetic properties of investigated compounds.

    The used methodology correctly predicts the expected overall magnetisation trend of the 1:12 phases, with the values of the NdFe$_{12-y}$Ti$_{y}$ compounds being larger than those of the ZrFe$_{12-y}$Ti$_{y}$ (y: $0 \leq y \leq 1$) compounds. Experimentally determined values for the saturation magnetisation $\mu_{0}M_{S}$ of the NdFe$_{11}$Ti compound are in the range of 1.38 to 1.70 T \cite{p45, p54, p56, p60} depending on the measurement temperature and method. The calculated values using DFT (DFT+$U$) are 1.68 (1.69)~T. In case of ZrFe$_{11}$Ti and (Zr,Nd)Fe$_{11}$Ti compounds there is no direct literature to compare. The $\mu_{0}M_{S}$ values calculated by DFT (DFT+$U$) are 1.51~T for ZrFe$_{11}$Ti and 1.60 (1.62) T for (Zr,Nd)Fe$_{11}$Ti. A comparison of the value of (Zr,Nd)Fe$_{11}$Ti with a similar compound such as (Zr$_{0.3}$Nd$_{0.7}$)(Fe$_{0.75}$Co$_{0.25}$)$_{11.5}$Ti$_{0.5}$ with the value of $\mu_{0}M_{S}$ of 1.63~T \cite{FU24} shows a good agreement. 
    
    Next to the magnetisation calculations promising $|BH|_{max}$ values were obtained for (Zr,Nd)Fe$_{11}$Ti, whose DFT (DFT+$U$) value is 510 kJ/m$^{3}$ (525 kJ/m$^{3}$). These values are higher than that of Sm$_{2}$Co$_{17}$B with 294~kJ/m$^{3}$ \cite{10}. Moreover, the quaternary compound (Zr,Nd)Fe$_{11}$Ti exceeds the values of 509~kJ/m$^{3}$ and 490~kJ/m$^{3}$ for (Y,Nd)Fe$_{11}$Ti and (Ce,Nd)Fe$_{11}$Ti \cite{step} as calculated from Erdmann \textit{et al.}. It must be noted that even higher $|BH|_{max}$ values can be obtained by using a lower Ti content in the quaternary compound (Zr,Nd)Fe$_{11.5}$Ti$_{0.5}$. The calculated values of the (Zr,Nd)Fe$_{11.5}$Ti$_{0.5}$ compound are 600 and 603 (DFT+$U$) kJ/m$^{3}$. With these values, the quaternary Zr-containing compound clearly exceeds the value of Nd$_{2}$Fe$_{14}$B with 512 kJ/m$^{3}$ \cite{10}. The substitution of Co into the (Zr,Nd)Fe$_{11}$Ti compound at 7.7~at.\%, giving the (Zr,Nd)Fe$_{10}$CoTi compound, results in a slight increase in saturation magnetization and a minimal decrease in the $|BH|_{max}$ value.

    In addition, mean field approximation (MFA) was used to calculate the Curie temperature, where a systematic overestimation was observed in the ferromagnetic (FM) state. Nevertheless, the values obtained are in quantitative agreement with the literature. As Miyake \textit{et al.} \cite{p71} reported, better results are obtained for the REFe$_{11}$Ti compounds using the local moment disordered (LMD) state which can be supported by NdFe$_{12-y}$Ti$_{y}$ compounds (y: $0 \leq y \leq 1$). In case of the ZrFe$_{12-y}$Ti$_{y}$ (y: $0 \leq y \leq 1$) compounds, the FM states appear to be more applicable. Note that FM-based calculations tend to overestimate (margin of error for the ternary compounds is about 45\%). The calculated Curie temperatures of the Nd-lean quaternary compound (Zr,Nd)Fe$_{11}$Ti are 783 (FM) and 652 (LMD)~K. In consequence, this compound has a higher $T_{C}$ than Nd$_{2}$Fe$_{14}$B (588 K). The $T_{C}$ can further be increased by Co substitution. According to the calculations an increase of about 85~K with a substitution of about 7.7 at\% Co from (Zr,Nd)Fe$_{11}$Ti to (Zr,Nd)Fe$_{10}$CoTi is achievable.

    For NdFe$_{11}$Ti, the exchange correlation function based on the generalised gradient approximation (GGA) fails to calculate the magnetocrystalline anisotropy constant $K_{1}$, as already reported in the literature \cite{p53, p60}. This is also the reason why the Local Spin Density Approximation (LSDA) has been used to obtain more accurate data. As the Zr concentration increases, GGA+$U$ supplies better results. The addition of Zr (Co) leads to an increase (small decrease) of the magnetocrystalline anisotropy. The found effects of Co substitution are supported by the work of Suzuki \textit{et al.} \cite{FU24}. From $K_{1}$ and $\mu_{0}M_{S}$, the hardness factor $\kappa$ can be calculated, which indicates the potential of a material to develop into a permanent magnet for which $\kappa > 1$ is desired. The 50\% Nd-lean quaternary compound (Zr,Nd)Fe$_{11}$Ti gives a theoretical $\kappa$ value of 0.97, classifying it as a semi-hard magnet. In order to obtain a higher $\kappa$ value, nitrogen substitution is a suitable option, as shown in previous works \cite{step, FU22}.
    
    In summary, the thesis shows that the replacement of the critical RE element Nd by more abundant Zr in ThMn$_{12}$ compounds provides possible and promising RE-TM magnets, which have comparable or only slightly lower magnetic properties than the currently utilized Nd-Fe-B magnets. Thus, the new quaternary (ZrNd)Fe$_{24-y}$Ti$_{y}$ compounds can be considered as a potential alternative to the Nd-Fe-B magnets. In parallel, the findings of this work suggest the investigation of the suggested quarternary ZrNdFe$_{24-y}$Ti$_{y}$ hard/semi-hard magnets, in order to confirm the theoretical calculations via experimental measurements.
    
    As an outlook to this study, further investigation of other yet uninvestigated elements, besides Zr, using a combination of DFT and KKR calculations can be considered for substitution with Nd atoms. Beyond that, an addition or another substitution to Zr, such as nitrogenation, may be possible to compensate for the weaknesses of the (ZrNd)Fe$_{24-y}$Ti$_{y}$ compounds. An alternative approach could involve the expansion of research to encompass other established magnetic structures, such as the 2:14 or 2:17 phases, in order to explore the potential decrease in RE elements within these compounds. The quaternary compounds studied have similar T$_{C}$'s as the Nd-Fe-B magnets, which makes them very attractive from a technological point of view. Subsequently, they could be used in almost the same range of applications as the Nd-Fe-B magnets, when slightly lower magnetic properties are acceptable.
    
\newpage
\thispagestyle{plain}
\addcontentsline{toc}{section}{References}
\printbibliography

\newpage
\thispagestyle{plain}
\addcontentsline{toc}{section}{List of Abbreviations}
\paragraph{\Large{List of Abbreviations}}\mbox{}\vspace{0.5cm}\\
RE - Rare-earth \\
TM - Transition metal \\
$m_{tot}$ - Total magnetic moment \\
\textit{\textmu}$_{0}$ - Magnetic constant \\
\textit{\textmu}$_{B}$ - Bohr magneton \\
$H$ - Magnetic field \\
$H_{a}$ - Anisotropy field \\
$H_{c}$ - Coercivity field \\
$H_{d}$ - Internal magnetic field \\
$M$ - Magnetisation \\
$M_{r}$ - Remanence magnetisation \\
$M_{S}$ - Magnetisation saturation \\
$M(T)$ - Finite temperature magnetisation \\
$T_{C}$ - Curie temperature \\
$|BH|_{max}$ - Maximum energy product \\
MCA - Magnetocrystalline anisotropy \\
MAE - Magnetocrystalline anisotropy energy \\
$\Delta _{MAE}$ - Anisotropy energy \\
$K_{1}$ - Anisotropy constant 1$^{st}$ order \\
$\kappa$ - Hardness factor \\
HF - Hartree-Fock \\
post-HF - post Hartree-Fock \\
DFT - Density functional theory \\
DFT +\textit{U} - Density functional theory with applied Hubbard \textit{U} correction \\
DMFT - Dynamical mean field theory \\
LDA - Local Density Approximation \\
LSDA - Local Spin Density Approximation \\
GGA - Generalised Gradient Approximation \\
PW91 - Functional by Perdew and Wang \\
PBE - Functional by Perdew, Burke and Ernzerhof \\
PAW - Projector Augmented Wave \\
USPP - Ultrasoft Pseudopotentials \\
VASP - Vienna \textit{ab initio} Simulation Package \\
KKR - Korringa-Kohn-Rostoker \\
SCF - self-consistent field \\
FM - Ferromagnetic \\
LMD - Local Moment Disorder \\
DLM - Disorder local moment \\
MFA - Mean-field approximation \\
ASA - Atomic sphere approximation \\
CPA - Coherent potential approximation \\
MJW - Moruzzi, Janak and Williams \\
BCC - Body-Centered Cubic \\
HCP - Hexagonally close packed \\
TB-LMTO-ASA - Tight-binding linear-muffin-tin-orbital atomic-sphere-approximation \\
DOS - Density of states \\
XRD - X-Ray diffraction \\
PPMS - Physical property measurement system \\
VSM - Vibrating sample magnetometer \\

\newpage

\thispagestyle{plain}
\addcontentsline{toc}{section}{Erklärung}
\paragraph{Erklärung}$~$ \vspace{1cm}\\
Hiermit versichere ich an Eides statt, dass ich diese Arbeit selbstständig verfasst und keine
anderen als die angegebenen Quellen und Hilfsmittel benutzt habe. Außerdem versichere
ich, dass ich die allgemeinen Prinzipien wissenschaftlicher Arbeit und Veröffentlichung,
wie sie in den Leitlinien guter wissenschaftlicher Praxis der Carl von Ossietzky Universität
Oldenburg festgehalten sind, befolgt habe.
\vspace{2cm}
\\
\begin{tabular}{@{}l@{}}\hline
Nico Yannik Merkt
\end{tabular}

\end{document}